\documentclass[]{aastex631}

\usepackage{amsmath}
\usepackage{float}
\usepackage{afterpage}
\afterpage{\clearpage}

\graphicspath{{./}{figures/}}

\NewPageAfterKeywords

\begin{document}

\title{An Imaging and Spectroscopic Exploration of the Dusty Compact Obscured Nucleus Galaxy Zw~049.057 \footnote {Based in part on observations obtained with the Southern African Large Telescope (SALT)}}

\author{J. S. Gallagher}
\altaffiliation{Department of Physics, Macalester College, 1600 Grand Ave, St. Paul, MN 55105, USA}
\affiliation{Department of Astronomy, University of Wisconsin-Madison, 475 N Charter Street, Madison, WI 53706, USA}

\author{R. Kotulla}
\affiliation{Department of Astronomy, University of Wisconsin-Madison, 475 N Charter Street, Madison, WI 53706, USA}

\author{L. Laufman}
\affiliation{School of Physics and Astronomy, University of Minnesota 116 Church Street S.E., Minneapolis, MN 55455, USA}

\author{E. Geist}
\affiliation{Department of Astronomy, University of Wisconsin-Madison, 475 N Charter Street, Madison, WI 53706, USA}

\author{S. Aalto}
\affiliation{Department of Space, Earth and Environment, Chalmers University of Technology, Onsala Space Observatory, 439 92 Onsala, Sweden}

\author{N. Falstad}
\affiliation{Department of Space, Earth and Environment, Chalmers University of Technology, Onsala Space Observatory, 439 92 Onsala, Sweden}

\author{S. K\"{o}nig}
\affiliation{Department of Space, Earth and Environment, Chalmers University of Technology, Onsala Space Observatory, 439 92 Onsala, Sweden}

\author{J. Krause}
\affiliation{Space Telescope Science Institute, 3700 San Martin Dr, Baltimore, MD 21218 USA}

\author{G. C. Privon}
\altaffiliation{Department of Astronomy, 530 McCormick Road, University of Virginia, Charlottesville, VA 22904, USA and Department of Astronomy, University of Florida, P. O. Box 112055, Gainesville, FL 32611, USA}
\affiliation{National Radio Astronomy Observatory, 520 Edgemont Rd, Charlottesville, VA 22903, USA}

\author{C. Wethers}
\affiliation{Department of Space, Earth and Environment, Chalmers University of Technology, Onsala Space Observatory, 439 92 Onsala, Sweden}

\author{A. Evans}
\affiliation{Department of Astronomy, 530 McCormick Road, University of Virginia, Charlottesville, VA 22904, USA}
\affiliation{National Radio Astronomy Observatory, 520 Edgemont Rd, Charlottesville, VA 22903, USA}

\author{M. Gorski}
\affiliation{Center for Interdisciplinary Exploration and Research in Astrophysics, Northwestern University, 
1800 Sherman Ave, Evanston, IL 60201, USA}

\begin{abstract}
Zw~049.057 is a moderate mass, dusty, early-type galaxy that hosts a powerful compact obscured nucleus (CON, L$_{FIR,CON} \geq$10$^{11}$~L$_{\odot}$). The resolution of HST enabled  measurements of the stellar light distribution and characterization of dust features. Zw~049.057 is inclined with a prominent three zone disk; the R$\approx$ 1kpc star forming inner dusty disk contains molecular gas, a main disk with less dust and an older stellar population, and a newly detected outer stellar region at R$>$6~kpc with circular isophotes. Previously unknown polar dust lanes are  signatures of a past minor merger that could have warped the outer disk to near face-on. Dust transmission measurements provide lower limit gas mass estimates for dust features. An extended region with moderate optical depth and M$\geq$ 2$\times$10$^8$~M$_{\odot}$ obscures the central 2~kpc. Optical spectra show strong interstellar Na~D absorption with a constant velocity across the main disk, likely arising in this extraplanar medium. Opacity measurements of the two linear dust features, pillars, give a total mass of $\geq$10$^6$~M$_{\odot}$, flow rates of $\geq$2~M$_{\odot}$~yr$^{-1}$, and few Myr flow times. Dust pillars are associated with the CON and are visible signs of its role in driving large-scale feedback. Our assessments of feedback processes suggest gas recycling sustains the CON. However, radiation pressure driven mass loss and efficient star formation must be avoided for the AGN to retain sufficient gas over its lifespan to produce substantial mass growth of the central black hole.
\end{abstract}


 \section{Introduction}\label{introduction}
 
The development of the baryonic components of galaxies involves the coevolution of central super massive black holes (SMBHs) with the stellar galactic bodies. These processes involve feedback from active galactic nuclei (AGN) that can remove gas via a variety of processes  in concert with mass inflows that feed growth of the central black holes and power AGNs. Galaxies with “compact obscured nuclei”  or CONs are systems with extraordinary central concentrations of dusty interstellar matter within radii of $\lesssim$100~pc \citep[see][]{Evans01,Sakamoto10,Aalto15,Martin16,Privon17,Scoville17,Aalto20,Falstad21,Donnan23,Gorski23}.  Gas columns in CONs are $\sim$10$^{25}$~cm$^{-2}$ which obscures nuclear emission across the spectrum, extending from  x-rays to millimeter wavelengths. Therefore CONs offer insights into nuclear growth and feedback under extreme conditions where the immediately surrounding gas reservoirs are unusually dense and massive. Due to their high luminosities, CONs expel dense, dusty gas that absorbs background galaxy light. Spatially extended dusty gas produced by these dramatic feedback processes can be observed with the Hubble Space telescope (HST)  in absorption at gas column densities and spatial scales that are difficult to access via other methods.

The CON host Zw~049.057 is an inclined early-type galaxy with an inner dusty star-forming disk \citep{Martin88,Scoville00,AlonsoH06}. This galaxy was cataloged on blue-sensitive photographic plates by Zwicky, who characterized it as a compact early-type galaxy. Later ground-based imaging and spectroscopy led to the classification of Zw~049.057 as a non-interacting starburst galaxy \citep[e.g., ][]{Poggianti00,Hattori04,AlonsoH06,Stierwalt13,Larson16}. Observations of the global H{\rm\,I} 21~cm line profile revealed that Zw~049.057 belongs to the rare group of LIRGs where H{\rm\,I} is in absorption \citep{Baan87,Mirabel88,Morganti18}. Following the IRAS detection of Zw~049.057 as a luminous infrared galaxy (LIRG) \citep{Soifer87}, the galaxy attracted interest in terms of its molecular content, including early detection as an OH maser source \citep{Baan87}. Improved millimeter and submillimeter observations demonstrated the extreme nature of the central molecular gas component of Zw~049.057 \citep{Baan87,Planesas91,Baan08}. Similarly a variety of observations established that the central molecular medium contains high density molecular matter \citep{Mangum08,Papadopoulos12,Mangum13,Baan17,HerreroIllana19} which includes a warm component with temperatures of $\approx$150~K \citep{Magnum13b,Petric18}, along with evidence for even hotter dust components \citep{GonzalezAlfonso19,Baba22}.

High angular resolution observations with mm/submm interferometers provide superb information on properties of the dense, bright nuclear regions, and demonstrate that emission from CONs provide a substantial fraction of the host galaxies' bolometric luminosities \citep{Costagliola15,Privon17,GonzalezAlfonso19,Falstad21,GarciaBernete22,Baba22}. Due to the extreme opacities of CONs, the nature of their power sources is unclear and AGN, extreme starbursts, or gravitational energy from gas accreting into the CON are options \citep{Aalto15,Aalto19,Aalto20,Gorski24}. Although classic  ultraluminous infrared galaxy (ULIRG) CONs, such as that in Arp~220, are associated with ongoing mergers the evolutionary channels leading to CONs in LIRGs are not fully understood. About 30\% of ULIRGs and 10-20\% of the surveyed luminous infrared galaxies (LIRGs) contain CONs \citep{Falstad21,Donnan23,Nishimura24}.  In this paper we present a study of the optical structure of the LIRG CON Zw~049.057 (CGCG~049-057, PGC054330; D=59~Mpc,  L$_{FIR} \approx$11.5~$\pm0.1$ L$_{\odot}$) \citep{Aalto15,Falstad15}. Due to its orientation and relative proximity,  Zw~049.057 is an excellent target for exploring connections between a CON and its host galaxy.

At optical wavelengths Zw~049.05 stands out due to the presence of prominent dust absorption features along the minor axis that were seen in the ground-based images by \cite{Martin88} and later confirmed using HST by \cite{Scoville00}. The minor axis dust features in Zw~049.057 contrast with most early-type disk galaxies (e.g., types S0 or S0/a) where dust absorption is largely confined to disks, either in the central regions or appearing as filaments due to disk warps \citep[e.g.,][]{Tran01,Keel15,Boizelle17}. The most prominent dust features, two narrow ``dust pillars"\footnote{We use the term ``pillar" to differentiate the linear central dust features in Zw~049.057 from more commonly observed dust filaments.}, connect to the central regions of Zw~049.057 \citep{Scoville00} tracing ongoing gas outflows and feedback associated with its nuclear region. Zw~049.057 is a CON host galaxy that also is in the midst of a galactic-scale dust storm.

Observations obtained with the {\it Herschel} observatory added to the initial evidence for a massive warm, dense central molecular component and associated molecular wind in Zw~049.057 \citep{Greve14,Rosenberg15,Kamenetzky17,Lu17}. Molecular line spectra obtained by \cite{Aalto15} revealed significant emission from vibrationally excited nuclear HCN emission from a warm, dense, and optically thick molecular medium which is the primary observational signature of a CON \citep[e.g.,][]{Imanishi13,Aalto15,Falstad21,Sakamoto21,Nishimura24}. \cite{Falstad15} also showed that the HCN emission arises from a very compact central region with a high gas column density, \cite{Falstad18} detected a nuclear molecular wind.  \cite{GonzalezAlfonso19} quantitatively interpreted the infrared and molecular line observations of Zw~049.057 with a radiative transfer model for the optically thick central dusty region. Their model yields a size of 15-25~pc for the deeply obscured region that provides $\sim$1/3 - 1/2 of L$_{FIR}$ in Zw~049.057. The nucleus is behind an H$_2$ column density of log$_{10}$(N(H$_2$) $\simeq 10^{24.8}$~cm$^{-2}$ that yields an atomic H gas column density of N$_H$ $\simeq$ 10$^{25.1}$~cm$^{-2}$ giving an ISM mass surface density towards the nucleus of an impressive $\sim$20~g~cm$^{-2}$. Sub-arcsecond radio continuum observations by \cite{Song22} penetrate the dense nuclear gas and dust envelope to reveal a compact continuum source that is consistent with the presence of an AGN within the Zw~049.057 CON.

\begin{figure}[]
   \centering
    \includegraphics[width=0.95\textwidth]{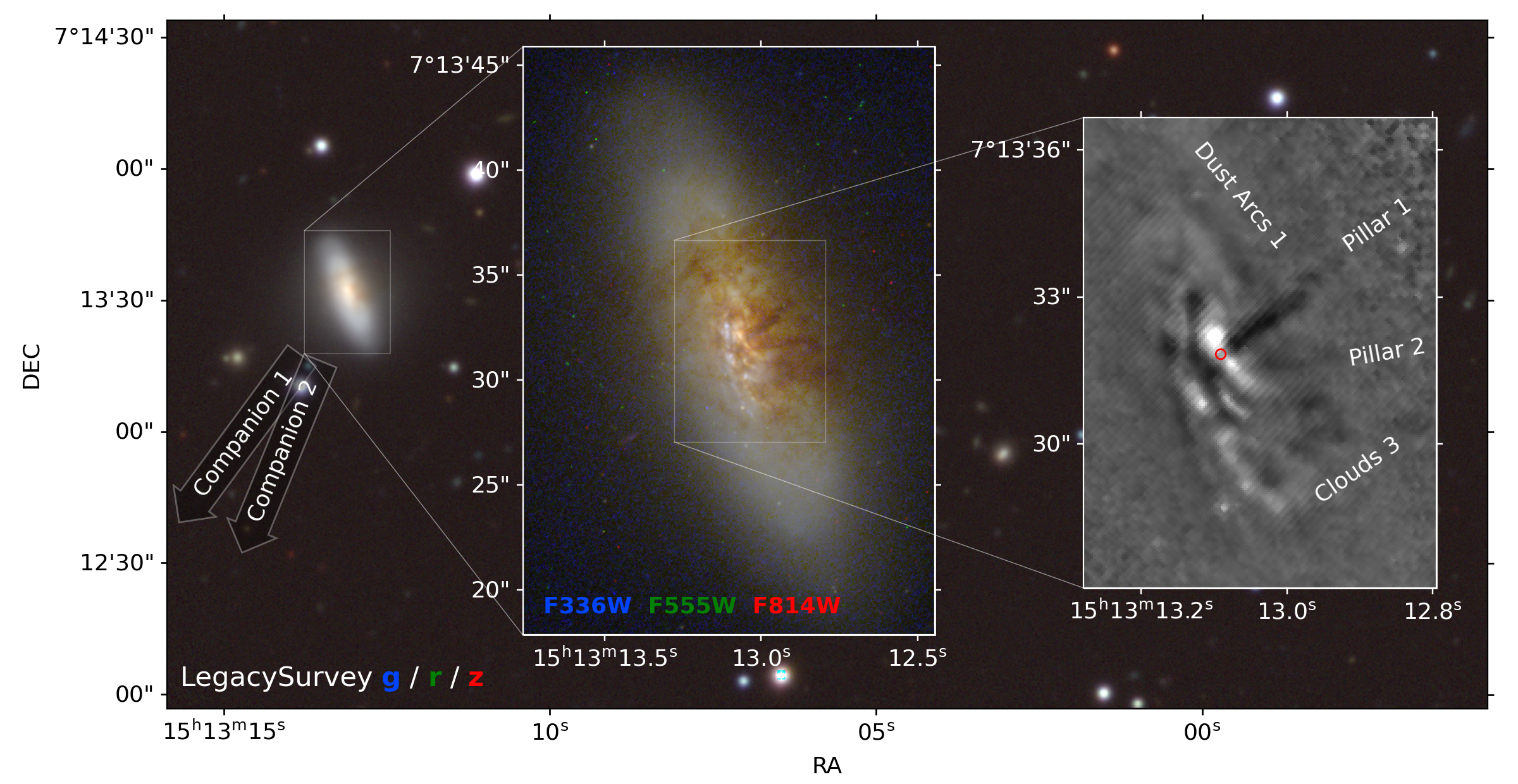}\\
    \includegraphics[width=0.95\textwidth]{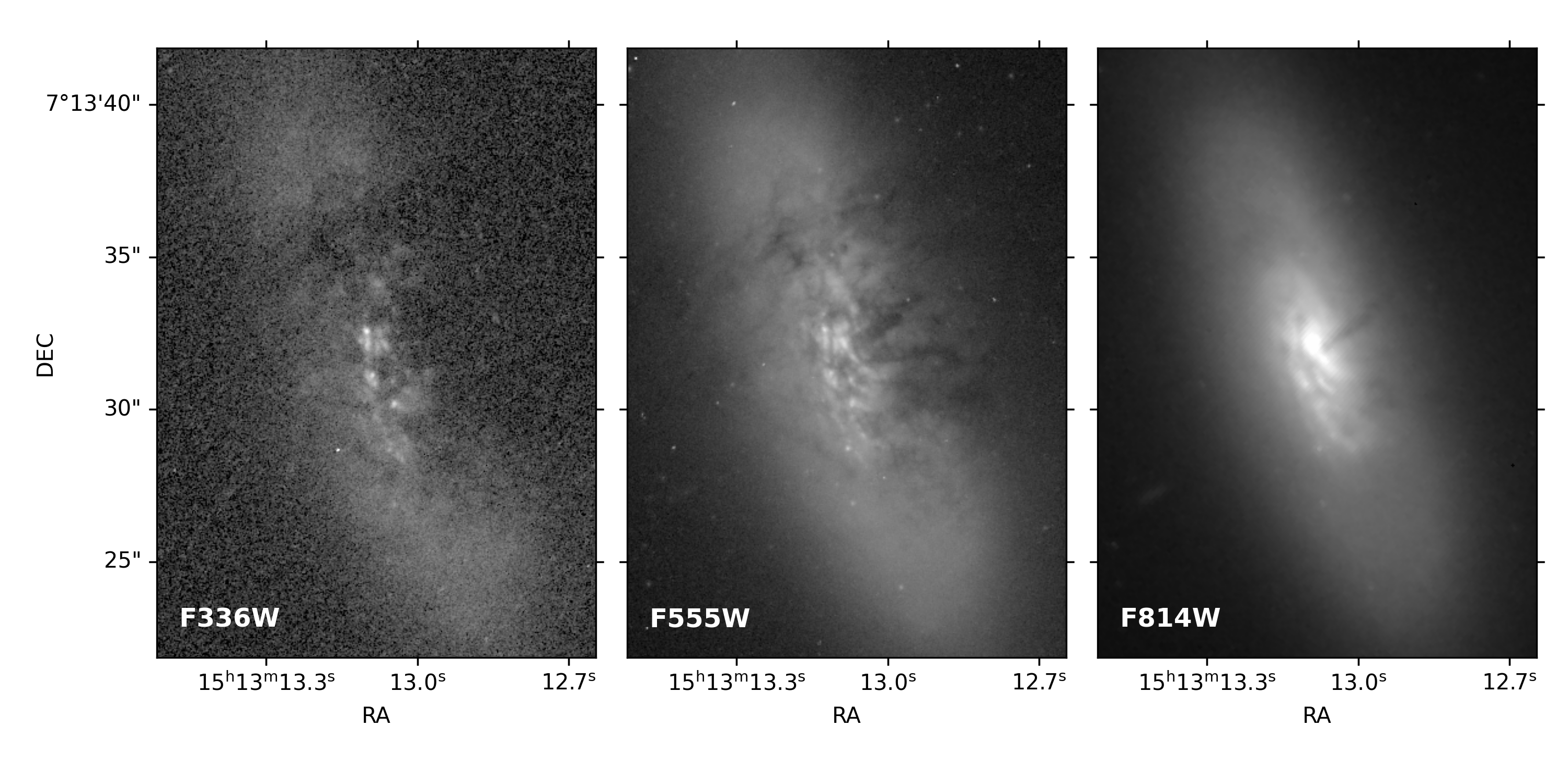}\hfill
   \caption{{\it Upper:} This figure illustrates our approach in analyzing the properties of Zw~049.057. The images from left to right are are from NOIRlab Legacy Survey's Sky Viewer (www.legacysurvey.org), our HST optical images, and a high-pass smoothed version of our F125W HST J-band image. The arrows on the left panel point to the southeasterly locations of the Zw~049.057 low luminosity companions. Dust Pillars-1 and -2 make up the distinctive dust ``V". The central image also shows the polar dust ring band crossing to the north of the central region. Our approach in this paper is to work from larger to smaller spatial scales, starting with the environment and global properties of Zw~049.057 and ending with characteristics of the nuclear region. The location of the CON is denoted by the red circle. {\it Lower:}  Images in 3 filters illustrate effects of the decrease in optical depth with increasing wavelength. Diffuse dust complexes obscure much of the center of the galaxy in the HST F336W  U$_{HST}$ filter but are largely transparent in the F814W I$_{HST}$ filter.}
   \label{fig:zw049_general}
\end{figure}

We selected Zw~049.057 for a detailed exploration of the properties of a CON host galaxy due to its proximity and extensive set of existing multi-wavelength observations. Our study focused on defining relationships between the Zw~049.057 CON and its galactic surroundings by assessing the properties of the extensive dust absorption features using images obtained with the Wide Field Camera 3 (WFC3) on the Hubble Space Telescope (HST). By mapping the distribution of dust absorption in structures extending outwards from the central regions, we investigated connections between the sub-kpc molecular outflows from the CON and larger kpc scale gas structures outlined by dust absorption. These data also allowed us to search for and find evidence of a recent minor merger. A spectrum obtained with the South African Large Telescope (SALT) enabled us to undertake an initial study of the kinematics of ionized gas in the dusty circumnuclear star forming region. The spatial outwards-inwards organization of the paper is visually outlined in the upper panel of Figure~\ref{fig:zw049_general} that includes labels for several of the major dust absorption features.
 
 In \S \ref{sec:observations} we review the observational material. The structure of the stellar components and stellar population properties of Zw~049.057 are discussed along with its local environment in \S \ref{sec:structphotom}. In \S \ref{sec:dust} properties of the dust features are analyzed.  Section \ref{sec:centerspec} describes the emission line properties of the inner region of Zw~049.057 where a high star SFR disk component is present. Section \ref{sec:nucleus} explores properties of the nuclear region and the associated giant ``dust pillars"--linear dust absorption features--that extend above the disk of Zw~049.057 \citep[see][]{Scoville00}. 
 A discussion of our results in the context of evolution and feedback associated with the CON in Zw~049.057 is in \S \ref{sec:discussion} that lead to our conclusions in \S \ref{sec:conclusions}.
 
 \section{Observations}\label{sec:observations}

 \subsection {Hubble Space Telescope WFC3 Imaging}\label{hstobs}
 
At ground-based resolution optical images confirm Zw~049.057 is an S0/a system with asymmetrical dust obscuration above the disk to the north-northwest. The high brightness central region stands out and is probably the basis for the Zwicky classification as a compact galaxy. Unfortunately this system is sufficiently distant that ground-based, seeing-limited images only provide a broad overview of its structure. We therefore obtained images of Zw~049.057 on March 5, 2017 using the Wide Field Camera 3 (WFC3) on the Hubble Space Telescope (HST) as a part of program GO-14278, P.I. J. S. Gallagher. At the distance of Zw~049.057, the WFC3 optical full width at half maximum (fwhm) point source resolution of 0.07~arcsec corresponds to a linear scale of 20~pc. The observations are summarized in Table~\ref{tab:wfc3obs}. The observations in this table are available via MAST at \dataset[DOI:10.17909/GXFM-6061]{https://doi.org/10.17909/GXFM-6061}. Images in the 3 optical filters are shown in Figure~\ref{fig:zw049_general}. Using HST, we can study the interstellar medium as traced by dust obscuration at gas coluumn densities of $\sim$10$^{21}$~cm$^{-2}$ on spatial scales where molecular and dust emission is too faint to be readily measured with millimeter/submillimeter interferometers.\footnote{The James Webb Space Telescope offers the potential to map inner dust near- and mid-infrared emission with subarcsecond resolution for regions emitting PAH's or where dust is sufficiently warm (most likely to occur near the nucleus and in starbursting zones)}. 

\begin{deluxetable}{ccrc}[H]
\tablecaption{WFC3 HST Observations of Zw~049.057 \label{tab:wfc3obs}}
\tablehead{
\colhead{Filter} & \colhead{Channel} & \colhead{T$_{expose}$ (s)} & \colhead{Data Set}
}
\startdata
F336W & WFC3/UVIS & 2550 & IDA603010 \\
F555W & WFC3/UVIS & 960  & IDA603020 \\
F814W & WFC3/UVIS & 1000 & IDA603030 \\
F125W & WFC3/IR &  306   & IDA603040 \\
\enddata
\end{deluxetable}  

We compared the coordinates on the original reduced WFC3 image frames to the Gaia reference frame using stars on the WFC3 chip 1 where the galaxy was observed and in the WFC3/IR channel. This comparison indicated coordinate offsets of several tenths of an arcsecond that are corrected in the latest versions of the WFC3/UVIS channel Zw~049.057 data in the MAST 
archive. Data with updated coordinates allowed us to determine the location of the nucleus and other features and to calibrate astrometric positions in the F125W image, where the archival image has offsets from the Gaia reference frame. This comparison also showed that the original NICMOS of \cite{Scoville00} images had a position offset of $\sim$0.6~arcseconds measured relative to WFC3 images with updated coordinates.

\subsection{Spectroscopy}\label{SALTspec}
Red region long-slit spectra along the major axis of Zw~049.057 were obtained with the Robert Stobie Spectrograph (RSS) on the Southern African Large Telescope (SALT) \citep[see,][]{Buckley08} on UT dates March 15 and 28, 2019. The 1.5~arcsecond width slit was set to position angles of 18$^\circ$ for the major axis. We obtained two exposures of 1100~s with the PG1800 grating yielding a spectral resolution of 3100 at H$\alpha$ derived from gaussian fits to night sky OH lines with an angular scale of 0.253~arcsec per 2x2 binned pixel. We estimate the seeing to have been $\sim$1.3~arcsec.

The SALT RSS PySALT pipeline \footnote{http://pysalt.salt.ac.za/} provided initial processed versions of the spectra to remove instrumental signatures \citep{Crawford10}.  Using iraf, we combined the exposures, transformed the spectra to remove slit curvature and applied a wavelength scale. We independently reduced each of the 3 CCDs in the RSS detector. Sky subtraction was done by selecting and combining regions above and below the area containing the signal from the galaxy which were subtracted from the signal area. 

We checked our RSS wavelength calibrations against sky emission lines using wavelengths tabulated by \cite{Osterbrock96}. We incorporated wavelength offsets derived from the sky lines of -0.19$\pm$0.05~\AA\ for the H$\alpha$ region and 0.23~\AA\ for the Na~D lines. We applied heliocentric corrections to our observed velocities using the IRAF rvcorrect routines.

 \section{Structure of the Stellar Body}\label{sec:structphotom}

\subsection{The Disk}\label{sec:thedisk}

Figure \ref{fig:zw049_general} shows the complex optical structure of Zw~049.057. The main stellar disk has red colors and is crossed by multiple dust absorption features. Despite the dust, the inner disk is distinguished by enhanced surface brightness within R$_{major} \approx$3 arcsec (860~pc). The inner disk is the site of intense star formation producing strong emission lines in optical spectra (\S \ref{sec:centerspec}) and in narrow-band imaging \citep{Poggianti00,AlonsoH06}. A UV-bright arm traces the eastern side of the inner disk. An inclination of 70$^{\circ}$ follows under the assumption that the main disk has an intrinsic axis ratio of b/a=0.2 \citep[see][]{Holmberg58}. The analysis by \cite{Scoville00} based on NICMOS images provides information on the properties of the inner star forming disk. Their F222M NICMOS data revealed the heavily obscured nuclear region at the center of the outer disk. Molecular material with a mass of M$_{molec}$=1.6$\pm0.2 \times$10$^9$~M$_{\odot}$ derived from CO 1-0 observations appears to be concentrated in the central part of the galaxy \citep{HerreroIllana19}. These observations do not fully sample the dense gas associated with the CON, but instead refer to gas in the central parts of the disk.  We discuss the star forming central disk of Zw~049.057 in more detail in \S \ref{sec:centerspec}.

\begin{figure}[]
    \centering
    \includegraphics[width=0.50\textwidth]{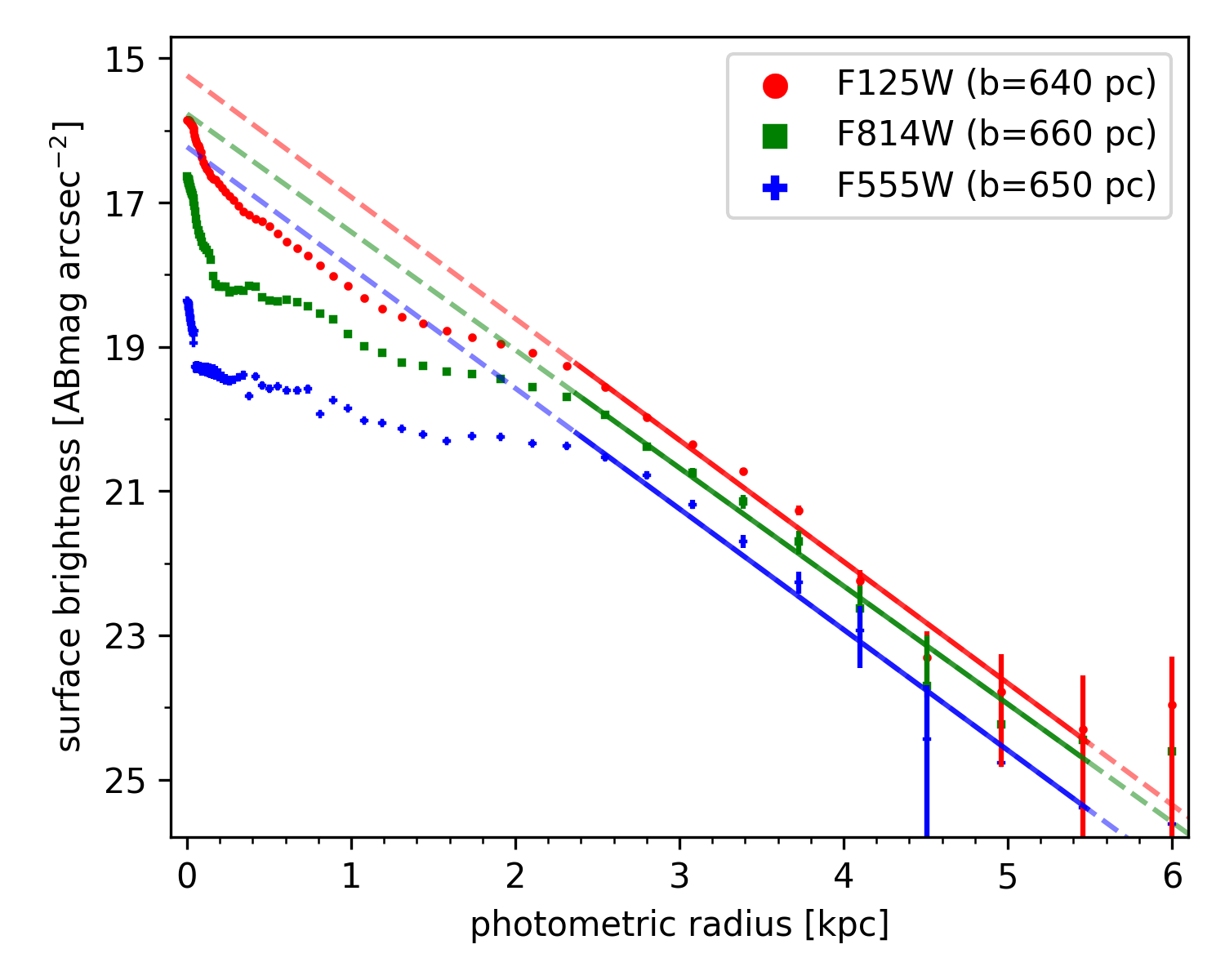}
    \includegraphics[width=0.47\textwidth]{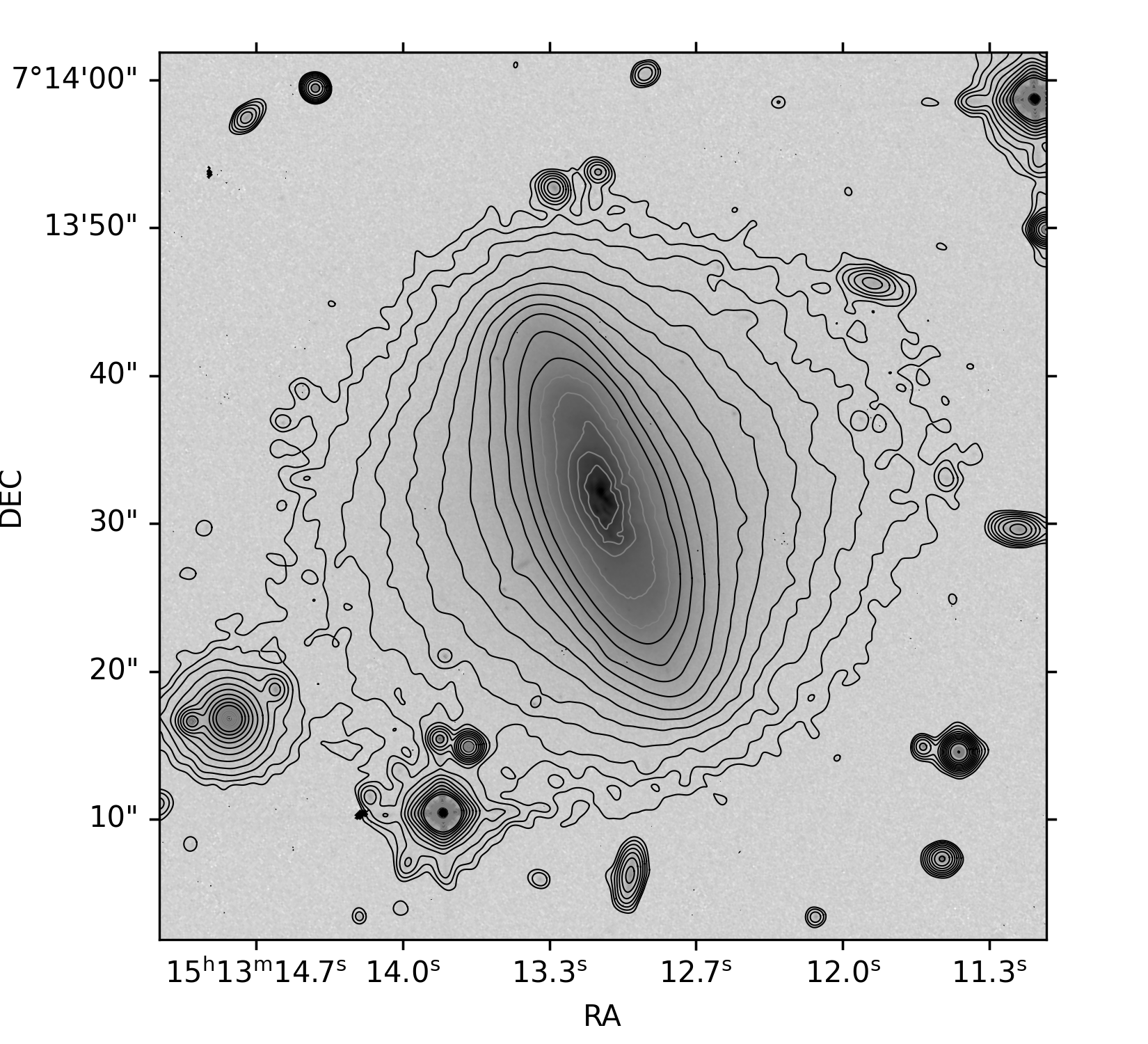}\hfill
    \caption{Results are shown from our isophotal analysis.{\it Left:}Radial intensity plots are shown from ellipse fits to the HST images. The less obscured outer disk regions show an exponential profile with a scale length of 2.2~arcseconds 640~pc). The effects of dust obscuration results in all of the profiles overshooting the observed intensities within R$\lesssim$2~kpc. The flatter intensity gradient at 6~kpc occurs in the region where the round outer light distribution begins to appear (see \S3.2). {\it Right:} Our WFC3 F125W image scaled to show the inclines inner disk of Zw~049.057 with a major axis position angle of 200$^{\circ}$ and nearly circular outer structure.
The outermost isophotes are at $\mu_{F125W, AB}$ $\simeq$ 19~mag~arcsec$^{-2}$. We suggest that this feature
is due to the presence of a warped outer disk associated with the interaction that produced
the northern polar dust lane. The inner isophotes display the observed b/a=0.4 axis ratio that we used to derive the inclination} 
    \label{fig:zw049-isophotes}
\end{figure}

Figure~\ref{fig:zw049-isophotes} shows results from ellipse fits to our the intensity distributions in our HST images. The main disk (i.e. the region beyond a radius of roughly 1~kpc) is exponential with a scale length of 640~pc and major axis at a position angle of 200$\pm$2$^{\circ}$. The outer or main disk is symmetric with a relatively smooth optical structure but intensities within r$\lesssim$1~~kpc are strongly affected in the optical bands by the ongoing galactic-scale dust storm. However, it is clear that the main disk in Zw~049.057 extends for 9 radial exponential scale lengths and in this sense is not compact. The substantial central dust obscuration in Zw~049.057 leads to difficulties in obtaining unique model fits to the galaxy's image.  
We therefore characterized the size of Zw~049.057 by determining the effective (half-light) radius R$_e$ = 0.92~kpc from an isophotal fit to the 3.6~$\mu$m \emph{Spitzer Space Telescope} (SST) archival image.\footnote{IRSA/Spitizer Enhanced Images Product, P.I. Mazzarella, Program 3672, doi.org/10.26131/irsa433} The measured R$_e$ for Zw~049.057 decreases with increasing wavelength as the level of dust obscuration of the bright central region declines, so the 3.6~$\mu$m image provides the most reliable estimate of R$_e$.

The NASA Extragalactic Database gives m$_{AB}$(3.6~$\mu$m)=13.54 $\pm$0.1, consistent with our photometry of an archival SST 3.6~$\mu$m image. For D=59~Mpc this yields M$_{AB}(3.6)=-20.2$ and L$_{3.6} =$3$\pm$0.3$\times$10$^{10}$~L$_{\odot}$. Adopting a stellar  mass-to-light ratio of $\gamma_{3.6}$, the stellar mass in Zw~049.057 is M$_* \approx$1.5 $\times$10$^{10}$($\gamma_{3.6}$/0.5)~M$_{\odot}$, consistent with the values found by \cite{U12} and a factor of 2.5 greater than the mass derived by \cite{Leroy19}. Zw~049.057 is a moderate mass galaxy, and we adopt 1.5 $\times$10$^{10}$~M$_{\odot}$ for its stellar mass in the remainder of this paper. Stellar masses are especially useful for comparisons between galaxies as fundamental parameters and quantities that can be measured using a variety of techniques. 

The photometric and structural properties of Zw~049.057 are summarized in Table \ref{tab:1}. The far-infrared luminosity of Zw~049.057 of log(L$_{IR}) = 11.5$ exceeds the stellar luminosity by a factor of 10. Since the bolometric luminosity in the dusty Z~049.057 system is essentially the same as L$_{IR}$, then the observed stellar mass to light ratio of Zw~049.057 is $\gamma_{tot,obs}$ =M$_*$/L$_{IR}$ $\approx$0.05. This  value of $\gamma_{tot,obs}$ is similar to values observed in extremely obscured ULIRGs, e.g., $\gamma_{tot,obs}$ $\approx$0.04 for Arp~220 \citep[see][]{Chandar23},  but is more extreme than the $\gamma_{tot,obs} \approx$0.2 for the starbursting M82 system, which has 1/3 the stellar mass of Zw~049.057 \citep{Leroy19}.  

\begin{deluxetable}{lcl}[]
\tablecaption{Photometery of Zw~049.057\label{tab:1}}
\tablehead{
\colhead{Property} & \colhead{Value} & \colhead{Notes}
}
\startdata
Magnitudes & U$_{336}$(0) $\lesssim$16.6 V$_{555}$(0)=14.56 I$_{814}$(0)=14.52 J$_{125}$(0)=14.93 & STMAG,  this study \\
log(L$_{\lambda}$/L$_{\odot \lambda}$) & HST-V: 9.6, HST-I: 9.8, HST-J:10.1, SST-3.6$\mu$m 10.5 & this study \\
Effective radius in kpc at $\lambda$ & HST-J 1.3; SST 3.6$\mu$m\, 0.9 & SST less affected by dust  \enddata
\tablecomments{Our HST V,I,J photometry has uncertainties at the 0.05 magnitude level and is in good agreement with ground-based data. The HST-U-band magnitude is an upper limit because the outer galaxy is not detected due to reduced surface brightness sensitivity in the F336W filter. The Galactic extinction correction is based on A$_V$=0.116 and tabulated values for each filter from the NASA Extragalactic Database.}
\end{deluxetable} 
 
\subsection{The Outer Stellar Component}\label{outerhalo}

While the main body of Zw~049.057 displays a disky structure, at fainter levels of $\mu_{F125W,STMAG} \geq$ 25~mag~arcsecond$^{-2}$ the isophotes trend towards being circular. This outer feature, illustrated in Figure~\ref{fig:zw049-isophotes}, is also detected in the optical HST and ground-based images. Photometry along the south-eastern major axis of the halo shows an approximately exponential light distribution with a scale length of $\approx$5~arcsec (1.4~kpc). 

The 3-dimensional form of the outer stellar region with circular isophotes is not clear. One possibility is a warped disk where the outer disk's line of nodes lies close to the current major axis of Zw~049.057. A warped outer disk would naturally account for the approximately exponential light distribution. Alternatively this material could be a diffuse spheroidal  stellar halo centered on the galaxy. If the symmetric light component is 
from a stellar halo, then this structure has a low degree of central mass concentration.  Reflected light from the dusty outflow is another possibility, although this is inconsistent with the opening angles of the observed dust features.Therefore we favor the warped, disky stellar component interpretation. 

Our images show the presence of polar dust arcs. Zw~049.057 experienced a recent minor merger that can induce disk warps \citep[e.g.,][]{Sparke09} and is in the late stage of a merger where the galaxy has not yet recovered its co-planar structure (see discussion in \S6). Alternatively, a stellar halo could have an  origin from stars lost during mergers. However, we do not detect the shells or streams in the round stellar component that are expected to persist for several Gyr after mergers  \citep{Oh08,Eliche18,Karademir19}. The absence of these features may be an observational limitation stemming from a combination of resolution and sensitivity, issues that could be addressed via deep imaging with the JWST.

Based on the present evidence of a merger that produced the polar dust lane (see \S \ref{sec:dust}), we prefer the warped disk model while recognizing that this interpretation requires we are observing Zw~049.057 from a special perspective. 
Given the likelihood that Zw~049.057 is the product of a galaxy-galaxy interaction, we searched the NASA ADS for companions. This revealed the apparently undisturbed pair of dwarf galaxies WISEA J151341.81+065603.6 and WISEA J151407.76+065504.3 at redshift velocities of v$_{helio}$= 3913~km~s$^{-1}$ and 3896 km~s$^{-1}$, respectively, with a projected distance of 300~kpc as 
the only nearby neighbors. For a relative peculiar velocity of 100~km~s$^{-1}$, typical of a small galaxy group, any interaction with the companions would have occurred more than 3~Gyr in the past.  Similarly nearby galaxies, including the companions, are not detected in H{\rm\,I} 21~cm line emission by the ALFALFA survey \citep{Haynes18}. Zw~049.057 may be an example of a galaxy that became isolated in its local environment through a merger with its neighbor, the isolation by annexation effect described by \cite{Fulmer17}.

 \subsection{Color Maps}\label{sec:stellarpops}

\begin{figure}[]
   \centering
   \centering   
    \includegraphics[width=0.99\textwidth]{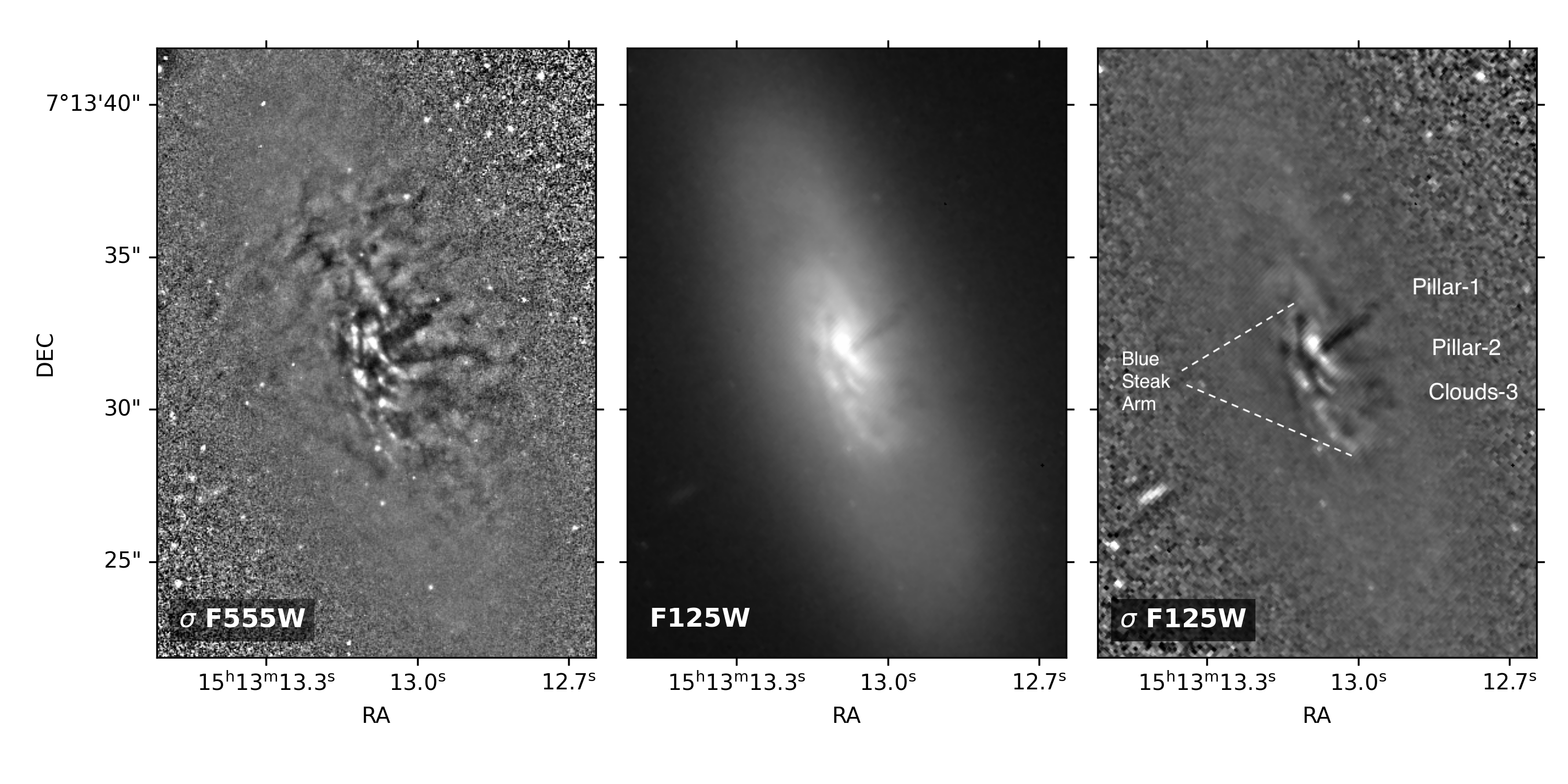} \hfill
    \includegraphics[width=0.44\textwidth] {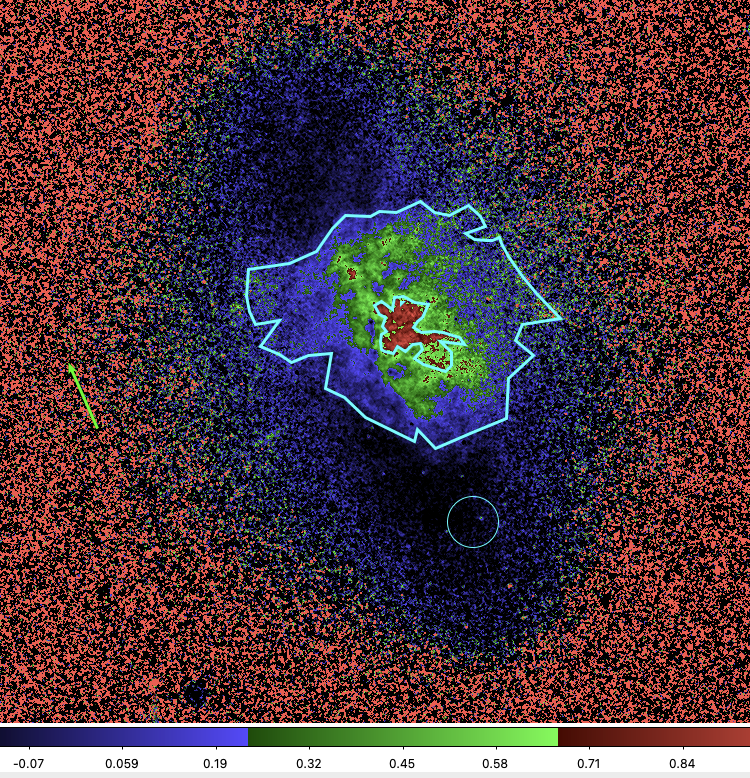} \hfill
    \includegraphics[width=0.53\textwidth]{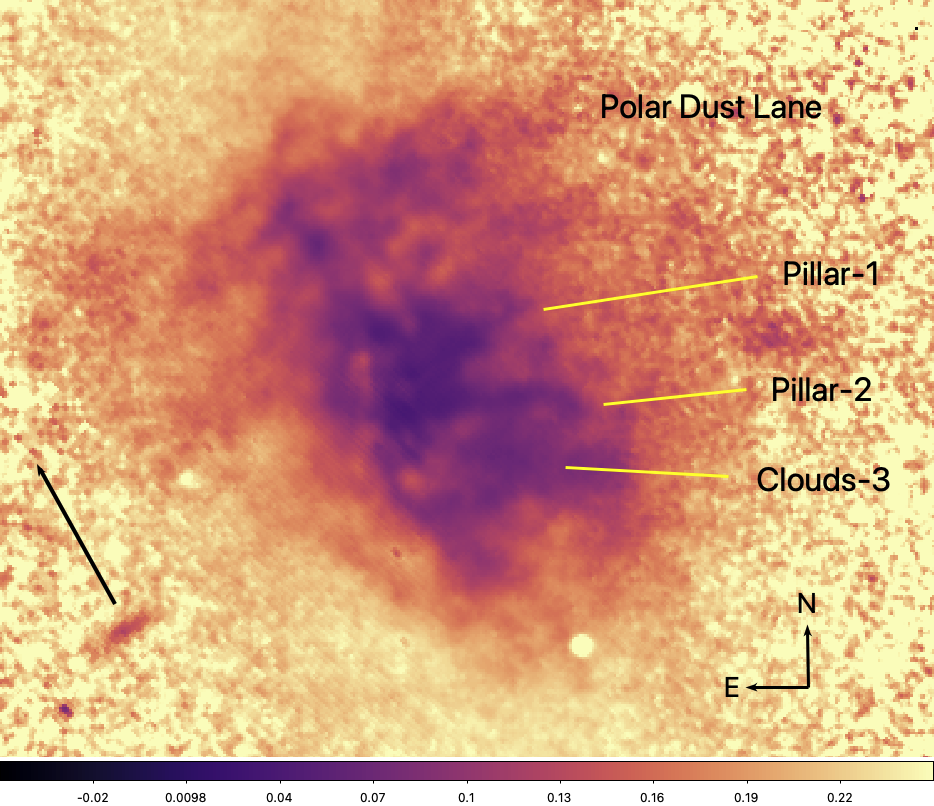}
   \caption{ {\it Top:} Large scale dust absorption structures in Zw~049.057 are shown in our HST WFC3/UVIS F555W and smoothed, background subtracted ($\sigma$) images in the F125W filter. Major features are marked; see text for details. The pair of dust pillars that make up the Zw~0049.057 dust "V" are present in our HST F125W image and at longer wavelengths as shown by \citep{Scoville00}. Pillar-1 stands out in the F125W image consistent with its inferred high optical depth. Pillar-2 to the south is narrower and less pronounced. The asymmetric ``blue streak" arm is labeled. {\it Bottom Left:} The distribution of the observed color (V-I)$_{STMAG}$=m(F555W)$_{ST}$-m(814W)$_{ST}$  shows the central reddening due to the polar ring and extraplanar interstellar matter. Deeper inner dust obscuration is associated with the pillars and outflows on the west side of the galaxy. The outer contour defines the ``central region" in \S \ref{sec:dustcenter} where we measured the average diffuse dust absorption. The inner contour outlines the more heavily obscured region that was replaced by the surroundings for the diffuse region opacity measurement. The color map is at the original pixel scale smoothed by a. $\sigma$=1 pixel gaussian. The circle shows an example of a disk region color measurement of (V-I)$_{STMAG}$=-0.16$\pm$ 0.07 while the deepest absorption around the nucleus has a color of (V-I)$_{STMAG} =$ 0.9.  {\it Bottom Right:} The ratio of the F555W to F125W images in the observed  e$^{-1}$~s$^{-1}$ matched to the PSF of the F125W data and smoothed with a $\sigma$=1 pixel Gaussian to show regions with a range in optical dust optical depths. Major feature are labeled. The arrow has a length of 3.5~arcsec in both figures. A background galaxy is located below the arrow.}  
   \label{fig:zw049_vijcolor}
\end{figure}

We used the HST images to make color maps of Zw~049.057 shown in Figure~\ref{fig:zw049_vijcolor}.  These maps are especially important in revealing the extensive presence of complex dust absorption in the middle of Zw~049.057. The F336W image has low signal-to-noise, and we therefore could not reliably produce a color map that avoided severe detection biases. The colors F555W to F125W NIR image is presented in terms of the ratio of observed count rates where the F555W were rebinned and convolved to match the pixel scale and approximate point spread resolution of the F125W data. The central colors of Zw~049.057 are driven by dust absorption. We discuss the properties of the dust in \S \ref{sec:dust} below.

The F336W image in Figure~\ref{fig:zw049_general} displays combined effects of high levels of dust opacity and intense star formation. This complexity is illustrated by the properties of the head of the ``blue streak" arm located at J2000 15:15:13.14 +07:13:32.6 that is clearly present in the F336W images and more clearly present at longer wavelengths (see Figure \ref{fig:zw049_pillars_color}). We photometered the knot at the northern head of this feature using a 0.13~arcsecond radius aperture giving m$_{555,STMAG}$=21.7, M$_{555}$ $\approx$ -12m and (U-V)$_{STMAG,obs}$= 0.8 or close to the color of the Sun. Even this UV-bright region is significantly reddened, 
but the quantitative obscuration level is difficult to assess because the F555W image includes a larger background contribution from the surroundings than the F336W image. The measured  (U-V)$_{STMAG,obs}$ therefore is an upper limit to the intrinsic color of this region.

The disk beyond the central heavily obscured zone has a roughly constant observed color of (V-I)$_{STMAG}$= -0.16 or ((V-I)$_{Vega} \approx$ 1.1) in the bluer and therefore less extincted regions in the main stellar disk. We attempted to model the stellar population in a relatively blue patch of the main disk using GALEV \citep{Kotulla2010} 
with photometric data from all of the bands other than the low quality F336W data. The resulting models spanned a considerable range from a moderately young population with modest dust extinction to a multi-Gyr population with lower extinction. Thus, if the main disk comparison regions are relatively dust free, then the disk contains old stellar populations. A full assessment of the stellar composition of the Zw~049.057 disk will need to await the analysis of our MUSE spectroscopic data (C. Wethers {\it et al.} in preparation).

\section{Interstellar Obscuration}\label{sec:dust}

\subsection{The Dust Absorption Structures}\label{sec:duststructures}

The variety of dust features on large scales are shown in Figure \ref{fig:zw049_vijcolor}.  The dust features in Zw~049.057 can be classified into several categories (see Figures \ref{fig:zw049_general} and \ref{fig:zw049_vijcolor} for labels):
\begin{itemize}
\item A complex of moderate opacity dust covers most of the central regions of Zw~049.057 with a geometry that is indicative of extraplanar material.
\item Dust bands that are slightly convex inwards crossing north and south of the galaxy center are polar ring signatures. 
\item Dust structures within inner the disk are concentrated within a radius of roughly 1~kpc and have curved shapes roughly following the elliptical pattern of of the disk isophotes.
\item Multiple dust features structures are observed to extend from the star forming inner disk. A prominent example is the Clouds-3 complex spanning the south-southwest of the galaxy center (see Figures \ref{fig:zw049_vijcolor} and \ref{fig:map} and \S \ref{sec:sub-discuss_struct}).
\item Clearly defined linear dust features, Pillars-1 and -2,  discovered by \cite{Scoville00}, are stand out due to their high opacities and and are discussed in \S \ref{sec:polarflows}. The pillars converge in the area of the nucleus.
\item Faint, approximately linear dust lanes are primarily observed in the F555W on the east side of Zw~049.057 extending outwards from the star forming disk towards a projected point in the eastern halo (see Figure~\ref{fig:zw049_vijcolor}). The origins of these features is not clear.
\end{itemize}

A full accounting of the properties of dust absorption would take into account the physical natures of radiation fields, dust scattering, the geometry of the system, and the physical properties of the dust \citep[see][for reviews]{Calzetti01,Gordon21,Popescu21}, information that we do not have. Our objective therefore is to apply simple pure absorption models to make initial estimates of dust opacities and associated gas masses. In the next sections we present our analysis of the major dust features within Zw~049.057.

\subsection{Dust Properties}\label{sec:dustproperties}

The difficulties associated with measuring dust optical depths and deriving approximate gas masses from dust obscuration levels are well known \citep[e.g.,][]{Calzetti97,Kylafis05,White00,Natale14,Calzetti21,Keel23}. If the dust lies outside of the stellar system, then the foreground dust screen ($fg$) approximation for purely absorption opacity can be used. In the foreground screen case, an absorber with optical depth $\tau_{fg}$ is in front of the luminosity sources giving I/I(0)=e$^{-\tau_{fg}}$.  This model yields a lower bound to the dust absorption optical depth for cases where the dust lies within the stellar bodies of galaxies. For example, the basic equation for radiative transfer shows that in an optically thick system with constant spatial distribution of opacity and luminosity density I/I$_0 \rightarrow S$ where $S$ is the source function, while the foreground screen predicts I/I$_0 \rightarrow$ 0.  Ignoring scattering and unresolved spatial variations in optical depths along the observed sight lines also leads to underestimates of dust opacities. Our measurements of dust optical depths derived from foreground screen models therefore yield lower limits to the true dust optical depths.

\subsection{Diffuse Dust Absorption}\label{sec:dustcenter}

The spatial structure of diffuse dust across the center of Zw~049.057 indicates that this material likely lies in a 3 dimensional structure that extends above the plane of the stellar disk and includes the northern polar dust lane. This is an area that is dominated by feedback from the galaxy's center and polar material. Therefore, measurements of the mass of interstellar matter in this region provides information on the scale of reservoirs associated with out-of-plane gas flows.

We selected the central dust component selected based on reddening levels in the color map shown in Figure \ref{fig:zw049_vijcolor}) that includes the selection boundaries. Two regions in the main disk of Zw~049.057 provided an estimate of the unreddened (V-I)$_{STMAG}$= -0.16$\pm0.07$ for the unreddened color of ZW~049.057. The calculated value of the optical $\tau_{fg}$ for a foreground dust screen came from our map of the (V-I)$_{STMAG}$ in Figure~\ref{fig:zw049_vijcolor}. We used the standard relationship m(obs)$_{\lambda}$ = m(0)$_{\lambda}$+A$_{\lambda}$ where the optical depth is $\tau_{\lambda} = 1.086$A$_{\lambda}$ and an extinction law $\tau_V$=$\tau_{\lambda}$($\lambda [nm]/555)^{-x}$ with x=1.3 \citep{Keel23}  to find the extinction and convert color excesses into the obscuration levels and optical depths.This process assumes the stellar populations in the main disk are typical of the galaxy as a whole. Therefore we would underestimate the extinction and optical depths in any regions with bluer colors than those of the main disk. In estimating the central reddening, we removed the inner most highly obscured central region of Zw~049.057 from consideration and replaced the obscuration in this varied zone with the mean reddening from the surroundings   The central region also is complex with some dust likely to be embedded or have variations in optical depths, which also leads to underestimates of optical depths. This procedure yielded a conservative  $\tau_{V,fg}$ $\geq$  0.8 for the central zone of Zw~049.057. 

Adopting a Galactic relationship of N$_H$ =1.9$\times$ 10$^{21}$ H-atoms~cm$^{-2}$A$_V$, $\tau_V$= 1.087A$_V$  and a helium mass fraction of 0.27 \citep{Draine11}, the foreground screen model gave M$_{gas,out}$ $\geqslant$ 10$^8$~M$_{\odot}$ and $\overline{N_H}$ $\geqslant$ 1.4 $\times$10$^{21}$~atoms~cm$^{-2}$ for the extraplanar gas seen in absorption across the central 8$\times$10$^6$~pc$^2$ of Zw~049.057 (see also Table~\ref{tab:dustpillars}). Since Zw~049.057 is likely to have a symmetric structure, the total mass of diffuse gas that accounts for the unseen hemisphere is twice our observed value, or  M(gas)$_{diffuse}$ $\geqslant$2 $\times$ 10$^8$~M$_{\odot}$.

H{\rm\,I} 21~cm absorption measurements by \cite{Mirabel88} give a column density of N$_H$=6.6$\times$ 10$^{21}$(T$_s$/100~K) where T$_s$ is the H{\rm\,I} spin temperature. The agreement between the H{\rm\,I} data and dust absorption measurements is reasonable given the considerable uncertainties, including the probability that the  H{\rm\,I} absorption is observed against the compact nuclear radio source where the ISM is likely to be densest. Therefore we adopt M$_{gas,extraplanar} \simeq 2 \times$10$^8$~M$_{\odot}$ for the important spatially extended central gas reservoir that equals $\approx$10\% of the molecular gas mass in Zw~049.057 \citep{HerreroIllana19}. Since most of the dusty gas is observed at projected heights of about 1~kpc, we do not find clear evidence for an escaping galactic wind. Extraplanar gas is likely to be in a dynamic state with the mass balance set by the competition between injection from the galaxy center and remnant merger material in polar orbits versus loss of gas as dissipation allows material to settle back to the disk.

\subsection{Embedded Slab Dust Pillar Absorption Models}\label{sec:pillarmodels}

\begin{figure}[]
    \centering
    \includegraphics[width=0.90\textwidth]{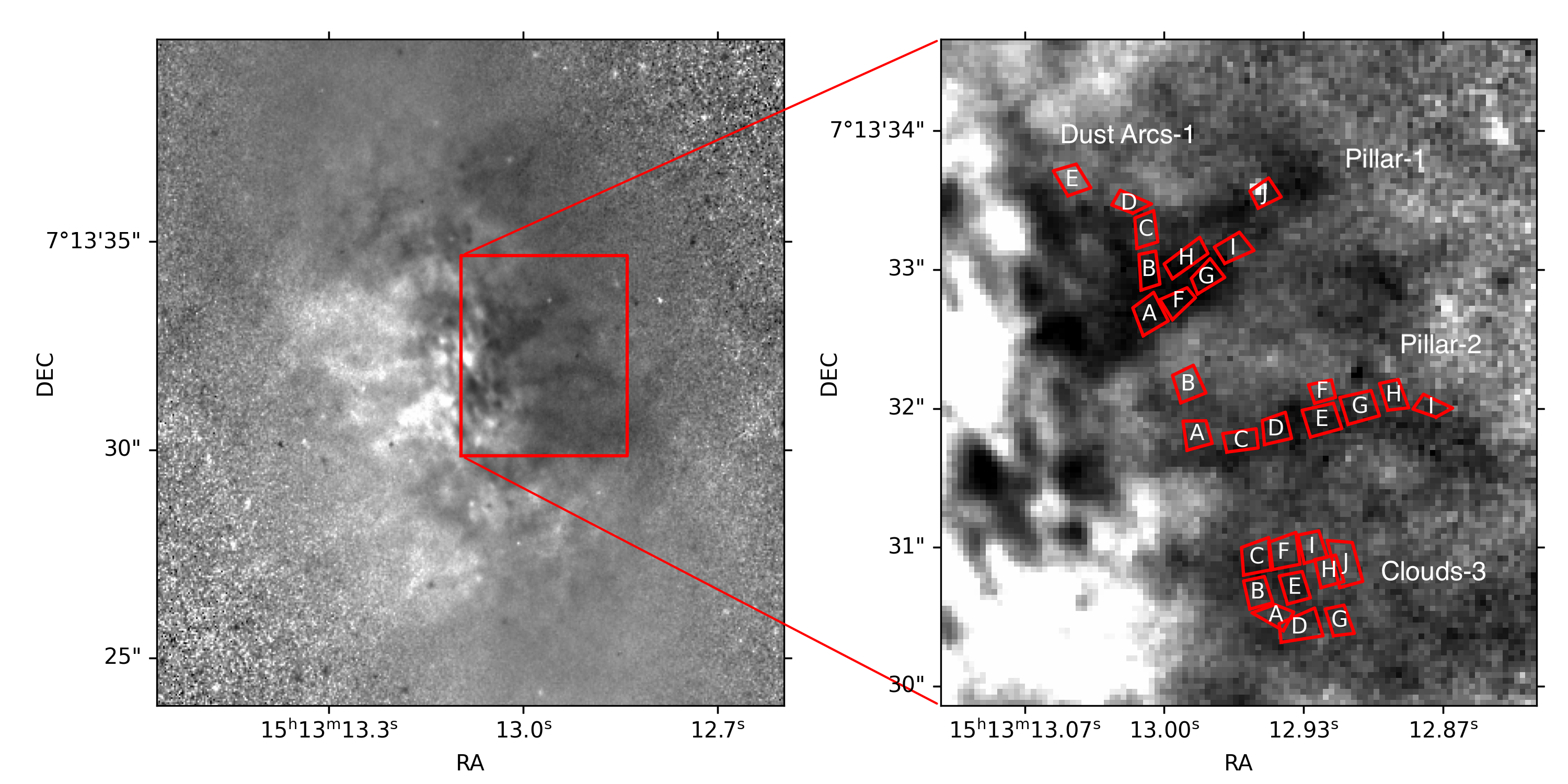}\hfill
    \
    \includegraphics[width=0.40\textwidth]{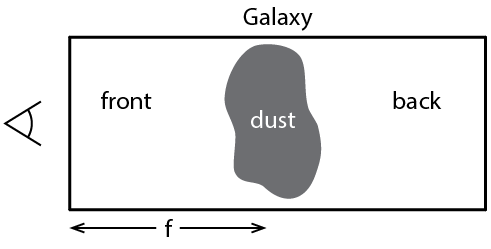}
    \caption{{\it Upper Left:} The Zw~049.057 F555W image divided by its mirror image along the major axis to distinguish the discreet dust absorption features. {\it Upper Right:} Map of the regions of interest. Pillar-1 is the top grouping, Pillar-2 in the middle, and Cloud-3 at the bottom of the figure. The boxes labeled with letters define the regions where photometric results are listed in Table~\ref{tab:dustpillars}. {\it Lower Center:} Schematic model of an idealized pillar embedded in a galaxy with constant stellar volume emissivity. Here $f$ denotes the fraction of the normalized length of the "galaxy" (i.e. $f=1$ refers to the entire column through the galaxy) to where the uniform opacity cloud is located. This basic embedded cloud model assumes that the width of the dust cloud is small compared to the length of the stellar region. A dark cloud at the surface would be at $f=0$ and the diagram shows a dark cloud at $f=1/2.$.} 
    \label{fig:map}
\end{figure}

The linear dust ``pillars" that make up the dust ``V" in Zw~049.057 are sufficiently well-defined to be treated as discreet features. For simplicity in calculating optical depths and other parameters, we empirically modeled the pillars as rectangular solid prisms of dusty material embedded  within a uniform stellar medium. Consistent with this model, we selected polygonal subregions to pick out the absorption peaks, and analyzed dust opacities via a simplified embedded cloud dust model. Taking advantage of the low extinction above the disk in the northwest and the system's relatively high degree of symmetry, we made an image of the galaxy that is flipped around the disk's major axis.  Dividing the galaxy by its flipped image provides an estimate of the galaxy background and the levels of obscuration in the dusty southwestern regions (Figure \ref{fig:map}) relative to the northeast side. In a manner similar to that adopted by \cite{White00}, we used this image to estimate the dust transmission levels and thus opacities in the labeled regions in the F555W and F814W filters from the flipped WFC3 HST images in Figure~\ref{fig:map}. If the east side of Zw~049.057 is somewhat obscured, then this approach underestimates the opacity of the pillars. As can be seen in Figure \ref{fig:map} the dust dark nebulae above and to the west of the central disk plane are complex but reasonably well-defined. However, the dust structure in the central star-forming disk is too confused for the flipped background method to be useful.

The presence of physically discreet dust absorption structures embedded within the stellar body of a system complicates the determination of dust opacities from measurements of the light transmission. For example, an optically thick cloud near the far side of an otherwise transparent stellar light distribution would have low contrast leading to an apparently high transmission; the foreground dust screen model would measure this as a region with low opacity. We can largely overcame this problem by taking advantage of the variations in opacity with wavelength. A high dust opacity cloud, even if deeply buried within the stellar body, will remain opaque to long wavelengths and such features can be found from their lack of variation in intensity with wavelength. \cite{Gallagher81}, for example,  applied this technique to identify molecular clouds in the dwarf galaxy NGC 185.

We approximated this situation with a simple discreet cloud model shown in Figure \ref{fig:map}. The model has 3 unknowns: the fractional location f of the cloud within the stellar system where f=0 on the front surface and $f\leq$1, the visual optical depth of the cloud $\tau_V$, and the extinction law that gives  $\tau_{lambda}$/$\tau_V$.  If we assumed a standard extinction law, which is appropriate for the diffuse absorption and applies in most other dusty early-type galaxies \citep[][ and references therein]{Finkelman12}, then we could solve for $f$ and $\tau_V$ from observations at two wavelengths:
   ${I_{\lambda,obs}}/{I_{\lambda,0}} = f + (1 - f) e^{-\tau_{\lambda}}$.

The linear sizes of the sampling regions in the pillars are $\sim$40-60~pc. Although the dust structures within the pillars are not well resolved, both pillars have a thin higher opacity component that appears to be surrounded by less opaque material. If an unresolved mix of dust optical depths exists, then our assumption of uniform opacity under weights the more opaque regions and our $\tau_V$ are lower limits \citep[e.g.,][]{Varosi99}.  An additional issue arises for locations where we estimate $\tau_V > 2$. At these optical depths the intensity ratios are $<$ 0.13 and our measurement uncertainties based on the uniformity of the background are $\pm$ 0.07 are too large to reliably measure such low ratios. The high density parts of the filaments are at the reliability limits of our data and require an analysis going beyond a simple models. Since our methods for deriving $\tau_V$ led to underestimates of optical depths, we once again obtained lower bound estimates of gas masses.

\subsection{Properties of the Dusty Polar Outflow Features}\label{sec:polarflows}

\subsubsection{Dusty Pillar-1}
\begin{figure}[]
    \centering
    \includegraphics[width=0.80\textwidth]{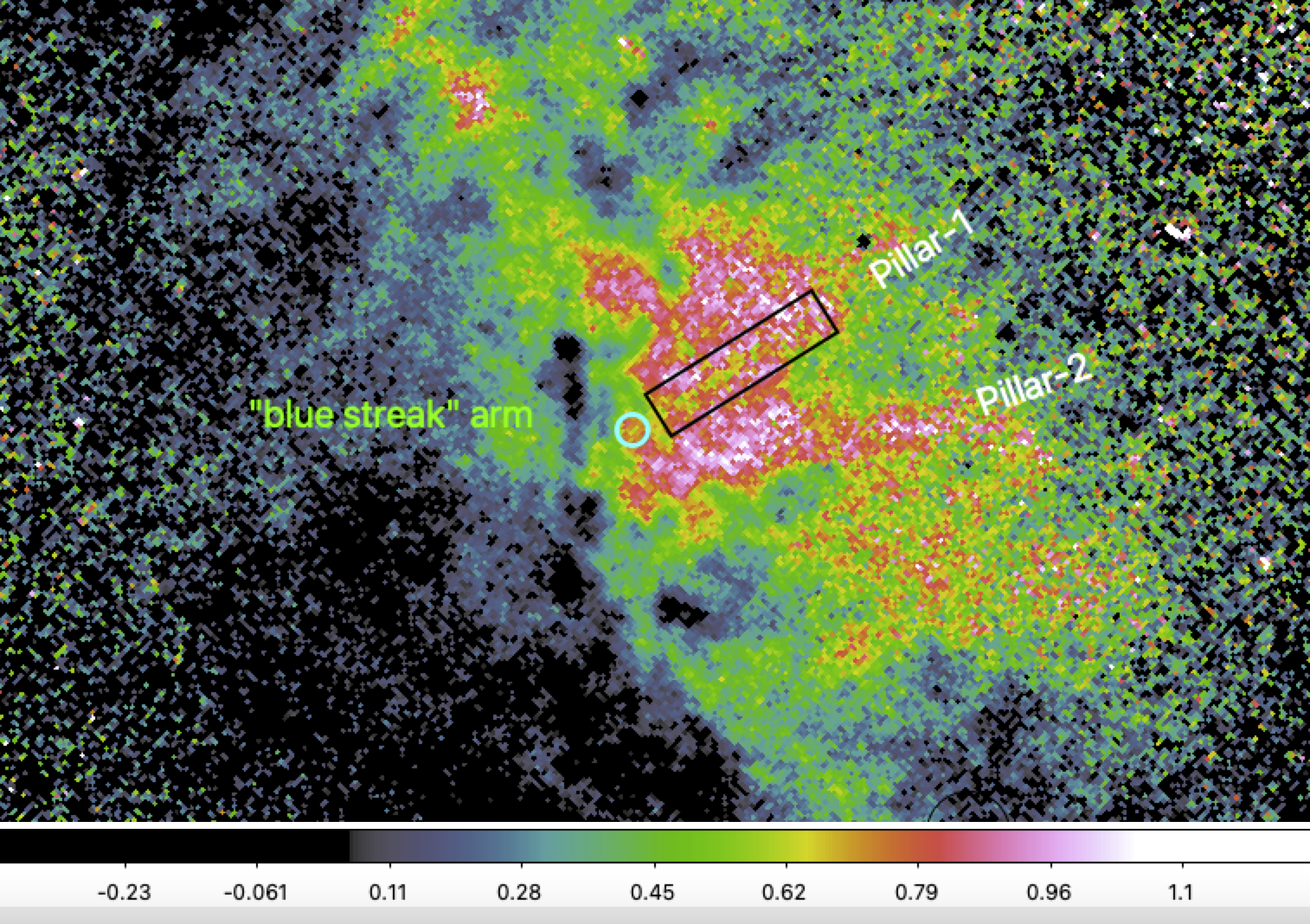}
\caption{The central region of Zw~049.057 is shown in a full resolution (V-I)$_{STMAG}$ image with north and east to the left. The location of Pillar-1 is marked by a cyan rectangle. Only the north edge of Pillar-1 is delimited by reddening, while the clumpy  Pillar-2 can be traced by its red colors.  The color at the location of the Pillar-1 spine is close to that of the surroundings, indicating a high optical depth that limits the transmission of reddened light. The ``blue streak" arm is also evident with the bright northern knot at its head. The 0.15~arcsecond circle shows the location of the CON as determined by radio continuum observations.}
  \label{fig:zw049_pillars_color}
\end{figure}

The distinctive dusty Pillar-1 in Figure \ref{fig:map} is oriented towards a position angle on the sky of 303$\pm$ 2$^{\circ}$ as measured from the galaxy center. Since the major axis position angle is 19$^{\circ}$, Pillar-1 is offset by 14$\pm$3$^{\circ}$ from the minor axis of the main disk. Pillar-1 has a maximum width of 110~pc near its base in the F555W image and narrows to 60~pc near the top of its coherent main structure. The width of Pillar-1 is not sharply defined since the opacity declines towards its northern side, away from the narrow, denser dust spine on the southern edge that has a width of $\leq$0.1~arcseconds (29~pc). The color map in Figure \ref{fig:zw049_pillars_color} shows that the spine of Pillar-1, even though clearly in strong absorption at near infrared wavelengths (see Figure~\ref{fig:zw049_pillars_color}, remarkably, does not display highly reddened colors. The projected length of Pillar-1 is 600~pc to the top of the linear main structure and it extends in a less organized way to near 750~pc.  If Pillar-1 is orthogonal to the disk, then its deprojected length is about 800~pc. Pillar-1 lies along the center of the minor axis L-band radio emitting region discussed by \cite{Falstad18} that  will be investigated in further detail by Lankhaar ({\it et al., in preparation}). The dust distribution along the length of Pillar-1 is not uniform, and includes a detached dark cloud beyond the termination of the main pillar (Pillar-1, Subregion-J of Figure~\ref{fig:map}). Our measurements of transmissions along Pillar-1 focus on the local absorption peaks.. 

Figure~\ref{fig:zw049_pillars_color} shows the (V-I)$_{STMAG,Obs}$ color map for the center of Zw~049.057. While the northern edge of Pillar-1 shows the expected reddened colors, its core has the color of the less reddened background. Evidently Pillar-1 is optically thick in the HST V- and I-bands so that it blocks most of the background light, leaving only the less obscured light from the foreground.  Pillar-1 fits with the simple model presented in the lower panel of Figure~\ref{fig:map} where the embedded cloud is optically thick. By analogy, the lack of the reddening along the Pillar-1 core suggests that it is not in the foreground of our sightline.  Therefore we estimated the obscuration along Pillar-1 using the simple embedded cloud obscuration model.

The results of this exercise gave a fractional depth within the luminous medium $f \approx$ 0.2-0.3 and $\tau_V$ $\approx$1-2 for Pillar-1. Table \ref{tab:dustpillars} lists the resulting  optical depths, gas mass estimates, and mean gas densities for the main clumps along Pillar~1. This analysis yielded  M$_{Pillar1,gas} \gtrsim$ 8 $\times$ 10$^5$ M$_{\sun}$ for the region projected above the disk of Zw~049.057. Due to our the neglect scattered/reflected light in our simple slab model and insensitivity to the optically thick spine and small scale substructures, we likely substantially underestimated the mass of gas in Pillar-1.  Even so we found the the visually striking, Pillar-1 only has $\gtrsim$ 0.5\% of the  $\sim 2 \times 10^8$ M$_{\odot}$ gas mass of the CON (see \S \ref{sec:nucleus}), and $\gtrsim$ 1\% of the extraplanar gas. Given the symmetry of the apparently associated L-band radio extension about the center of Zw~049.057 and considerable level of uncertainties, we adopted a total mass in both sides of Pillar-1 of $\gtrsim$2 $\times$10$^6$~M$_{\odot}$.

Pillar~1 appears to be spatially connected to the nuclear regions and likely to the CON in  Zw~049.057 \citep{Scoville00}.  If Pillar-1 represents a collimated outflow, then we can estimate a minimum lifetime of $t_{fil1}$ $\approx$ 3$\times$10$^6$(L$_{800 pc}$/V$_{300-outflow}$)~yr. The outflow velocity of 300~km~s$^{-1}$ is chosen to be in a regime where matter could escape and remain collimated from the nuclear region.  
If Pillar~1 is an outflow feature, it can be produced in a short time. 

Alternatively Pillar-1 could be a part of an approximately conical wall resulting from the fast nuclear outflow seen by \cite{Falstad19}. The pattern of dust reddening in Figure~\ref{fig:zw049_pillars_color} could be due to a dusty outflow with an opening angle of $\approx$30-40$^{\circ}$. The lifetime of a cone wall would be determined by the evolutionary time scale of the outflow which can be considerably longer than the linear flow time. However, if the pillar is an outflow cone wall, then it is surprising that the structure aligns with a radio continuum plume \citep[][, Lankhaar 2024]{Falstad18}. A linear shape well above the main disk also is unexpected for an outflow cone, since  expansion of the outflow will increase as the surrounding gas pressure drops, allowing the outflow to expand horizontally and break the linear cone walls \citep[e.g.,][]{TenorioTagle98}. Even so it is likely that a variety of gas outflows from the CON are largely responsible for the pillars and other complex dust structures above the center of Zw~049.057. 

\begin{deluxetable}{lcccccc}
\tablecaption{Obscuration and Gas Masses in Zw~049.057 Extraplanar Absorption Features\label{tab:dustpillars}}
\tablehead{
\colhead{Region} & \colhead{Adopted $\tau_V$} & \colhead{Area arcsec$^2$} & \colhead{Area pc$^2$} &\colhead{N$_H$}cm$^{-2}$ & \colhead {n$_H$} cm$^{-3}$ & \colhead{Estimated M$_{gas}$/M$_{\odot}$}
}
\startdata
Central Region & 0.8  & 97 & 8$\times$10$^6$ & 1.4$\times$ 10$^{21}$ & &10$^8$  \\
& & & & & & \\
Pillar-1: & & & & & \\
1A & 1.9$\pm$0.2 & 0.042 & 3500 & 3.1 $\times$ 10$^{21}$ &  17   & 1.2 $\times$ 10$^5$\\
1B & 1.8$\pm$0.2 & 0.034 & 2800 & 1.7 $\times$ 10$^{21}$ &  10  & 5.3 $\times$ 10$^4$\\
1F & $\geq$2     & 0.040 & 3300 & 2.2 $\times$ 10$^{21}$ &   13  & $\geq$8 $\times$ 10$^4$ \\
1G & 2.1$\pm$0.3 & 0.028 & 2300 & 1.7 $\times$ 10$^{21}$ &  11   & 4.3 $\times$ 10$^4$\\
1H & 1.3$\pm$0.2 & 0.042 & 3500 & 2.2 $\times$ 10$^{21}$ &   12   &  8.4 $\times$ 10$^4$\\
1I & 1.0$\pm$0.1 & 0.030 & 2500 & 1.9 $\times$ 10$^{21}$ &   12   & 5.0 $\times$ 10$^4$\\
1J & 0.8$\pm$0.1 & 0.028 & 2300 &1.4 $\times$ 10$^{21}$ &   9   &  3.5 $\times$ 10$^4$\\
Sum 1A-J &   &   &  &  & & 5 $\times$ 10$^5$ \\
Pillar-1 total & 1.7: & 0.320 & 2.6 $\times$10$^4$ & 2.9 $\times$ 10$^{21}$ &    & $\geq$ 8 $\times$ 10$^5$\\
 & & & & & & \\
Pillar-2-mean: & & & & &  \\ 
Sum 2A-2I & 1.1$\pm$0.1 & 0.25 & 2.0$\times$10$^6$ &1.9$\times$ 10$^{21}$ & $\sim$10  & 4 $\times$ 10$^5$ \\
& & & & & & \\
Clouds-3:& & & & & \\
3A & 2: & 0.029 & 2400 & 4 $\times$ 10$^{21}$ &    & 1.0 $\times$ 10$^5$\\
3B & 2.0$\pm$0.3 & 0.034 & 2700 & 3.4 $\times$ 10$^{21}$ &   & 1.0 $\times$ 10$^5$\\
3E  & 1.4$\pm$0.2 & 0.035 & 2900 & 2.4 $\times$ 10$^{21}$ &      & 7.5 $\times$ 10$^4$\\
3F  & 2: & 0.043 & 3500 & 4 $\times$ 10$^{21}$ &      & 1.5 $\times$ 10$^5$\\
3G  & 2.0$\pm$0.3 & 0.030 & 2500 & 3.4 $ \times$ 10$^{21}$ &    &  9.3  $\times$ 10$^4$\\
3H & $>$2 & 0.032 & 2500 & $>$4 $\times$ 10$^{21}$ &     & $>$ 1$\times$ 10$^5$\\
Sum 3A-H &   &   &    &   & & $>$ 5 $\times$ 10$^5$\\
& & & & & & \\
Dust Arc Examples: & & & & & \\
1C   & 1.6$\pm$0.1 & 0.030 & 2400 & 2.7 $\times$ 10$^{21}$ &       & 8.4 $\times$ 10$^4$\\
1D & 1.5$\pm$0.1 & 0.022 & 1800 &2.6 $\times$ 10$^{21}$ &    & 5.3 $\times$ 10$^4$\\
1E & 2: & 0.034 & 2800 & 1.7 $\times$ 10$^{21}$ &      & 1 $\times$ 10$^5$\\
\enddata
\tablecomments{Regions and the lettered Subregions are defined in Figure \ref{fig:map}. Opacities are based on the embedded screen model for Pillar~1, Clouds-3, and Dust Arcs. A foreground screen model was applied to Pillar-2. See text for details of the approximations used to derive optical depths. As discussed in the text, optical depths are likely to be lower limits, especially for Pillar-1. The ``Pillar-1 total" entry is from a rectangular region encompassing the main pillar.  Gas masses are derived using a Galactic conversion factor from dust opacities to gas columns as discussed in \S \ref{sec:dustcenter}. The gas volume densities n$_H$ are derived for Pillar-1  and Pillar-2 assuming each subregion is rectangular with a depth equal to the $\sqrt{area}$.}
\end{deluxetable}

\subsubsection{Dusty Pillar-2}

Pillar-2 has a longer projected length ($\approx$ 900~pc) and is narrower in the F555W image ($\approx$50~pc) than the more visible Pillar-1. Located at position angle 277$^{\circ}\pm$3$^{\circ}$, it emerges close to the disk's minor axis and is offset by $\approx$26$^{\circ}$ from Pillar~1. Unlike Pillar~1, the presence of Pillar-2 is clear in color maps where it is bounded by a ridge of obscuration and shows reddened colors that are constant along its length to within 0.2 magnitudes. Pillar-2 likely is located in the foreground of our line of sight with only a modest optical depth. Therefore we utilized the foreground screen model with our adopted power law opacity representation that we applied to the measured regions to derive the average optical depths and associated quantities in Table \ref{tab:dustpillars}. Pillar-2 has a total gas mass of $\gtrsim$ 6$\times$10$^5$~M$_{\odot}$. The dust clouds are more clearly clumped along Pillar-2 than in Pillar-1 with a spacing of $\approx$0.4~arcseconds (100~pc). These opacity peaks could be due to internal instabilities in the flow or time variations in the mass loss rate that would take place on 100~kyr time scales.

\subsubsection{The Two Pillars of Zw~049.057}\label{sec:two_pillars}

Both of the Zw~049.057 pillars originate from near the center of the galaxy and have substantial dust optical depths indicating that they mainly consist of  molecular gas (see\ref{sec:two_pillars}). Production of both pillars by a single collimated outflow that changes direction over time, as seen in the NGC~1377 molecular jet \citep{Aalto16,Aalto20}, would be difficult to achieve. A discreet jump in projected angle of 23$^{\circ}$ plus an additional offset due to the variation in the location of the two pillar along our line of sight would require large changes in the outflow direction.  Furthermore, since both pillars have continuous structures,  switches in outflow direction would need to occur extremely quickly, in much less than a flow time.  In an additional complication, Pillar-2 crosses the main disk of Zw~049.057 about 0.5~arcseconds (150~pc) to the south of the location of the CON. If Pillar-2 originates as a collimated structure in the CON, it would need to change direction soon after being launched to avoid connecting to the CON when observed in projection. 

Alternatively, Pillar-2 could be associated with a collimated outflow from a second nucleus or a product of a wall compressed by the fast molecular outflow along the major axis that is observed close to the nucleus (see Lankhaar {\it et al.}, 2024 in preparation). We discuss these possibilities in \S \ref{sec:discussion}. Due to these uncertainties about the nature of Pillar-2, we do not include the mass of a hidden hemisphere Pillar-2 in our mass outflow estimates.

\subsubsection{Extraplanar Dark Clouds and Arcs}

Properties of a sample of extraplanar dark clouds in Clouds-3 of Figure \ref{fig:map} and the Dust Arcs (examples labeled in the 1st region of the Figure and in Figure~\ref{fig:zw049_general}) are included in Table \ref{tab:dustpillars}. These features are relatively opaque and therefore likely to be massive. Extraplanar dust in the Clouds-3 region is dominated by a roughly circular group of clouds located about 500~pc above the disk. Our analysis based on the embedded absorbing cloud model indicated that the outer Dust Arcs clumps C, D, I, and J have lower opacity and are likely to be in the foreground. As shown in Table~\ref{tab:dustpillars} the core of the Cloud-3 complex is optically thick and our model suggests it is embedded at approximately the same distance within the stellar distribution of Zw~049.057 as Pillar-1. The Cloud-3 structure could consist of material ejected by the starburst or be a part of a disturbed inner polar gas ring. However, the moderately high optical depths for the Clouds-3 favor a connection to the dense interstellar medium of the starburst disk as observed in other high SFR galaxies (e.g., NGC~253 and NGC~891 \cite{Sofue94,Howk97}).

\section{SALT Spectra of the Zw~049.057 Disk}\label{sec:centerspec}

We obtained a spectrum along the major axis using SALT to explore the central star forming disk in  Zw~049.057. We limit our discussion here to the spatial distributions and kinematics of the emission lines and the Na~D interstellar absorption doublet. Optical spectroscopic properties of Zw~049.057 will be discussed in detail by C. Wethers {\it et al.} (in preparation) based on VLT-MUSE integral field spectroscopy.

Red optical spectra near the center of Zw~049.057 show strong nebular emission lines from H$\alpha$, [NII], and [SII]  \cite[e.g., ][]{Martin88,Poggianti00}, e.g., Sloan Digital Sky Survey (SDSS) spectra from the BOSS spectrograph gives an H$\alpha$ emission line equivalent width of 47~\AA. The 2~arcsec diameter BOSS fiber, however, encompasses only part of the inner, optically bright star forming disk. Based on the observed H$\alpha$ 
equivalent width the central region of Zw~049.057 could be in a starburst state of intense star formation \citep[see also][]{Poggianti00}. The SDSS BOSS spectrum also detects faint [OIII] emission that is indicative of ionization leakage from an obscured AGN as observed in the CON galaxy NGC~4418 by \cite{Wethers24}. The presence of an AGN is also supported by the Chandra X-ray detection of an obscured central source \citep{Lehmer10}, central radio compact radio source \citep{Song22} and radio continuum plume \citep{Falstad18}. Unfortunately while these measurements detect the AGN, they do not lead to reliable estimates of its luminosity.

\begin{figure}[]
   \centering
    \includegraphics[width=0.70\textwidth]{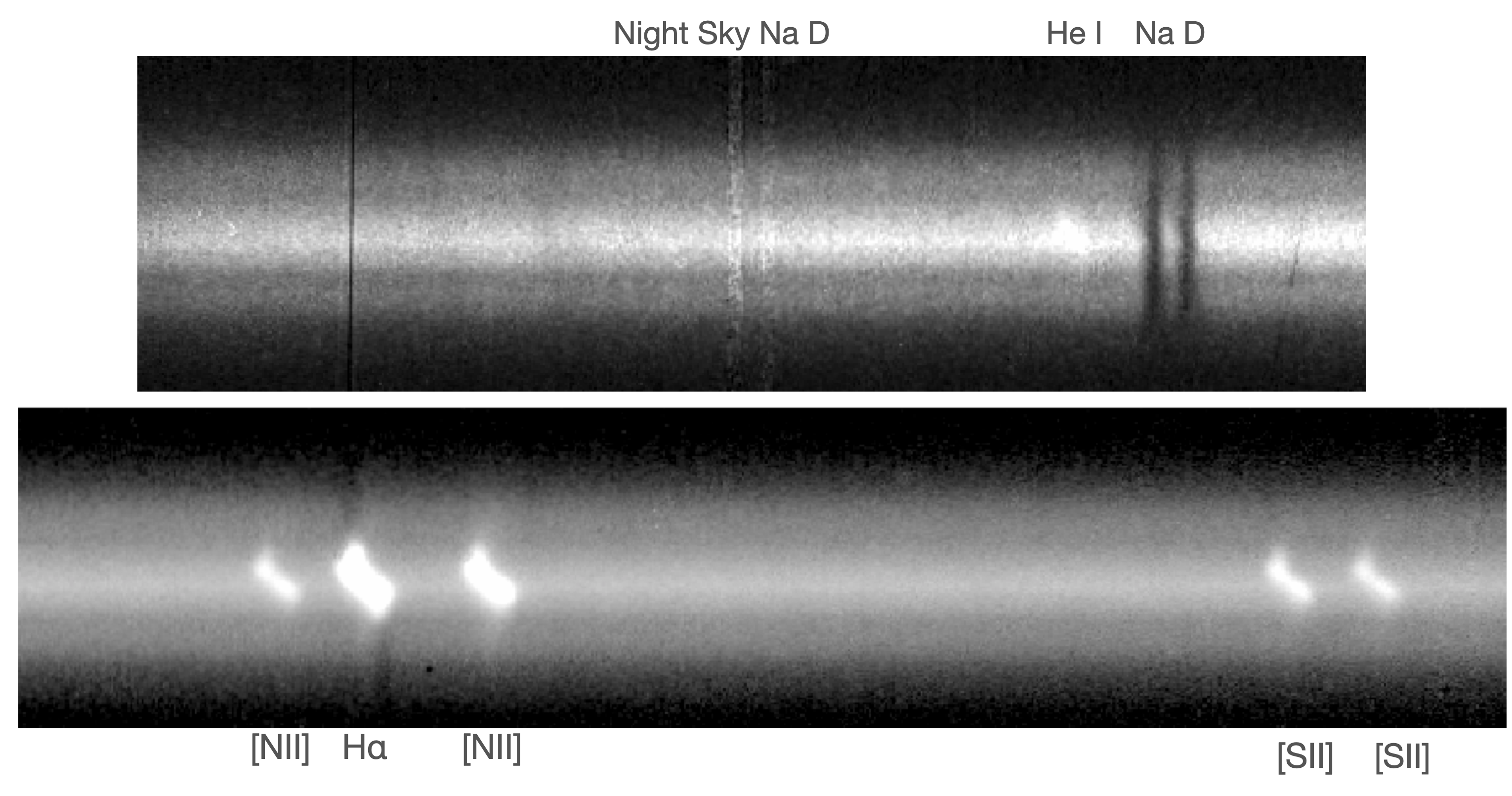}
   \caption{Illustrations of major axis spectra obtained with RSS on SALT. The upper panel shows the linear Na~D absorption lines and weak He~I $\lambda$ 5876~\AA\, emission that is inclined due to rotation. Faint stellar absorption lines showing pronounced rotation can be seen bluewards of the Na~D lines. The spectrum covers a 35~arcsecond region along the slit. We plan a full study of the stellar kinematics of Zw~049.057 using integral field spectra from MUSE (Wethers {\it et al.}, in preparation). The lower panel shows the strong nebular lines from [NII], H$\alpha$, and the [SII]. These features show rotation in the same sense as stellar absorption H$\alpha$ lines in the outer disk. Diffuse emission [NII] emission is visible beyond the central star forming  disk. The spatial width of the lower spectrum also is 35~arcseconds.} 
   \label{fig:zw049_spec_examples}
\end{figure}

Emission in the inner regions of Zw~049.057 is seen in red region spectra from SALT/RSS and  includes spatially extended emission lines of He~I, [NII], H$\alpha$, and [SII] (see Figure~\ref{fig:zw049_spec_examples}). All of the emission lines from the central zone of 
Zw~049.057 are tilted in wavelength, presenting the classic appearance of a rotating nuclear emission ring. Based on our spectrum, the diameter of the inner star forming disk is $\approx$6~arcseconds (1.7~kpc) as derived from the length of the linear spectrum that is produced by the rotating ring of emission. The emission line intensities are reduced in the heavily obscured nuclear region over a diameter of $\leq$430~pc based on the estimated seeing and our 1.5 arcsecond slit width. 

\begin{figure}[]
   \centering
    \includegraphics[width=0.90\textwidth]{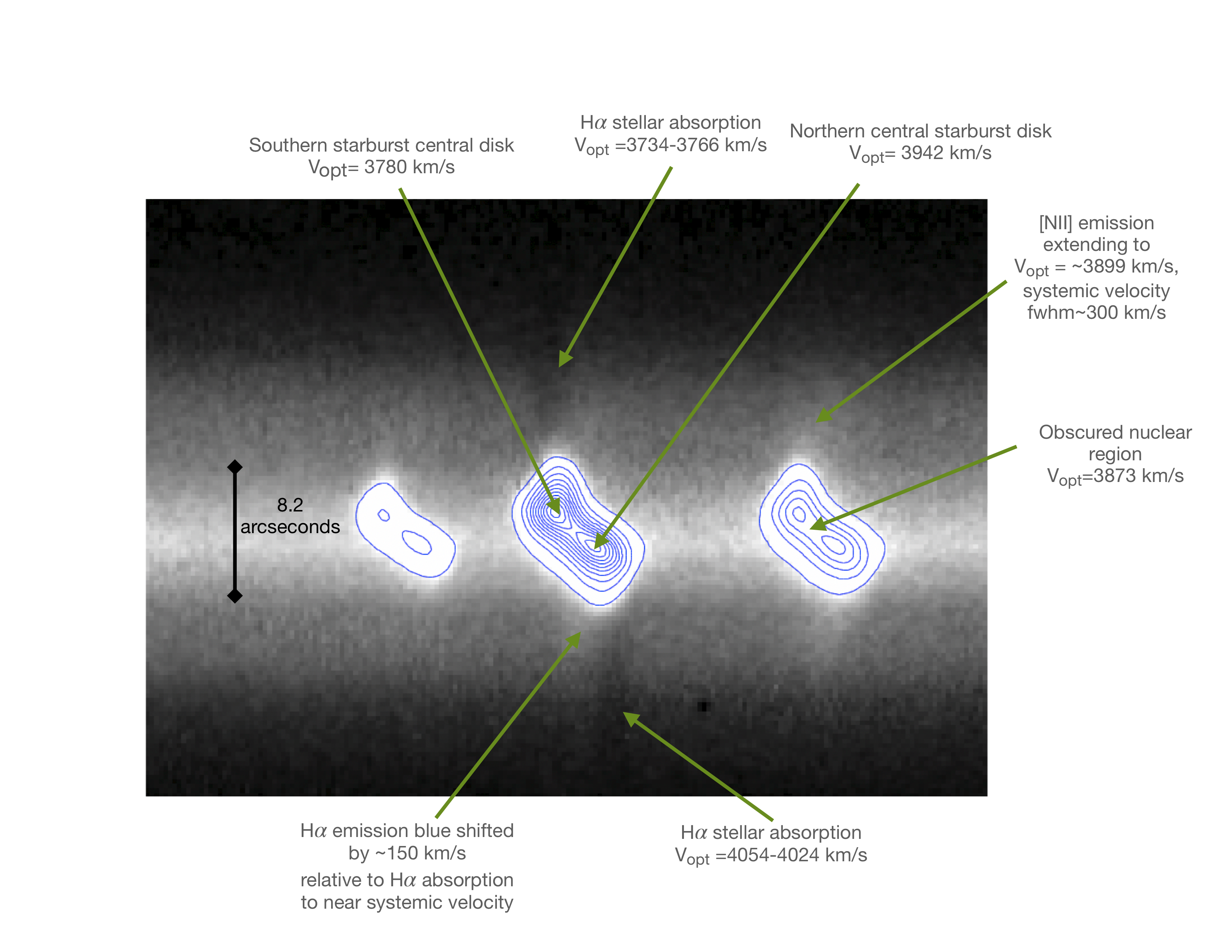}
   \caption{The SALT/RSS spectrum along the major axis of Zw~049.057 shows the presence of 3 main components of the system: the inner high SFR disk, H$\alpha$ absorption from the main disk, and diffuse emission at R$\gtrsim$3~arcseconds from gas near the systemic velocity. Red and blue wings of the [NII] $\lambda$ 6583~\AA, line at locations away from the center are due to the superposition of extended faint emission located at the systemic velocity of Zw~049.057.  The CON is located within the obscured nuclear region at a scale well below the angular resolution of the spectra.} 
   \label{fig:zw049_spec_summary}
\end{figure}

The H{\rm\,II} spectra of Zw~049.057 in the H$\alpha$ region are shown in Figure~\ref{fig:zw049_spec_summary}. These data illustrate the velocity shift across the disk and the emergence of faint blue and red emission line wings near the edge of the inner disk. The H$\alpha$ emission intensity comfortably exceeds that of [NII] throughout the central star forming zone, with a typical observed ratio of approximately 2:1. Accounting for underlying H$\alpha$ stellar absorption would further increase the H$\alpha$ strength; the H$\alpha$ to [NII] emission line intensity ratio is typical of stellar ionization. Beyond the central disk, the equivalent widths of the emission lines rapidly decline. For example, the peak equivalent width of [N II] $\lambda$ line is -15 to -17~\AA\ across the star forming zone, but drops to less than half of this value at radii 3~arcsec beyond the inner disk. The change in emission intensity relative to the stellar background change is associated with a drift of the emission towards the systemic velocity.

The actively star forming region in Zw~049.057 is several times larger than the typical galactic central molecular zones (CMZs, $\sim$200-400~pc). The $\sim$-25~\AA\ equivalent width of H$\alpha$ emission averaged over the star forming disk from our spectra indicates active star formation across the inner disk region  \citep[see also][]{Poggianti00}.  Radio data, however, indicate that the SFR beyond the CON-dominated core is not sufficiently high to be a major contributor to the bolometric luminosity and does not clearly qualify as a starburst on the basis of the H$\alpha$ emission equivalent width unless dust absorption is reducing its value by a factor of at least 4 \citep{Bergvall16}. The L-band continuum flux from the NVSS is 53.8~mJy in a 45~arcsecond beam \citep{Condon98} while \cite{Baan08} find a flux of 45.7 mJy in a 0.5~arcsecond beam. assuming that the radio source is not variable, we assigned an L-band flux of $\lesssim$ 10~mJy to the star forming disk beyond the nuclear region. Using the calibration by \citep{Kennicutt12}, the L-band radio measurements gave a SFR of $\lesssim$ 3~M$_{\odot}$~yr$^{-1}$, which provides approximately 10\% of the L$_{FIR}$ from Zw~049.057, while the optically hidden CON and its surroundings within r = 100~pc dominate the luminosity from Zw~049.057.

Figure~\ref{fig:zw049_spec_summary} shows how the kinematics of the H{\rm\,II} emission changes exterior to the star forming disk region and arises from a possibly diffuse ionized gas component centered near the systemic velocity, similar to the situation observed in the CON host galaxy NGC~4418 \citep{Ohyama19}. Our observed heliocentric velocity near the center of the star forming disk is v$_{helio} =$ 3881$\pm$8~km~s$^{-1}$ while the centroid of the velocity  across this disk gives 3888$\pm$10~km~s$^{-1}$. The latter measurement is less affected by dust, and we adopt v(sys)$_{helio} =$3888$\pm$10~km~s$^{-1}$. This is consistent with the velocities observed in molecular lines by \cite{Aalto15} who assumed v$_{helio}$=3900~km~s$^{-1}$ for their study and the v$_{helio}$=3898~km~s$^{-1}$ found by \cite{Mirabel88} from H{\rm\,I} absorption.  The absence of emission at the disk rotation velocity in the outer stellar disk beyond the central starburst zone and the lack of young stellar complexes in our F336W image show the main disk of Zw~049.057 does not support active star formation.

The projected velocity width along the major axis of the star forming disk and into the main disk is $\Delta v = $150$\pm$10~km~s$^{-1}$. Correcting for inclination gives a rotation speed of $V_{rot}$=160$\pm$10~km~s$^{-1}$. This is substantially lower than the $\Delta$v$_{0.5}$=200-250~km~s$^{-1}$ measured globally from the full CO~1-0 line width at zero intensity by \citep{HerreroIllana19}, who also find a characteristic double-peaked emission profile from a rotating disk with the intensity peaks at velocities close to the optical measurements.  The the CO~2-1 rotation curve in the inner 2-arcsec by \cite{Falstad18} has a peak rotation velocity of $\approx$150~km~s$^{-1}$ with turbulent gas and outflowing material contributing to the broader line wings. The molecular line data therefore agree with the our emission line galactic rotation speed in the disk of Zw~049.057.

We explored emission line widths using the [NII] $\lambda$6583 line because it is not affected by underlying stellar absorption. In the outer part of the star forming disk the [NII] fwhm are measured to reach to nearly 300~km~s$^{-1}$ albeit with modest reliability due to the faintness of the line. Even in the central position the [NII] line profile is only slightly asymmetric so the line is well fit by a Gaussian with a fwhm=180~km~s$^{-1}$. Our exploratory spectra therefore do not show obvious signs of a fast ionized wind from the CON, but we emphasize that this region is heavily dust-obscured and we are observing along the major axis.

\subsection{Kinematics of Interstellar Sodium}

Remarkably strong Na~D interstellar absorption lines are detected across 20~arcseconds (5.7~kpc) of the major axis of Zw~049.057. These lines are relatively narrow, with a full width at half maximum derived from Gaussian fits to the line core of 130~ km~s$^{-1}$. The Na~D lines are centered at the galaxy's systemic heliocentric radial velocity with 
v$_{helio}$=3895$\pm$8~km~s$^{-1}$ (see Figure~\ref{fig:zw049_spec_examples}). Absorption from Na~D does not show the disk rotation but is constant in velocity across the major axis to within 50~km~s$^{-1}$. Therefore it differs from the blue shifted, asymmetric or P-Cygni Na~D line profiles observed in cool galactic winds \citep{Martin05,Cazzoli16,Baron20}. It also differs from the situation in the CON galaxy NGC~4418 where the Na~D absorption contains a rotational signature \citep{Ohyama19}.  The Na~D absorptions in Zw~049.057, however, are similar in velocity width and radial velocity to the the diffuse [NII] emission detected beyond its starburst disk and the H{\rm\,I} absorption studied by \cite{Mirabel88}. These features trace a multi-phase interstellar gas component that does not display the rotational kinematics of the disk and is not part of a well-defined galactic outflow.

\section{The Nuclear Environment}\label{sec:nucleus}

\begin{figure}[]
   \centering
    \includegraphics[width=0.50\textwidth]{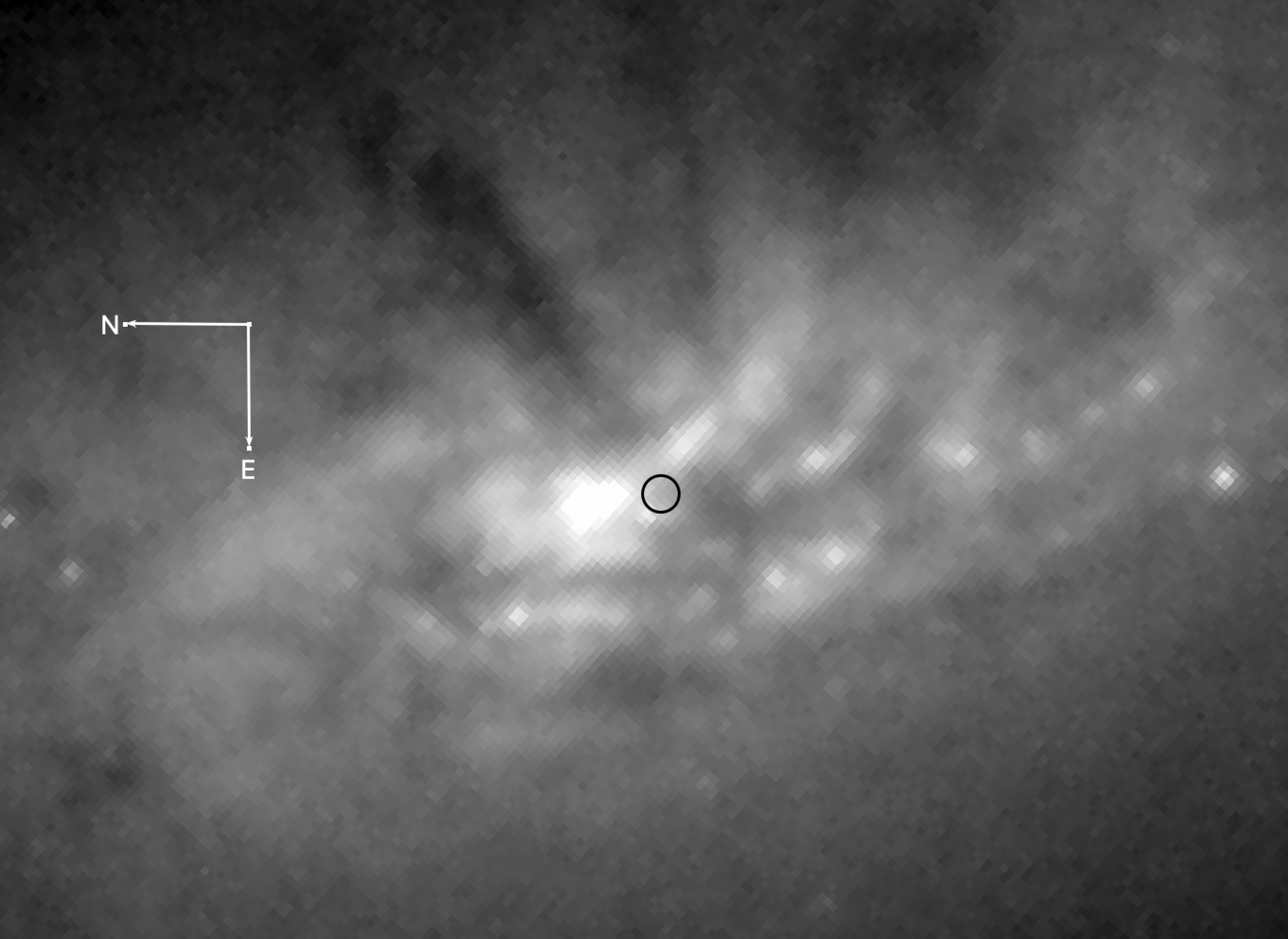} \hfill
   \caption{The location of the nucleus is shown with a circle with a radius of 0.1~arcseconds on the F814W image of Zw0~49.057. Dust Pillar-1 shows the high opacity dust spine on its southern edge. Pillar-1 is perpendicular to the bright arc that appears to be in front of the nucleus. This luminous arc is titled with respect to the disk plane of Zw~049.057. Dust Pillars-1 and -2 would be projected to meet $\approx$1~arcsecond east-southeast of the nucleus.The central disk contains a long, spiral arm-like feature (the blue streak in color maps) that extends from the nucleus in the north to the south-southwest. Knots along the arm are star forming sites, some of which are visible in our F336W image. The compass arms have lengths of 1~arcsecond.} 
   \label{fig:zw049_814wnucleus}
\end{figure}

The gas reservoir within the Zw~049.057 CON can be estimated from the model by \cite{GonzalezAlfonso19}. Assuming a column density of 10$^{24.8}$~cm$^{-2}$, a covering factor for the gas of $\eta$~steradians, and Galactic H/He, the nuclear reservoir mass is,
M$_{nuc,gas} \simeq$ 3.5 $\times$ 10$^8$ (R/R$_{20pc}$)$^2$ ($\eta/4\pi$) M$_{\odot}$.
For a CON we expect $(\eta/4\pi) \approx$ 1, with some leakage as shown by the presence of [OIII] line emission, and adopt M$_{nuc-gas}$= 3.5 $\times$ 10$^8$ M$_{\odot}$. The CON is well supplied with gas to support star formation, outflows, and/or power an AGN for significant time scales. It is fully obscured until at least millimeter wavelengths, complicating efforts to determine the nature of the power sources within the CON \citep[e.g.,][]{Aalto15}.. 

HST NICMOS-2 imaging obtained by \cite{Scoville00} showed the structure of the central star forming disk, including the region containing the invisible CON and demonstrated that the CON is located below the highest opacity dust pillar, a result that we confirm. The central region of the star forming disk appears to be relatively symmetric without a bar, but the modest resolution in the near infrared combined with the presence of dust limit our ability to detect inner stellar substructures. A single, well-defined spiral arm extends to a radius of 3.5~arcsec (1~kpc) along the southern major axis, which is seen as the "blue streak" in the optical images (see Figures \ref{fig:zw049_general}, and \ref{fig:zw049_pillars_color}). Bright U-band knots lying along this arm show that its visibility is enhanced by star formation even though this highly visible arm is substantially obscured by dust. The presence of an an asymmetric arm structure fits with the possibility of a relatively recent minor merger where the interloper recently entered into the central region of Zw~049.057 \citep[e.g.,][]{Mihos94,Laine99,Bournaud05}.

Using HST images with celestial coordinates updated with data from Gaia, we can locate the hidden CON on our images. Figure~\ref{fig:zw049_814wnucleus} shows the nucleus on our F814W image where we adopted a position of 15:13:13.092 +07:13:31.84 that is the location of the 690~GHz continuum peak found by \cite{Falstad18}. The circle has a radius of 0.1 arcsec that allows for residual positional offsets from the Gaia reference frame. As expected, the position of the CON is close to the lower boundary of Pillar-1 (see Figure \ref{fig:zw049_814wnucleus}). However, Pillar-2 is offset from the CON, which is located  to the northeast of the projected crossing point between Pillar-1 and Pillar-2. The Pillar crossing point of the two pillars is also near the convergence of two curved disk dust lanes located to the south and west of the center. This region may be part of a multi-phase polar wind from the center of Zw~049.057 where a molecular outflow was observed by \citep[][see also Lakhar {\it et al.} 2024, in preparation]{Falstad18}. Multiphase winds containing ionized and molecular components appear to be a common feature in LIRG CONs \citep{Ohyama19,Luo22,Wethers24}.

The CON's position places it near the center of a luminous arc-like feature that becomes more pronounced at longer wavelengths. As shown in Figure~\ref{fig:zw049_814wnucleus}, the luminous arc towards the CON is tilted with respect to the inner disk by approximately 20$^{\circ}$. Whether this angular offset is physical or due to varying levels of dust opacity is not clear. The brightness of this arc can be understood if it is an optically thick region that contains active star formation, in which case the surface brightness is determined by the source function, or ratio of stellar emissivity to dust opacity.

\section{Discussion}\label{sec:discussion}

\subsection{Zw~049.057: Structure \& Interactions}\label{sec:sub-discuss_struct}

Zw~049.057 is the main member of a small galaxy group. Its only neighbors are two undisturbed dwarfs at projected distances of $\sim$0.3~Mpc. We classify the stellar galaxy as an S0pec system on the basis of its disky structure, outer disk colors that are consistent with an older stellar population, and lack of spiral arms in its main stellar disk. The peculiar notation stems from the strong effects of extraplanar dust and presence of the dusty but high brightness star-forming inner disk. The main stellar disk has a scale length of $\approx$0.65~kpc, a half-light radius of 0.9~kpc, and, in agreement with previous work, a stellar mass of 1.5$\times$10$^{10}$~M$_{\odot}$. The most unusual stellar feature in Zw~049.057 is the nearly round outer stellar light distribution that appears at a radius of about 6~kpc. The lack of a large scale bar in the Zw~049.057 stellar system supports the hypothesis that its CON originated from an external interaction rather than internal secular evolution.

The CMZ and nucleus of Zw~049.057 are highly obscured by dust that is associated with a massive molecular interstellar medium. Our SALT spectra demonstrate that optically visible emission arising from beyond the CON and CMZ is brightest outside of the inner few hundred pc, and is in a rotating disk with V$_{rot}$ = 160~km~s$^{-1}$ that extends to r$\approx$1~kpc. Assuming a spherical mass distribution, the dynamical mass within r=1~kpc is  6$\times$10$^9$~M$_{\odot}$ or $\sim$40\% of the total stellar mass, consistent with the results from the SST photometric stellar mass estimates. This region is relatively gas-rich with M$_{molec}$/M$_{dyn}$ $\sim$0.3. Most of the structure within the star forming disk is due to dust, aside from the single, well-defined ``blue streak" arm that from extends the center to a radius of near 1~kpc. We do not detect a nuclear stellar bar, but such a feature could be hidden by dust. 

\subsubsection{Evidence for a Minor Merger}

Our HST images reveal a previously unknown polar ring component of the interstellar medium in Zw~049.057, previously classified as a non-interacting galaxy. The highest opacity part of the apparently star-free dusty polar ring crosses to the north of the galaxy center while a second, fainter crossing is present to the south. Polar ring structures are rare and originate from galaxy-galaxy interactions or mergers.  Given the absence of nearby companions, the presence of a polar ring demonstrates that the evolution Zw~049.057 was perturbed by a merger \citep[see, e.g,][for a discussion of similar issues for the ``spindle galaxy" NGC~2685]{Jozsa09}. Models show that when a galaxy disk is present, material from a merging companion  experiences tidal torques \citep[e.g.,][]{Arnaboldi03} These lead to twists towards the stellar disk plane and can feed material into the central zones of the galaxy \citep[e.g.,][]{Jozsa09,Sparke09}. Unfortunately, since the time scale for the disruption of polar rings can be as long as $\sim$Gyr \citep{Mapelli15}, the presence of a polar ring does not usefully constrain the time span since the merger occurred in Zw~049.057. 

A past merger also offers an explanation for the the circular outer isophotes in Zw~049.057.  An interaction within an intruder can produce rapid angle changes in the orbital plane of disk stars either through warping of preexisting stars or the addition of tidally stripped material from the intruder \citep{Sparke09}. Color measurements of the outer disk would help to distinguish its origin. Material from a  lower mass intruder is expected to be more metal-poor and thus bluer than the colors of the co-planar outer main disk of Zw~049.057, but our data are too shallow to provide this information. The central concentration of dense interstellar matter,  ``blue streak" inner arm, and possible misalignment of the central region containing the CON also can be understood in the context of a merger. The high degree of overall symmetry of the outer stellar body of Zw~049.057 further suggests that this was a minor merger that did not lead to strong disruption of the stellar disk.  As a minor merger, Zw~049.057 fits with general trends observed for dusty early-type galaxies. Most of these systems show evidence for the accretion of dusty gas, frequently from mergers with gas-rich dwarfs onto previously gas deficient early-type galaxies \citep[e.g.,][]{Kaviraj12,Finkelman12,Shabala12,Alatolo13,Glass22}. In this case the intruding galaxy should have entered its final stage, arriving into the center of  Zw~049.057, possibly producing the ``blue streak" arm. The time scale for this final stage in the merger process are set by dynamical processes and are in the range of 10-100s of Myr \citep{Laine99}.

Zw~049.057 therefore joins the classic LIRG CON galaxies NGC~4418 \citep{Boettcher19} and IC~860 \citep{Luo22} as products of mergers or past interactions. Mergers or interactions that compactify the interstellar medium in galaxies offer a channel to produce dense central gas concentrations that could be favorable for formation of CONs. In this sense the classic CONs in LIRGs could be considered to be an extension of the more frequent ULIRG CONs that are in merging galaxies. However, not all CONs in LIRGs are necessarily in galaxies that experienced significant recent interactions; e.g. the case of the non-interacting galaxy ESO~320-G030 \citep{GonzalezAlfonso21}.  Furthermore, the LIRG CON hosts fundamentally differ from most of the ULIRGs. Classic LIRG CONs are in early-type galaxies in the late stages of minor mergers versus ongoing mergers between massive galaxies in ULIRGs. The emergence of the CONs in the classic LIRGs appears to be delayed with respect to the peak of the merger activity.  Zw~049.057 and the other two classic LIRG CON mergers fit with the pattern for the activation of nuclei during late phase mergers in early-type galaxies \citep{George17,Sazanova21,Smercina22}. The current information therefore places the classic LIRG CONs in the sequence of minor-merger/interaction-induced star formation and nuclear activity that exists broadly among the population of moderate mass early-type galaxies in the local universe.

\subsection{The Dusty ISM}

\subsubsection{Central Dust Obscuration}

Dust absorption features within and above the central disk are the defining optical features of Zw~049.057. This material with a mass of approximately 2$\times$10$^8$~M$_{\odot}$ is likely to be largely in the form of H{\rm\,I} and the source of the observed H{\rm\,I} 21~cm absorption line. This medium is observed in the optical Na~D and diffuse [NII] emission lines that do not show evidence for significant rotational support. The Na~D, [NII] and H{\rm\,I} absorption lines have modest velocity dispersions of $\sigma$=50-60~km~s$^{-1}$ but no significant peculiar velocity offset from the central heliocentric velocity of 3890$\pm$ 10~km~s$^{-1}$ that we find for Zw~049.057. We do not detect an outflowing ionized galactic wind, but our data are limited because of the slit position lies along the major axis.

Since the polar ring is seen nearly edge on across the disk, it provides a possible source for the low velocity width neutral gas component. However, since the Na~D absorption covers $\approx$5.7~kpc of the main disk, diffuse gas in the polar ring would need to be more spatially extended than the polar dust lanes that are crossing the disk to produce a low velocity gradient.\footnote{A SALT spectrum that we obtained through the center of Zw~049.057 along the minor axis also lacks evidence for rotation in the Na~D absorption line as might be expected if the absorbing gas is in a polar ring.} This H{\rm\,I}-rich ISM component appears to be an extraplanar medium supported by turbulence  \citep[e.g.,][]{Boettcher19}, a point to be more fully explored once data on the full velocity field of the extraplanar neutral medium become available from the Wethers {\it et al.  in preparation} MUSE data.  Alternatively, non-rotating gas could be associated with the nearly circular outer stellar distribution that we interpret as a face-on disk component produced by warping during the past interaction. However, in this case it is not clear how or why the gas would have a high velocity dispersion and retain face-on kinematics when seen in projection against the moderately inclined main stellar disk of Zw~049.057

Even though the diffuse interstellar matter component of Zw~049.057 contains only about 10\% of the gas mass, its diffuse structure makes it an effective $\tau_V$ $\sim$0.8 obscuring screen across the central galaxy.  The injection of modest amounts of dusty gas into extended regions around galaxies is an efficient path to ultraviolet obscuration that does not require large amounts of energy or mass.  That the diffuse material has low or strongly misaligned angular momentum with respect to the disk is interesting in terms of its potential to feed matter into the CON that requires the collection substantial masses of low specific angular momentum gas into a radius of $\lesssim$100~pc. Since the medium is turbulent, dissipation is occurring, so the Zw~049.057 extraplanar medium is not likely to be a static structure. The possibility exists that gas with low or misaligned angular momentum falling back onto the disk from the extraplanar diffuse interstellar medium plays a role in fostering inflows that sustain the Zw~049.057 CON. 

\subsubsection{Dust and the Starburst}

Dust clumps are frequently present in the inner disk of Zw~049.057 and dusty structures with a variety of sizes and structures also extend above the star forming disk. The dense regions in our Clouds-3 location reach heights of $\approx$ 500~pc beyond the Zw~049.057 mid-plane and contain masses of $\sim$1 $\times$10$^5$~M$_{\odot}$. These dark clouds are likely associated with concentrations of massive stars as is seen in other galaxies with active star formation \citep[e.g.,][]{Sofue94,Howk97}. While lower opacity connections to the dense clouds reach higher elevations above the disk, it appears that the vertical extent of dusty gas ejected by young stars at r$>$100~pc is limited. This material is unlikely to be sustained above the disk and is likely to fall back into the disk in cycles driven by the local intensity of star formation. Despite the high gas densities, no evidence for ongoing star formation is found in our HST images of the extraplanar dusty regions.  

The low level extensions of dusty gas above the Zw~049.057 starburst, the dust clouds and arcs, combine with the molecular medium in the disk to block most of the light from the starburst. This effect can be seen in Figure~\ref{fig:zw049_814wnucleus} where multiple clumps dust clouds and lanes block most of the moderately inclined inner disk. The inner disk contributes to the high opacity of the galaxy, helping to reduce its optical/ultraviolet luminosity and lead to this system being an ultraviolet drop-out with (U$_{F336W,STMAG}$-V$_{F555W,STMAG}$) $\sim$ 2.  The presence of the actively star forming  disk is only quantitatively detected via the H$\alpha$ region emission in Zw~049.057, and even in this case the level of the star forming activity is difficult to estimate from the optical emission lines due to dust obscuration.

\subsection{Dust Pillar Outflows}

The pair of dust pillars in Zw~049.057 are direct evidence that the outflow from the nuclear region of Zw~049.057 feeds material out of the disk plane, extending to vertical distances of $\approx$ 1~kpc. The Zw~049.057 dust pillars are optically thick into the infrared and our analysis of their dust optical depths indicates they mainly consist of molecular material. Lower bound mass estimates for the gas in the two pillars based on simple dust absorption models are $\sim$1$\times$ 10$^6$~M$_{\odot}$ or $<$1\% of the total ISM mass. 

The physical nature of the pillars is not yet fully understood.  Lakhaar {\it et al.} (2024 in preparation) and \cite{Falstad18} show that C- and L-band radio radio continuum emission is present in the direction of Pillar-1, indicating that an outflow is present similar to the types found in other galaxies with gas-rich nuclei \citep[e.g.][]{Alatalo14,Luo22}. The radio continuum flux rises to lower frequencies and is likely to have a significant non-thermal emission component. The radio-molecular jet correlation in Zw~049.057 then suggests an outflow from an AGN that produces a non-thermal radio jet. If the pillars are individual outflows, then they remain remarkably well collimated over their lengths, a situation that is observed with ALMA by \cite{Aalto16} and \cite{Aalto20} in jets emerging from the dusty nucleus of the early-type galaxy NGC~1377. This requires a fast outflow velocity that may be detected by future molecular line observations. In this scenario, the presence of two outflow pillars requires two sources, e.g., possibly the presence of a second nucleus with an associated nuclear outflow or bi-directional outflows from a single nuclear region. 

Alternatively, the  dust ``V" extending from the center of Zw~049.057 towards the west could mark walls of interstellar gas compressed by the inner fast molecular wind along the minor axis recently mapped by Lankhaar {\it et al.} (2024).  If both pillars are part of a outflow cone, then ideally the point of the  cone's ``V" should be at the origin of the out-flowing wind. However, the centerline of the ``V" is slightly offset by $\approx$5-10$^{\circ}$ from the pole of the disk. An extension of the directions of Pillars -1 and -2 meets approximately 0.5~arcsecond (150~pc) to the south-southeast of the CON. If a conical structure is present, it does not have a simple symmetric structure.   The fast molecular wind observed from the CON in HCN is at PA = 105$^{\circ}$ and therefore lies between the two dust pillars. However, this region located between the pillars does not show clear signs of extra dust extinction (see Figures~\ref{fig:zw049_vijcolor} and \ref{fig:zw049_pillars_color}). Evidently the fast molecular wind either does not have a substantial dust optical depth or is not present at distances of $\gtrsim$100~pc from the CON where the HST imaging becomes useful. While a model where the pillars are associated with collimated outflows from the nuclear region of ZW~049.057 provides a natural explanation of their unusual structures, the presence of a nuclear molecular outflow along the minor axis of Zw~049.057 producing compressed walls cannot be excluded, especially for Pillar-2.

Assuming the pillars are collimated flows and that symmetry applies such that Pillar-1 emerge on both sides of Zw~049.057, the pillars contain $\gtrsim$2$\times$10$^6$~M$_{\odot}$. As discussed in \S4.5.2 for a collimated outflow,  $\dot{M}_{pillar} \simeq M_{pillar}/t_{flow}$ and an outflow velocity of 300~km~s$^{-1}$ leads to an estimate 
of a pillar mass feeding rate of $\dot{M}_{pillar} \geq$2~M$_{\odot}$~yr$^{-1}$. This value is a lower bound because we are not sensitive to optically thick regions, such as the spine in Pillar-1, that can carry much of the mass \citep[see][]{Aalto20}. If the wider opening angle, more diffuse, and faster molecular outflow traced by HCN escapes from the nuclear region, then mass loss rates from the CON will be considerably higher than those based our analysis of dust optical depths in the pillars.

In Zw~049.057, the termination of the nuclear outflows may not simply result from the velocity of the injected material, e.g., as in a ballistic model. First, the outflows lengths may reflect time scales during which the nuclear region ejects collimated flows. Second, gas in the diffuse medium or in polar orbits is present above the plane leading to shocks that could disrupt the collimated outflows. This process could provide a mechanism to feed energy into the diffuse neutral medium that would be needed to sustain its dispersion of 50-60~km~s$^{-1}$. It also would offer a process whereby gas from the nucleus return to the disk where its low specific angular momentum can stimulate inflows to the CON.
The relationship between nuclear outflow(s) and the H{\rm\,I}-dominated diffuse interstellar medium that covers 6.8 $\times$10$^6$~pc$^2$ in the center of Zw~049.057 merits further study in the context of galactic level gas recycling \citep{Aalto19}.  If this region is fed by the CON, even an outflow carrying $\dot{M}$ = 2-3~M$_{\odot}$~yr$^{-1}$ could produce the diffuse extraplanar ISM in $\sim$50~Myr, provided that the terminated outflows  expand to fill a nearly galaxy-wide volume.  

The lifetime of the CON in Zw~049.057 depends on the how long the nucleus can retain its shroud of dense interstellar matter. The length of Pillar-1 provides a minimum CON lifetime of a few million years. The upper limit on the CON lifetime depends on the still unknown balance over time between gas outflows and inflows. However, CON lifetimes of $\sim$100~Myr are possible if gas is efficiently recycled between the surrounding interstellar medium and CON \citep[e.g.,][]{Gorski24}. \cite{Falstad15} and Lakhaar {\it et al.} (2024) present evidence that the required gas inflows may exist for keeping the Zw~049.057 CON fueled for more than an internal gas loss time span.  Our observations bear on this issue by showing the possibility of a outflow-fallback recycling mechanism associated with the pillars in Zw~049.057, potentially aided by infall from the past merger and the associated presence of diffuse interstellar matter that has low or strongly misaligned angular momentum.

\subsubsection{The Nucleus}

The CON, as expected, is hidden from optical view. Its projected location lies within a luminous arc that could be the edge of a CMZ disk. If so, the tilt of the region suggests an offset in angle from the main disk of Zw~049.057 potentially associated with the merger. However, due to the presence of extensive dust absorption in the optical and near-infrared, this possibility should be confirmed by high angular resolution observations at longer wavelengths. The luminous appearance this arc shows that the region supports star formation, as is observed in other dust arcs in  the central disk of Zw~049.057. 
Even though the compact CON containing an AGN \citep{Song22} is completely obscured, its presence in Zw~049.057 is indirectly revealed at optical wavelengths by its connected kpc-scale, highly structured dust pillars in Zw~049.057. The presence of filamentary or collimated outflows from a nuclear region on kpc scales is a signature that can be used to survey for candidate classical CONs in early-type galaxies. 

The combination of high power density and dust opacity are key features of CONs. In Zw~049.057 these factors lead to the low ratio of M$_*$/L$_{IR}$ $\approx$ 0.05, suggesting that low M$_*$/L$_{IR}$ also may be a way to search for current epoch CONs in samples of dust obscured galaxies. This approach can be especially powerful when combined with mid-IR spectroscopy as discussed by \cite{GarciaBernete22} as well as the more standard molecular spectroscopy \citep[e.g.,][]{Falstad21}. The M$_*$/L$_{IR}$ selection method, however, may not apply to young star forming dusty galaxies seen at moderate-to-high redshifts. The discrimination between the contribution from a rapidly star forming population and a CON is likely to be less well defined in galaxies where dominant young stellar populations drive down the critical stellar M$_*$/L$_{bol}$ ratio and where compact galaxy sizes facilitate heavy levels of dust obscuration. For these systems measurement sensitive to line emission from ``greenhouse" regions with high submillimeter optical depths and high power densities are required to search for CONs.

\subsubsection{The Central Back Hole}

The high central molecular mass concentrations in combination with the high power outputs of CONs present a dynamic environment where central super massive black hole (SMBH) growth is possible. Unfortunately the  division of the CON's luminosity between recently formed stars, gravitational potential energy released by infalling gas, and accretion onto a SMBH is not known \citep{Gorski23}. As an extreme example, if 90\% of the luminosity from Zw~049.057 is from black hole accretion, then the SMBH accretion rate would be about 0.06/$\eta$~M$_{\odot}$~yr$^{-1}$ where $\eta$ is the SMBH efficiency factor for the conversion of mass into energy. Whether this results in substantial growth depends on the current mass of the central black hole and the lifetime of the CON phase. Presumably the nucleus will exit from the CON phase once its internal gas supply is depleted. An AGN in the Zw~049.057 \citep[e.g.,][]{Lehmer10,Song22} producing 0.9L$_{FIR}$ from accretion onto a SMBH might have the potential to sustain itself for ~10$^8$~yr provided that mass loss via winds is very modest. In this optimistic situation the central supermassive black hole (SMBH) would increase in mass by $\sim$6 $\times$10$^6$~M$_{\odot}$.  Issues include the time span over which this level of luminosity can be sustained by the AGN and the ability of the accreting SMBH to operate at sufficiently high luminosity. These factors in turn constrain the amount of mass growth by a SMBH during a CON evolutionary phase. 

A complication arises if the central AGN in Zw~049.057 follows the trends observed between the black hole Eddington factor $\lambda_{Edd} = L_{AGN}/L_{Edd, BH}$ and central gas column density found for AGN by \cite{Ricci2022}. Due to the increased opacity of dust, the Eddington limit is reduced and we would expect  log($\lambda_{Edd}) \lesssim -1.5$. This limit is set by luminosities where radiation pressure from the accreting  can readily eject the surrounding dusty gas \citep[e.g.,][]{Ricci2017,Ishibashi18,Venanzi20}. A SMBH producing 90\% of the luminosity in Zw~049.057 following this dust radiation pressure constraint needs to have M$_{bh}$ $\sim$10$^8$~M$_{\odot}$ and so would experience insignificant growth in mass even during a 100~Myr CON evolutionary time span.

This requirement is in tension with the likely SMBH mass. Adopting an upper limit to the bulge mass for Zw~049.057 based on our M$_{dyn}$-M$_{molec}$=4.5 $\times$ 10$^9$~M$_{\odot}$, and applying the bulge mass-SMBH mass calibration of \cite{Saglia16}  gives M$_{SMBH}$ $\lesssim$3 $\times$ 10$^{7}$~M$_{\odot}$ for Zw~049.057. The more modest mass SMBH that is expected in Zw~049.057, when constrained by radiation pressure limits on $\lambda_{Edd}$, would provide much less than half the  L$_{bol}$ in Zw~049.057.  Significant black hole mass growth via accretion in Zw~049.057 is most likely for low mass central black holes with M$\lesssim$10$^7$~M$_{\odot}$ operating at high $\lambda_{Edd}$ for 100~Myr time spans and avoiding catastrophic radiation-driven dusty gas loss.  By analogy with accretion onto massive protostars \citep[e.g.,][]{Krumholz05,Kuiper18,Gorski24}, several factors in Zw~049.057 could allow a high Eddington factor and significant SMBH mass growth without producing excessively rapid gas loss. For example, the dust envelope in the CON is unlikely to be spherical and homogeneous, spatially extended [OIII] optical emission is a signature of AGN ionizing radiation escaping through holes in the CON \citep[e.g.,][]{Wethers24}, a substantial molecular wind is observed that carries away momentum, the mass of gas in the CON can help to stabilize the system, and inflowing gas can produce ram pressure \citep[see][]{Aalto20,Gorski24}. Measurements of the mass distributions within the nucleus of Zw~049.057 and other CONs can test the high mass SMBH CON model and lead to better constraints on the evolutionary connections between CONs and SMBH growth.

\subsection{Evolution} 

\subsubsection{CON Formation}

Even though minor mergers are associated with CONs, the evolutionary processes that lead to CONs are not understood.  We considered the possibility that CONs could preferentially form in galaxies with massive nuclear star clusters and black holes. Alternatively, nucleus-nucleus interactions are known to produce CONs in major merger ULIRGs, and are a natural outcome when both galaxies have nuclei \citep{Downes98}. In minor mergers, the intruder will be low mass dwarf galaxy that is less likely to contain nuclei \citep{Matthews02,Reines13}, and when present, nuclei are not normally surrounded by a dense molecular medium. However, the offset dual pillars in Zw~049.057 allow for the possibility that a second nucleus is present and fostering the central concentration of gas around the original, more massive nucleus of Zw~049.057. 

In the merger scenario, early-type LIRGs such as Zw~049.057 are caught in a late phase of the merger process with a starburst in progress. Unfortunately, a simple correlation is not seen between merger phase and properties of the CONs in the 3 classic LIRG CONs NGC~4418, IC~860, and Zw~049.057. Instead variations exist within this small LIRG-CON sample between the dominance of the CON relative to its immediate surroundings. For example, NGC~4418 contains a CON but displays a strong post-starburst spectrum beyond its nuclear region \citep{Varenius14,Ohyama19,Boettcher20}. Its evolution may be in a different stage and possibly driven by different processes, such as gas accretion from a companion rather than a minor merger \citep{Boettcher20}. IC~860 has developed a strong outflow and its stellar body to be more highly disturbed than Zw~049.057, suggesting it is in an earlier phase of a merger than Zw~049.057 \citep[e.g.,][]{Luo22}. 

These differences might be understood if CONs form rarely during minor mergers, e.g., via an unusual channel such as interactions between dual nuclei, but once present have lifetimes in the 100~Myr range. Under this assumption CONs can be present throughout late merger/early post-merger stages, similar to the pattern for AGN to emerge in post-burst galaxies \citep{Yesuf14,Li19}.  The converse position would be for CONs to frequently form in LIRGs but to be short-lived. In this model the CON fraction in LIRGs of 10-20\% \citep{Falstad21} is proportional to their lifetimes relative to the $\sim$100~Myr  LIRG lifetimes  \citep{Marcillac06,Pereira15,Cortijo17}. A CON such as that in Zw~049.057 could readily live for 10-20~Myr after which masses in outflows would naturally lead to the rapid depletion of its cloaking gas. Short-lived CONs  would experience limited gas recycling, potentially in combination with star formation, and would not have time to contribute significantly to accretion-driven mass growth of a central SMBH.

If the limit imposed by radiation pressure must be met to be a CON, then any pre-CON nuclei  might also need to be sufficiently massive to overcome rapid gas loss due to radiation pressure as the CON forms. This condition could be achieved by a sufficiently massive black hole with a low $\lambda_{Edd}$ aided by a massive nuclear star cluster.  Therefore, the presence of unusually massive galaxy nuclei is a possible factor in the production of luminous CONs through their ability to retain dusty gas while operating at high luminosity.

\subsubsection{Time Scales}

The inner stellar disk is the site of optically detectable star formation in Zw~049.057 (see \S\ref{sec:centerspec} and \cite{Poggianti00,Scoville00, AlonsoH06}. Star formation in in early-type disk galaxies frequently is concentrated in the inner regions \citep{Rathore22}, so Zw~049.057 follows the usual pattern. Radio flux measurements show the spatially extended young stellar population has a SFR of about 3~M$_{\odot}$~yr$^{-1}$ that contributes only about 10\% the luminosity of Zw~049.057. The molecular mass of 1.5 $\times$10$^9$~M$_{\odot}$~yr$^{-1}$ in the inner disk gives a gas exhaustion time of 0.5~Gyr. Zw~049.057 is a marginal starburst if we limit our consideration to the inner star forming disk well beyond the CON.  However, for a normal stellar initial mass function, a total SFR of 24~M$_{\odot}$~yr$^{-1}$ is required to produce half of the luminosity from Zw~049.057, as, for example, assumed by \cite{GonzalezAlfonso19}. Since radio data show the bulk of the bolometric luminosity from Zw~049.057 arises from r$\lesssim$100~pc, most of the star formation in Zw~049.057 could be within this region that is dominated by the CON.  Without a CON, Zw~049.057 would be a normal early-type galaxy with active star formation that would lack the luminosity to be classified as a LIRG. 

Unfortunately, the division between power supplied by a central AGN and star formation within the CON cannot yet be disentangled. While the central SMBH cannot absorb a significant fraction of the CON's mass,  star formation is an especially important gas sink. For example, a CON region SFR of 20~M$_{\odot}$~yr$^{-1}$ with a normal stellar initial mass function would exhaust the Zw~049.057 CON's 3.5 $\times$ winds from a rapidly star 10$^8$~M$_{\odot}$ gas supply in about 17~Myr.  Winds from a rapidly star forming CON will further reduce its lifetime. Increasing the lifetime of our hypothetical high luminosity, star forming CON to 100~Myr would require extremely efficnet recycling and involvement of $\geq$50\% of the gas in the galaxy to be processed into stars. This extreme situation would leave a very massive, $\sim$10$^9$~M$_{\odot}$~yr$^{-1}$,  post-burst stellar population within r$\lesssim$100~pc. This extreme situation, however, is unlikely for a variety of reasons including issues  with angular momentum, although the stellar  masses in the centers of compact galaxies such as M32 approach this level \citep{Janz16}.

Once an early-type LIRG such as Zw~049.057 reaches the post-CON stage it may show little evidence of having been in a minor merger, and probably would have the stellar properties of a post-burst, early-type galaxy. Any remnants of extreme star formation or dense gas associated with the CON would be very centrally concentrated, the HI content is unlikely to be high, and the previously obscured nucleus may emerge as an AGN. If star formation were an important factor in the CON's evolution, the nuclear star cluster and its surroundings would be abnormally massive. In the case where the CON does not support extreme star formation, these properties are consistent with the established trends among post-burst early-type systems \citep[e.g.,][]{Kannappan04,Stark13,Sazanova21,Smercina22}. CON LIRGs are one evolutionary channel associated with processes leading to early-type post-burst galaxies but in ways where the central SMBH and nuclear star cluster may have experienced a profound and active period of star formation, mass growth, and feedback.

\section{Conclusions}\label{sec:conclusions}

Optical and near infrared imaging with HST and a major axis H$\alpha$ region spectrum taken with SALT reveal a variety of interesting features in the CON hosting LIRG Zw~049.057. As one of the nearest examples of a CON, the properties of this system provide new insights into the nature of interactions between a powerful CON and its host galaxy and associated feedback.

\raggedbottom

\begin{itemize}

\item Our discovery of a polar dust ring demonstrates that Zw~049.057, like several other LIRG CONs, is the product of a minor merger that led to the central concentration of the molecular interstellar medium in the inner kpc of Zw~049.057.

\item  Diffuse dust absorption with $\tau_V \approx$0.8,  corresponding to a gas mass of $\gtrsim$2 $\times 10^8$~M$_{\odot}$, screens the central $\sim$2~kpc$^2$ of Zw~049.057 at optical and shorter wavelengths.  This complex, turbulent medium has a multi-phase structure including ionized gas, material injected from the CON and from the star-forming disk, as well as interstellar matter associated with the polar merger.  

\item Na~D interstellar absorption observed along the major axis does not participate in the rotation of the Zw~049.057 disk and kinematically resembles the gas observed in HI 21~cm absorption. The Na~D absorption traces a reservoir of low angular momentum gas that could promote gas inflows to the center of the galaxy and thereby feed the CON.

\item High equivalent widths of the H$\alpha$ and [N II] emission lines indicate a recent enhancement in star formation activity at (r $\gtrsim$300~pc) that also leads to injections of dusty gas to near kpc heights above the disk plane. However, the disk SFR of 3~M$_{\odot}$~yr$^{-1}$ from beyond the nuclear region does not suffice to contribute significantly to the L$_{FIR}$ =10$^{11.5}$~L$_{\odot}$ from Zw~049.057.

\item Two distinctive linear ``dust pillars" extend from the region of the CON to kpc heights above the disk. Pillar-1 originates from the projected position of the CON while Pillar-2 that lies close to the galaxy's minor axis is offset from Pillar-1 by about 100~pc in projection as well as in angle and depth along the line of sight. Pillar-1 has properties of a collimated outflow coming from the CON.  Pillar 2 could arise from a curved outflow from the CON, a second unobserved nucleus, or the compressed wall formed by the fast nuclear molecular outflow. The collimated outflow model for pillars gives mass flow rates from the CON region of $>$2~ M$_{\odot}$~yr$^{-1}$. Independent of the specifics of their origins, dust pillars are optically visible signatures of ongoing large scale mechanical and mass feedback from the hidden CON in Zw~049.057. The presence of dust pillar absorption features emanating from galaxy nuclear regions offers a way to find candidate CONs via optical imaging.

\item The termination of both dust pillars into more diffuse structures within heights of $\lesssim$1~kpc indicates that gas ejected from the nuclear region does not readily escape the galaxy. Confinement of ejected gas to the inner galaxy is favorable for gas recycling between the CON and the wider interstellar medium. A minimum CON lifetime of 3~Myr comes from our estimates of flow times along the dust pillars, while the maximum CON lifetime is not constrained.

\item  The region within r $\approx$100~pc of the CON provides most of the luminosity from Zw~049.057. The compact size and radio properties of the CON suggests the presence of a powerful central AGN, possibly in combination with intense, compact star formation. The powerful nuclear region contains only a small fraction of the galaxy's stellar mass, leading to the low M$_*$/L$_{FIR}$ = 0.05 in Zw~049.057. A low M$_*$/L$_{FIR}$ ratio is potentially a useful marker for the presence of candidate CONs in LIRGs.

\item If young stars are a significant power source in the CON, then a SFR of $\sim$20~M$_{\odot}$~yr$^{-1}$ is required to produce most of the observed L$_{FIR}$ from Zw~049.057. Such a high SFR would lead to rapid growth of a massive stellar complex within a r$\lesssim$100~pc radius region and rapid depletion of the CON's gas supply. 

\item Radiation pressure on dust can be a factor in CONs. In the case of a spherical dust distribution, radiation pressure limits the SMBH Eddington factor and maximum luminosity of the AGN to levels well below the observed L$_{FIR}$. This constraint could be overcome by the gravitation from a dust-enveloped SMBH with M$_{SMBH}$ $\sim$10$^8$~M$_{\odot}$, that would be unusually massive for a galaxy like Zw~049.057, or by a CON structure such that a central AGN avoids the radiation pressure limit. In either of these two cases the SMBH then could accrete with a high Eddington factor thereby enabling the AGN to provide much of L$_{FIR}$  for Zw~049.057, and possibly promoting substantial growth if the SMBH has a mass in the 10$^6$~M$_{\odot}$ range.

\end{itemize}

\begin{acknowledgements}
We thank our colleagues and especially Kazushi Sakamoto and Nick Scoville for informative discussions of CONs and their relationships to properties of Zw~049.057. Our appreciation goes to Françoise Combes for providing important suggestions for improving the manuscript and to the referee for helpful reviews.
Based on observations made with the NASA/ESA Hubble Space Telescope, obtained [from the Data Archive] at the Space Telescope Science Institute, which is operated by the Association of Universities for Research in Astronomy, Inc., under NASA contract NAS 5-26555. These observations are associated with program \#HST-GO-14728. Support for program HST-GO-14728 was provided by NASA through a grant from the Space Telescope Science Institute, which is operated by the Association of Universities for Research in Astronomy, Inc., under NASA contract NAS 5-26555. J.S.G., J. K., and L. L. are also thankful for funding of this research provided by the University of Wisconsin-Madison College of Letters and Science. RK gratefully acknowledges partial funding support from the National Aeronautics and Space Administration under project 80NSSC18K1498, and from the National Science Foundation under grants No 1852136 and 2150222. EG gratefully acknowledges  funding support from  the National Science Foundation under grant No 1852136. S. A., S. K., and C.W.  gratefully acknowledge funding from the European Research Council (ERC) under the European Union's Horizon 2020 research and innovation programme (grant agreement No. 789410). Some of the observations reported in this paper were obtained with the Southern African Large Telescope under program 2018-2-SCI-31 (PI:  J. Gallagher). The NSF’s NOIRLab is operated by the Association of Universities for Research in Astronomy (AURA) under a cooperative agreement with the National Science Foundation. Database access and other data services are provided by the Astro Data Lab. The Legacy Surveys consist of three individual and complementary projects: the Dark Energy Camera Legacy Survey (DECaLS; NOAO Proposal ID 2014B-0404; PIs: David Schlegel and Arjun Dey), the Beijing-Arizona Sky Survey (BASS; NOAO Proposal ID 2015A-0801; PIs: Zhou Xu and Xiaohui Fan), and the Mayall z-band Legacy Survey (MzLS; NOAO Proposal ID 2016A-0453; PI: Arjun Dey). DECaLS, BASS and MzLS together include data obtained, respectively, at the Blanco telescope, Cerro Tololo Inter-American Observatory, NSF's National Optical Infrared Astronomy Research Laboratory (NOIRLab); the Bok telescope, Steward Observatory, University of Arizona; and the Mayall telescope, Kitt Peak National Observatory, NOIRLab. The Legacy Surveys project is honored to be permitted to conduct astronomical research on Iolkam Du’ag (Kitt Peak), a mountain with particular significance to the Tohono O’odham Nation. This research has made use of the NASA/IPAC Infrared Science Archive, which is funded by the National Aeronautics and Space Administration and operated by the California Institute of Technology.  The National Radio Astronomy Observatory is a facility of the National Science Foundation operated under cooperative agreement by Associated Universities, Inc.
\end{acknowledgements}

{\it Software:} python \citep{Astropy13}, saoimage ds9 \citep{Joye03}, pySALT \citep{Crawford17}, iraf \citep{Tody86,Tody93}

{\it Facilities:} SALT-Robert Stobie Spectrograph \citep{Kobulnicky03}, HST-WFC3, STScI/MAST, IRSA/SEIP NOIRlab Legacy Explore, NASA-NED \citep{Helou91}, NASA ADS {\citep [see][]{Kurtz00}

\bibliography{zw049_dust_aasv2_dftfin}

\begin{thebibliography}{}
\expandafter\ifx\csname natexlab\endcsname\relax\def\natexlab#1{#1}\fi
\providecommand{\url}[1]{\href{#1}{#1}}
\providecommand{\dodoi}[1]{doi:~\href{http://doi.org/#1}{\nolinkurl{#1}}}
\providecommand{\doeprint}[1]{\href{http://ascl.net/#1}{\nolinkurl{http://ascl.net/#1}}}
\providecommand{\doarXiv}[1]{\href{https://arxiv.org/abs/#1}{\nolinkurl{https://arxiv.org/abs/#1}}}

\bibitem[{{Aalto} {et~al.}(2015){Aalto}, {Mart{\'\i}n}, {Costagliola},
  {Gonz{\'a}lez-Alfonso}, {Muller}, {Sakamoto}, {Fuller},
  {Garc{\'\i}a-Burillo}, {van der Werf}, {Neri}, {Spaans}, {Combes}, {Viti},
  {M{\"u}hle}, {Armus}, {Evans}, {Sturm}, {Cernicharo}, {Henkel}, \&
  {Greve}}]{Aalto15}
{Aalto}, S., {Mart{\'\i}n}, S., {Costagliola}, F., {et~al.} 2015, \aap, 584,
  A42, \dodoi{10.1051/0004-6361/201526410}

\bibitem[{{Aalto} {et~al.}(2016){Aalto}, {Costagliola}, {Muller}, {Sakamoto},
  {Gallagher}, {Dasyra}, {Wada}, {Combes}, {Garc{\'\i}a-Burillo}, {Kristensen},
  {Mart{\'\i}n}, {van der Werf}, {Evans}, \& {Kotilainen}}]{Aalto16}
{Aalto}, S., {Costagliola}, F., {Muller}, S., {et~al.} 2016, \aap, 590, A73,
  \dodoi{10.1051/0004-6361/201527664}

\bibitem[{{Aalto} {et~al.}(2019){Aalto}, {Muller}, {K{\"o}nig}, {Falstad},
  {Mangum}, {Sakamoto}, {Privon}, {Gallagher}, {Combes}, {Garc{\'\i}a-Burillo},
  {Mart{\'\i}n}, {Viti}, {van der Werf}, {Evans}, {Black}, {Varenius},
  {Beswick}, {Fuller}, {Henkel}, {Kohno}, {Alatalo}, \& {M{\"u}hle}}]{Aalto19}
{Aalto}, S., {Muller}, S., {K{\"o}nig}, S., {et~al.} 2019, \aap, 627, A147,
  \dodoi{10.1051/0004-6361/201935480}

\bibitem[{{Aalto} {et~al.}(2020){Aalto}, {Falstad}, {Muller}, {Wada},
  {Gallagher}, {K{\"o}nig}, {Sakamoto}, {Vlemmings}, {Ceccobello}, {Dasyra},
  {Combes}, {Garc{\'\i}a-Burillo}, {Oya}, {Mart{\'\i}n}, {van der Werf},
  {Evans}, \& {Kotilainen}}]{Aalto20}
{Aalto}, S., {Falstad}, N., {Muller}, S., {et~al.} 2020, \aap, 640, A104,
  \dodoi{10.1051/0004-6361/202038282}

\bibitem[{{Alatalo} {et~al.}(2013){Alatalo}, {Davis}, {Bureau}, {Young},
  {Blitz}, {Crocker}, {Bayet}, {Bois}, {Bournaud}, {Cappellari}, {Davies}, {de
  Zeeuw}, {Duc}, {Emsellem}, {Khochfar}, {Krajnovi{\'c}}, {Kuntschner},
  {Lablanche}, {Morganti}, {McDermid}, {Naab}, {Oosterloo}, {Sarzi}, {Scott},
  {Serra}, \& {Weijmans}}]{Alatolo13}
{Alatalo}, K., {Davis}, T.~A., {Bureau}, M., {et~al.} 2013, \mnras, 432, 1796,
  \dodoi{10.1093/mnras/sts299}

\bibitem[{{Alatalo} {et~al.}(2014){Alatalo}, {Nyland}, {Graves}, {Deustua},
  {Shapiro Griffin}, {Duc}, {Cappellari}, {McDermid}, {Davis}, {Crocker},
  {Young}, {Chang}, {Scott}, {Cales}, {Bayet}, {Blitz}, {Bois}, {Bournaud},
  {Bureau}, {Davies}, {de Zeeuw}, {Emsellem}, {Khochfar}, {Krajnovi{\'c}},
  {Kuntschner}, {Morganti}, {Naab}, {Oosterloo}, {Sarzi}, {Serra}, \&
  {Weijmans}}]{Alatalo14}
{Alatalo}, K., {Nyland}, K., {Graves}, G., {et~al.} 2014, \apj, 780, 186,
  \dodoi{10.1088/0004-637X/780/2/186}

\bibitem[{{Alonso-Herrero} {et~al.}(2006){Alonso-Herrero}, {Rieke}, {Rieke},
  {Colina}, {P{\'e}rez-Gonz{\'a}lez}, \& {Ryder}}]{AlonsoH06}
{Alonso-Herrero}, A., {Rieke}, G.~H., {Rieke}, M.~J., {et~al.} 2006, \apj, 650,
  835, \dodoi{10.1086/506958}

\bibitem[{{Arnaboldi} {et~al.}(1993){Arnaboldi}, {Capaccioli}, {Cappellaro},
  {Held}, \& {Sparke}}]{Arnaboldi03}
{Arnaboldi}, M., {Capaccioli}, M., {Cappellaro}, E., {Held}, E.~V., \&
  {Sparke}, L. 1993, \aap, 267, 21

\bibitem[{{Astropy Collaboration} {et~al.}(2013){Astropy Collaboration},
  {Robitaille}, {Tollerud}, {Greenfield}, {Droettboom}, {Bray}, {Aldcroft},
  {Davis}, {Ginsburg}, {Price-Whelan}, {Kerzendorf}, {Conley}, {Crighton},
  {Barbary}, {Muna}, {Ferguson}, {Grollier}, {Parikh}, {Nair}, {Unther},
  {Deil}, {Woillez}, {Conseil}, {Kramer}, {Turner}, {Singer}, {Fox}, {Weaver},
  {Zabalza}, {Edwards}, {Azalee Bostroem}, {Burke}, {Casey}, {Crawford},
  {Dencheva}, {Ely}, {Jenness}, {Labrie}, {Lim}, {Pierfederici}, {Pontzen},
  {Ptak}, {Refsdal}, {Servillat}, \& {Streicher}}]{Astropy13}
{Astropy Collaboration}, {Robitaille}, T.~P., {Tollerud}, E.~J., {et~al.} 2013,
  \aap, 558, A33, \dodoi{10.1051/0004-6361/201322068}

\bibitem[{{Baan} {et~al.}(2017){Baan}, {An}, {Kl{\"o}ckner}, \&
  {Thomasson}}]{Baan17}
{Baan}, W.~A., {An}, T., {Kl{\"o}ckner}, H.-R., \& {Thomasson}, P. 2017,
  \mnras, 469, 916, \dodoi{10.1093/mnras/stx895}

\bibitem[{{Baan} {et~al.}(1987){Baan}, {Henkel}, \& {Haschick}}]{Baan87}
{Baan}, W.~A., {Henkel}, C., \& {Haschick}, A.~D. 1987, \apj, 320, 154,
  \dodoi{10.1086/165531}

\bibitem[{{Baan} {et~al.}(2008){Baan}, {Henkel}, {Loenen}, {Baudry}, \&
  {Wiklind}}]{Baan08}
{Baan}, W.~A., {Henkel}, C., {Loenen}, A.~F., {Baudry}, A., \& {Wiklind}, T.
  2008, \aap, 477, 747, \dodoi{10.1051/0004-6361:20077203}

\bibitem[{{Baba} {et~al.}(2022){Baba}, {Imanishi}, {Izumi}, {Kawamuro},
  {Nguyen}, {Nakagawa}, {Isobe}, {Onishi}, \& {Matsumoto}}]{Baba22}
{Baba}, S., {Imanishi}, M., {Izumi}, T., {et~al.} 2022, \apj, 928, 184,
  \dodoi{10.3847/1538-4357/ac57c2}

\bibitem[{{Baron} {et~al.}(2020){Baron}, {Netzer}, {Davies}, \& {Xavier
  Prochaska}}]{Baron20}
{Baron}, D., {Netzer}, H., {Davies}, R.~I., \& {Xavier Prochaska}, J. 2020,
  \mnras, 494, 5396, \dodoi{10.1093/mnras/staa1018}

\bibitem[{{Bergvall} {et~al.}(2016){Bergvall}, {Marquart}, {Way}, {Blomqvist},
  {Holst}, {{\"O}stlin}, \& {Zackrisson}}]{Bergvall16}
{Bergvall}, N., {Marquart}, T., {Way}, M.~J., {et~al.} 2016, \aap, 587, A72,
  \dodoi{10.1051/0004-6361/201525692}

\bibitem[{{Boettcher} {et~al.}(2019){Boettcher}, {Gallagher}, \&
  {Zweibel}}]{Boettcher19}
{Boettcher}, E., {Gallagher}, J.~S., I., \& {Zweibel}, E.~G. 2019, \apj, 885,
  160, \dodoi{10.3847/1538-4357/ab4904}

\bibitem[{{Boettcher} {et~al.}(2020){Boettcher}, {Gallagher}, {Ohyama},
  {Varenius}, {Aalto}, {Falstad}, {K{\"o}nig}, {Sakamoto}, \&
  {Yoast-Hull}}]{Boettcher20}
{Boettcher}, E., {Gallagher}, John~S., I., {Ohyama}, Y., {et~al.} 2020, \aap,
  637, A17, \dodoi{10.1051/0004-6361/201834880}

\bibitem[{{Boizelle} {et~al.}(2017){Boizelle}, {Barth}, {Darling}, {Baker},
  {Buote}, {Ho}, \& {Walsh}}]{Boizelle17}
{Boizelle}, B.~D., {Barth}, A.~J., {Darling}, J., {et~al.} 2017, \apj, 845,
  170, \dodoi{10.3847/1538-4357/aa8266}

\bibitem[{{Bournaud} {et~al.}(2005){Bournaud}, {Jog}, \& {Combes}}]{Bournaud05}
{Bournaud}, F., {Jog}, C.~J., \& {Combes}, F. 2005, \aap, 437, 69,
  \dodoi{10.1051/0004-6361:20042036}

\bibitem[{{Buckley} {et~al.}(2008){Buckley}, {Barnes}, {Burgh}, {Crawford},
  {Cottrell}, {Kniazev}, {Nordsieck}, {O'Donoghue}, {Rangwala}, {S{\'a}nchez},
  {Sharples}, {Sheinis}, {V{\"a}is{\"a}nen}, \& {Williams}}]{Buckley08}
{Buckley}, D.~A.~H., {Barnes}, S.~I., {Burgh}, E.~B., {et~al.} 2008, in Society
  of Photo-Optical Instrumentation Engineers (SPIE) Conference Series, Vol.
  7014, Ground-based and Airborne Instrumentation for Astronomy II, ed. I.~S.
  {McLean} \& M.~M. {Casali}, 701407, \dodoi{10.1117/12.789438}

\bibitem[{{Calzetti}(1997)}]{Calzetti97}
{Calzetti}, D. 1997, \aj, 113, 162, \dodoi{10.1086/118242}

\bibitem[{{Calzetti}(2001)}]{Calzetti01}
---. 2001, \pasp, 113, 1449, \dodoi{10.1086/324269}

\bibitem[{{Calzetti} {et~al.}(2021){Calzetti}, {Battisti}, {Shivaei}, {Messa},
  {Cignoni}, {Adamo}, {Dale}, {Gallagher}, {Grasha}, {Grebel}, {Kennicutt},
  {Linden}, {{\"O}stlin}, {Sabbi}, {Smith}, {Tosi}, \& {Wofford}}]{Calzetti21}
{Calzetti}, D., {Battisti}, A.~J., {Shivaei}, I., {et~al.} 2021, \apj, 913, 37,
  \dodoi{10.3847/1538-4357/abf118}

\bibitem[{{Cazzoli} {et~al.}(2016){Cazzoli}, {Arribas}, {Maiolino}, \&
  {Colina}}]{Cazzoli16}
{Cazzoli}, S., {Arribas}, S., {Maiolino}, R., \& {Colina}, L. 2016, \aap, 590,
  A125, \dodoi{10.1051/0004-6361/201526788}

\bibitem[{{Chandar} {et~al.}(2023){Chandar}, {Caputo}, {Linden}, {Mok},
  {Whitmore}, {Calzetti}, {Elmegreen}, {Lee}, {Ubeda}, {White}, \&
  {Cook}}]{Chandar23}
{Chandar}, R., {Caputo}, M., {Linden}, S., {et~al.} 2023, \apj, 943, 142,
  \dodoi{10.3847/1538-4357/acac96}

\bibitem[{{Condon} {et~al.}(1998){Condon}, {Cotton}, {Greisen}, {Yin},
  {Perley}, {Taylor}, \& {Broderick}}]{Condon98}
{Condon}, J.~J., {Cotton}, W.~D., {Greisen}, E.~W., {et~al.} 1998, \aj, 115,
  1693, \dodoi{10.1086/300337}

\bibitem[{{Cortijo-Ferrero} {et~al.}(2017){Cortijo-Ferrero}, {Gonz{\'a}lez
  Delgado}, {P{\'e}rez}, {Cid Fernandes}, {Garc{\'\i}a-Benito}, {Di Matteo},
  {S{\'a}nchez}, {de Amorim}, {Lacerda}, {L{\'o}pez Fern{\'a}ndez}, \&
  {Tadhunter}}]{Cortijo17}
{Cortijo-Ferrero}, C., {Gonz{\'a}lez Delgado}, R.~M., {P{\'e}rez}, E., {et~al.}
  2017, \aap, 607, A70, \dodoi{10.1051/0004-6361/201731217}

\bibitem[{{Costagliola} {et~al.}(2015){Costagliola}, {Sakamoto}, {Muller},
  {Mart{\'\i}n}, {Aalto}, {Harada}, {van der Werf}, {Viti}, {Garcia-Burillo},
  \& {Spaans}}]{Costagliola15}
{Costagliola}, F., {Sakamoto}, K., {Muller}, S., {et~al.} 2015, \aap, 582, A91,
  \dodoi{10.1051/0004-6361/201526256}

\bibitem[{{Crawford}(2017)}]{Crawford17}
{Crawford}, S.~M. 2017, in Astronomical Society of the Pacific Conference
  Series, Vol. 512, Astronomical Data Analysis Software and Systems XXV, ed.
  N.~P.~F. {Lorente}, K.~{Shortridge}, \& R.~{Wayth}, 375

\bibitem[{{Crawford} {et~al.}(2010){Crawford}, {Still}, {Schellart}, {Balona},
  {Buckley}, {Dugmore}, {Gulbis}, {Kniazev}, {Kotze}, {Loaring}, {Nordsieck},
  {Pickering}, {Potter}, {Romero Colmenero}, {Vaisanen}, {Williams}, \&
  {Zietsman}}]{Crawford10}
{Crawford}, S.~M., {Still}, M., {Schellart}, P., {et~al.} 2010, in Society of
  Photo-Optical Instrumentation Engineers (SPIE) Conference Series, Vol. 7737,
  Observatory Operations: Strategies, Processes, and Systems III, ed. D.~R.
  {Silva}, A.~B. {Peck}, \& B.~T. {Soifer}, 773725, \dodoi{10.1117/12.857000}

\bibitem[{{Donnan} {et~al.}(2023){Donnan}, {Rigopoulou}, {Garc{\'\i}a-Bernete},
  {Pereira-Santaella}, {Alonso-Herrero}, {Roche}, {Aalto},
  {Hern{\'a}n-Caballero}, \& {Spoon}}]{Donnan23}
{Donnan}, F.~R., {Rigopoulou}, D., {Garc{\'\i}a-Bernete}, I., {et~al.} 2023,
  \aap, 669, A87, \dodoi{10.1051/0004-6361/202244937}

\bibitem[{{Downes} \& {Solomon}(1998)}]{Downes98}
{Downes}, D., \& {Solomon}, P.~M. 1998, \apj, 507, 615, \dodoi{10.1086/306339}

\bibitem[{{Draine}(2011)}]{Draine11}
{Draine}, B.~T. 2011, {Physics of the Interstellar and Intergalactic Medium}

\bibitem[{{Eliche-Moral} {et~al.}(2018){Eliche-Moral},
  {Rodr{\'\i}guez-P{\'e}rez}, {Borlaff}, {Querejeta}, \& {Tapia}}]{Eliche18}
{Eliche-Moral}, M.~C., {Rodr{\'\i}guez-P{\'e}rez}, C., {Borlaff}, A.,
  {Querejeta}, M., \& {Tapia}, T. 2018, \aap, 617, A113,
  \dodoi{10.1051/0004-6361/201832911}

\bibitem[{{Evans} {et~al.}(2003){Evans}, {Becklin}, {Scoville}, {Neugebauer},
  {Soifer}, {Matthews}, {Ressler}, {Werner}, \& {Rieke}}]{Evans01}
{Evans}, A.~S., {Becklin}, E.~E., {Scoville}, N.~Z., {et~al.} 2003, \aj, 125,
  2341, \dodoi{10.1086/374234}

\bibitem[{{Falstad} {et~al.}(2015){Falstad}, {Gonz{\'a}lez-Alfonso}, {Aalto},
  {van der Werf}, {Fischer}, {Veilleux}, {Mel{\'e}ndez}, {Farrah}, \&
  {Smith}}]{Falstad15}
{Falstad}, N., {Gonz{\'a}lez-Alfonso}, E., {Aalto}, S., {et~al.} 2015, \aap,
  580, A52, \dodoi{10.1051/0004-6361/201526114}

\bibitem[{{Falstad} {et~al.}(2018){Falstad}, {Aalto}, {Mangum}, {Costagliola},
  {Gallagher}, {Gonz{\'a}lez-Alfonso}, {Sakamoto}, {K{\"o}nig}, {Muller},
  {Evans}, \& {Privon}}]{Falstad18}
{Falstad}, N., {Aalto}, S., {Mangum}, J.~G., {et~al.} 2018, \aap, 609, A75,
  \dodoi{10.1051/0004-6361/201732088}

\bibitem[{{Falstad} {et~al.}(2019){Falstad}, {Hallqvist}, {Aalto}, {K{\"o}nig},
  {Muller}, {Aladro}, {Combes}, {Evans}, {Fuller}, {Gallagher},
  {Garc{\'\i}a-Burillo}, {Gonz{\'a}lez-Alfonso}, {Greve}, {Henkel}, {Imanishi},
  {Izumi}, {Mangum}, {Mart{\'\i}n}, {Privon}, {Sakamoto}, {Veilleux}, \& {van
  der Werf}}]{Falstad19}
{Falstad}, N., {Hallqvist}, F., {Aalto}, S., {et~al.} 2019, \aap, 623, A29,
  \dodoi{10.1051/0004-6361/201834586}

\bibitem[{{Falstad} {et~al.}(2021){Falstad}, {Aalto}, {K{\"o}nig}, {Onishi},
  {Muller}, {Gorski}, {Sato}, {Stanley}, {Combes}, {Gonz{\'a}lez-Alfonso},
  {Mangum}, {Evans}, {Barcos-Mu{\~n}oz}, {Privon}, {Linden},
  {D{\'\i}az-Santos}, {Mart{\'\i}n}, {Sakamoto}, {Harada}, {Fuller},
  {Gallagher}, {van der Werf}, {Viti}, {Greve}, {Garc{\'\i}a-Burillo},
  {Henkel}, {Imanishi}, {Izumi}, {Nishimura}, {Ricci}, \&
  {M{\"u}hle}}]{Falstad21}
{Falstad}, N., {Aalto}, S., {K{\"o}nig}, S., {et~al.} 2021, \aap, 649, A105,
  \dodoi{10.1051/0004-6361/202039291}

\bibitem[{{Finkelman} {et~al.}(2012){Finkelman}, {Brosch}, {Funes}, {Barway},
  {Kniazev}, \& {V{\"a}is{\"a}nen}}]{Finkelman12}
{Finkelman}, I., {Brosch}, N., {Funes}, J.~G., {et~al.} 2012, \mnras, 422,
  1384, \dodoi{10.1111/j.1365-2966.2012.20710.x}

\bibitem[{{Fulmer} {et~al.}(2017){Fulmer}, {Gallagher}, \&
  {Kotulla}}]{Fulmer17}
{Fulmer}, L.~M., {Gallagher}, J.~S., \& {Kotulla}, R. 2017, \aap, 598, A119,
  \dodoi{10.1051/0004-6361/201628070}

\bibitem[{{Gallagher} \& {Hunter}(1981)}]{Gallagher81}
{Gallagher}, J.~S., \& {Hunter}, D.~A. 1981, \aj, 86, 1312,
  \dodoi{10.1086/113012}

\bibitem[{{Garc{\'\i}a-Bernete} {et~al.}(2022){Garc{\'\i}a-Bernete},
  {Rigopoulou}, {Aalto}, {Spoon}, {Hern{\'a}n-Caballero}, {Efstathiou},
  {Roche}, \& {K{\"o}nig}}]{GarciaBernete22}
{Garc{\'\i}a-Bernete}, I., {Rigopoulou}, D., {Aalto}, S., {et~al.} 2022, \aap,
  663, A46, \dodoi{10.1051/0004-6361/202142749}

\bibitem[{{George}(2017)}]{George17}
{George}, K. 2017, \aap, 598, A45, \dodoi{10.1051/0004-6361/201629667}

\bibitem[{{Glass} {et~al.}(2022){Glass}, {Sansom}, {Davis}, \&
  {Popescu}}]{Glass22}
{Glass}, D. H.~W., {Sansom}, A.~E., {Davis}, T.~A., \& {Popescu}, C.~C. 2022,
  \mnras, 517, 5524, \dodoi{10.1093/mnras/stac3001}

\bibitem[{{Gonz{\'a}lez-Alfonso} \& {Sakamoto}(2019)}]{GonzalezAlfonso19}
{Gonz{\'a}lez-Alfonso}, E., \& {Sakamoto}, K. 2019, \apj, 882, 153,
  \dodoi{10.3847/1538-4357/ab3a32}

\bibitem[{{Gonz{\'a}lez-Alfonso} {et~al.}(2021){Gonz{\'a}lez-Alfonso},
  {Pereira-Santaella}, {Fischer}, {Garc{\'\i}a-Burillo}, {Yang},
  {Alonso-Herrero}, {Colina}, {Ashby}, {Smith}, {Rico-Villas},
  {Mart{\'\i}n-Pintado}, {Cazzoli}, \& {Stewart}}]{GonzalezAlfonso21}
{Gonz{\'a}lez-Alfonso}, E., {Pereira-Santaella}, M., {Fischer}, J., {et~al.}
  2021, \aap, 645, A49, \dodoi{10.1051/0004-6361/202039047}

\bibitem[{{Gordon}(2021)}]{Gordon21}
{Gordon}, K.~D. 2021, in Star Formation Rates of Galaxies, ed. A.~{Zezas} \&
  V.~Buat (Cambridge University Press), 96--111

\bibitem[{{Gorski} {et~al.}(2023){Gorski}, {Aalto}, {K{\"o}nig}, {Wethers},
  {Yang}, {Muller}, {Viti}, {Black}, {Onishi}, \& {Sato}}]{Gorski23}
{Gorski}, M.~D., {Aalto}, S., {K{\"o}nig}, S., {et~al.} 2023, \aap, 670, A70,
  \dodoi{10.1051/0004-6361/202245166}

\bibitem[{{Gorski} {et~al.}(2024){Gorski}, {Aalto}, {K{\"o}nig}, {Wethers},
  {Yang}, {Muller}, {Onishi}, {Sato}, {Falstad}, {Mangum}, {Linden}, {Combes},
  {Mart{\'\i}n}, {Imanishi}, {Wada}, {Barcos-Mu{\~n}oz}, {Stanley},
  {Garc{\'\i}a-Burillo}, {van der Werf}, {Evans}, {Henkel}, {Viti}, {Harada},
  {D{\'\i}az-Santos}, {Gallagher}, \& {Gonz{\'a}lez-Alfonso}}]{Gorski24}
---. 2024, arXiv e-prints, arXiv:2403.16759, \dodoi{10.48550/arXiv.2403.16759}

\bibitem[{{Greve} {et~al.}(2014){Greve}, {Leonidaki}, {Xilouris}, {Wei{\ss}},
  {Zhang}, {van der Werf}, {Aalto}, {Armus}, {D{\'\i}az-Santos}, {Evans},
  {Fischer}, {Gao}, {Gonz{\'a}lez-Alfonso}, {Harris}, {Henkel}, {Meijerink},
  {Naylor}, {Smith}, {Spaans}, {Stacey}, {Veilleux}, \& {Walter}}]{Greve14}
{Greve}, T.~R., {Leonidaki}, I., {Xilouris}, E.~M., {et~al.} 2014, \apj, 794,
  142, \dodoi{10.1088/0004-637X/794/2/142}

\bibitem[{{Hattori} {et~al.}(2004){Hattori}, {Yoshida}, {Ohtani}, {Sugai},
  {Ishigaki}, {Sasaki}, {Hayashi}, {Ozaki}, {Ishii}, \& {Kawai}}]{Hattori04}
{Hattori}, T., {Yoshida}, M., {Ohtani}, H., {et~al.} 2004, \aj, 127, 736,
  \dodoi{10.1086/381060}

\bibitem[{{Haynes} {et~al.}(2018){Haynes}, {Giovanelli}, {Kent}, {Adams},
  {Balonek}, {Craig}, {Fertig}, {Finn}, {Giovanardi}, {Hallenbeck}, {Hess},
  {Hoffman}, {Huang}, {Jones}, {Koopmann}, {Kornreich}, {Leisman}, {Miller},
  {Moorman}, {O'Connor}, {O'Donoghue}, {Papastergis}, {Troischt}, {Stark}, \&
  {Xiao}}]{Haynes18}
{Haynes}, M.~P., {Giovanelli}, R., {Kent}, B.~R., {et~al.} 2018, \apj, 861, 49,
  \dodoi{10.3847/1538-4357/aac956}

\bibitem[{{Helou} {et~al.}(1991){Helou}, {Madore}, {Schmitz}, {Bicay}, {Wu}, \&
  {Bennett}}]{Helou91}
{Helou}, G., {Madore}, B.~F., {Schmitz}, M., {et~al.} 1991, in Astrophysics and
  Space Science Library, Vol. 171, Databases and On-line Data in Astronomy, ed.
  M.~A. {Albrecht} \& D.~{Egret}, 89--106, \dodoi{10.1007/978-94-011-3250-3_10}

\bibitem[{{Herrero-Illana} {et~al.}(2019){Herrero-Illana}, {Privon}, {Evans},
  {D{\'\i}az-Santos}, {P{\'e}rez-Torres}, {U}, {Alberdi}, {Iwasawa}, {Armus},
  {Aalto}, {Mazzarella}, {Chu}, {Sanders}, {Barcos-Mu{\~n}oz}, {Charmandaris},
  {Linden}, {Yoon}, {Frayer}, {Inami}, {Kim}, {Borish}, {Conway}, {Murphy},
  {Song}, {Stierwalt}, \& {Surace}}]{HerreroIllana19}
{Herrero-Illana}, R., {Privon}, G.~C., {Evans}, A.~S., {et~al.} 2019, \aap,
  628, A71, \dodoi{10.1051/0004-6361/201834088}

\bibitem[{{Holmberg}(1958)}]{Holmberg58}
{Holmberg}, E. 1958, Meddelanden fran Lunds Astronomiska Observatorium Serie
  II, 136, 1

\bibitem[{{Howk} \& {Savage}(1997)}]{Howk97}
{Howk}, J.~C., \& {Savage}, B.~D. 1997, \aj, 114, 2463, \dodoi{10.1086/118660}

\bibitem[{{Imanishi} \& {Nakanishi}(2013)}]{Imanishi13}
{Imanishi}, M., \& {Nakanishi}, K. 2013, \aj, 146, 91,
  \dodoi{10.1088/0004-6256/146/4/91}

\bibitem[{{Ishibashi} {et~al.}(2018){Ishibashi}, {Fabian}, \&
  {Maiolino}}]{Ishibashi18}
{Ishibashi}, W., {Fabian}, A.~C., \& {Maiolino}, R. 2018, \mnras, 476, 512,
  \dodoi{10.1093/mnras/sty236}

\bibitem[{{Janz} {et~al.}(2016){Janz}, {Norris}, {Forbes}, {Huxor},
  {Romanowsky}, {Frank}, {Escudero}, {Faifer}, {Forte}, {Kannappan},
  {Maraston}, {Brodie}, {Strader}, \& {Thompson}}]{Janz16}
{Janz}, J., {Norris}, M.~A., {Forbes}, D.~A., {et~al.} 2016, \mnras, 456, 617,
  \dodoi{10.1093/mnras/stv2636}

\bibitem[{{Joye} \& {Mandel}(2003)}]{Joye03}
{Joye}, W.~A., \& {Mandel}, E. 2003, in Astronomical Society of the Pacific
  Conference Series, Vol. 295, Astronomical Data Analysis Software and Systems
  XII, ed. H.~E. {Payne}, R.~I. {Jedrzejewski}, \& R.~N. {Hook}, 489

\bibitem[{{J{\'o}zsa} {et~al.}(2009){J{\'o}zsa}, {Oosterloo}, {Morganti},
  {Klein}, \& {Erben}}]{Jozsa09}
{J{\'o}zsa}, G.~I.~G., {Oosterloo}, T.~A., {Morganti}, R., {Klein}, U., \&
  {Erben}, T. 2009, \aap, 494, 489, \dodoi{10.1051/0004-6361:200809372}

\bibitem[{{Kamenetzky} {et~al.}(2017){Kamenetzky}, {Rangwala}, \&
  {Glenn}}]{Kamenetzky17}
{Kamenetzky}, J., {Rangwala}, N., \& {Glenn}, J. 2017, \mnras, 471, 2917,
  \dodoi{10.1093/mnras/stx1595}

\bibitem[{{Kannappan}(2004)}]{Kannappan04}
{Kannappan}, S.~J. 2004, \apjl, 611, L89, \dodoi{10.1086/423785}

\bibitem[{{Karademir} {et~al.}(2019){Karademir}, {Remus}, {Burkert}, {Dolag},
  {Hoffmann}, {Moster}, {Steinwandel}, \& {Zhang}}]{Karademir19}
{Karademir}, G.~S., {Remus}, R.-S., {Burkert}, A., {et~al.} 2019, \mnras, 487,
  318, \dodoi{10.1093/mnras/stz1251}

\bibitem[{{Kaviraj} {et~al.}(2012){Kaviraj}, {Ting}, {Bureau}, {Shabala},
  {Crockett}, {Silk}, {Lintott}, {Smith}, {Keel}, {Masters}, {Schawinski}, \&
  {Bamford}}]{Kaviraj12}
{Kaviraj}, S., {Ting}, Y.-S., {Bureau}, M., {et~al.} 2012, \mnras, 423, 49,
  \dodoi{10.1111/j.1365-2966.2012.20957.x}

\bibitem[{{Keel} {et~al.}(2015){Keel}, {Maksym}, {Bennert}, {Lintott},
  {Chojnowski}, {Moiseev}, {Smirnova}, {Schawinski}, {Urry}, {Evans},
  {Pancoast}, {Scott}, {Showley}, \& {Flatland}}]{Keel15}
{Keel}, W.~C., {Maksym}, W.~P., {Bennert}, V.~N., {et~al.} 2015, \aj, 149, 155,
  \dodoi{10.1088/0004-6256/149/5/155}

\bibitem[{{Keel} {et~al.}(2023){Keel}, {Windhorst}, {Jansen}, {Cohen},
  {Summers}, {Holwerda}, {Bradford}, {Robertson}, {Ferrami}, {Wyithe}, {Yan},
  {Conselice}, {Driver}, {Robotham}, {Grogin}, {Willmer}, {Koekemoer}, {Frye},
  {Hathi}, {Ryan}, {Pirzkal}, {Marshall}, {Coe}, {Diego}, {Broadhurst},
  {Rutkowski}, {Wang}, {Willner}, {Petric}, {Cheng}, \& {Zitrin}}]{Keel23}
{Keel}, W.~C., {Windhorst}, R.~A., {Jansen}, R.~A., {et~al.} 2023, \aj, 165,
  166, \dodoi{10.3847/1538-3881/acbdff}

\bibitem[{{Kennicutt} \& {Evans}(2012)}]{Kennicutt12}
{Kennicutt}, R.~C., \& {Evans}, N.~J. 2012, \araa, 50, 531,
  \dodoi{10.1146/annurev-astro-081811-125610}

\bibitem[{{Kobulnicky} {et~al.}(2003){Kobulnicky}, {Nordsieck}, {Burgh},
  {Smith}, {Percival}, {Williams}, \& {O'Donoghue}}]{Kobulnicky03}
{Kobulnicky}, H.~A., {Nordsieck}, K.~H., {Burgh}, E.~B., {et~al.} 2003, in
  Society of Photo-Optical Instrumentation Engineers (SPIE) Conference Series,
  Vol. 4841, Instrument Design and Performance for Optical/Infrared
  Ground-based Telescopes, ed. M.~{Iye} \& A.~F.~M. {Moorwood}, 1634--1644,
  \dodoi{10.1117/12.460315}

\bibitem[{{Kotulla} {et~al.}(2010){Kotulla}, {Fritze}, {Weilbacher}, \&
  {Anders}}]{Kotulla2010}
{Kotulla}, R., {Fritze}, U., {Weilbacher}, P., \& {Anders}, P. 2010, {GALEV
  Evolutionary Synthesis Models}, Astrophysics Source Code Library, record
  ascl:1010.033.
\newblock \doeprint{1010.033}

\bibitem[{{Krumholz} {et~al.}(2005){Krumholz}, {McKee}, \&
  {Klein}}]{Krumholz05}
{Krumholz}, M.~R., {McKee}, C.~F., \& {Klein}, R.~I. 2005, \apjl, 618, L33,
  \dodoi{10.1086/427555}

\bibitem[{{Kuiper} \& {Hosokawa}(2018)}]{Kuiper18}
{Kuiper}, R., \& {Hosokawa}, T. 2018, \aap, 616, A101,
  \dodoi{10.1051/0004-6361/201832638}

\bibitem[{{Kurtz} {et~al.}(2000){Kurtz}, {Eichhorn}, {Accomazzi}, {Grant},
  {Murray}, \& {Watson}}]{Kurtz00}
{Kurtz}, M.~J., {Eichhorn}, G., {Accomazzi}, A., {et~al.} 2000, \aaps, 143, 41,
  \dodoi{10.1051/aas:2000170}

\bibitem[{{Kylafis} \& {Xilouris}(2005)}]{Kylafis05}
{Kylafis}, N.~D., \& {Xilouris}, E.~M. 2005, in American Institute of Physics
  Conference Series, Vol. 761, The Spectral Energy Distributions of Gas-Rich
  Galaxies: Confronting Models with Data, ed. C.~C. {Popescu} \& R.~J. {Tuffs},
  3--16, \dodoi{10.1063/1.1913911}

\bibitem[{{Laine} \& {Heller}(1999)}]{Laine99}
{Laine}, S., \& {Heller}, C.~H. 1999, \mnras, 308, 557,
  \dodoi{10.1046/j.1365-8711.1999.02712.x}

\bibitem[{{Larson} {et~al.}(2016){Larson}, {Sanders}, {Barnes}, {Ishida},
  {Evans}, {U}, {Mazzarella}, {Kim}, {Privon}, {Mirabel}, \&
  {Flewelling}}]{Larson16}
{Larson}, K.~L., {Sanders}, D.~B., {Barnes}, J.~E., {et~al.} 2016, \apj, 825,
  128, \dodoi{10.3847/0004-637X/825/2/128}

\bibitem[{{Lehmer} {et~al.}(2010){Lehmer}, {Alexander}, {Bauer}, {Brandt},
  {Goulding}, {Jenkins}, {Ptak}, \& {Roberts}}]{Lehmer10}
{Lehmer}, B.~D., {Alexander}, D.~M., {Bauer}, F.~E., {et~al.} 2010, \apj, 724,
  559, \dodoi{10.1088/0004-637X/724/1/559}

\bibitem[{{Leroy} {et~al.}(2019){Leroy}, {Sandstrom}, {Lang}, {Lewis}, {Salim},
  {Behrens}, {Chastenet}, {Chiang}, {Gallagher}, {Kessler}, \&
  {Utomo}}]{Leroy19}
{Leroy}, A.~K., {Sandstrom}, K.~M., {Lang}, D., {et~al.} 2019, \apjs, 244, 24,
  \dodoi{10.3847/1538-4365/ab3925}

\bibitem[{{Li} {et~al.}(2019){Li}, {French}, {Zabludoff}, \& {Ho}}]{Li19}
{Li}, Z., {French}, K.~D., {Zabludoff}, A.~I., \& {Ho}, L.~C. 2019, \apj, 879,
  131, \dodoi{10.3847/1538-4357/ab1f68}

\bibitem[{{Lu} {et~al.}(2017){Lu}, {Zhao}, {D{\'\i}az-Santos}, {Xu}, {Gao},
  {Armus}, {Isaak}, {Mazzarella}, {van der Werf}, {Appleton}, {Charmandaris},
  {Evans}, {Howell}, {Iwasawa}, {Leech}, {Lord}, {Petric}, {Privon}, {Sanders},
  {Schulz}, \& {Surace}}]{Lu17}
{Lu}, N., {Zhao}, Y., {D{\'\i}az-Santos}, T., {et~al.} 2017, \apjs, 230, 1,
  \dodoi{10.3847/1538-4365/aa6476}

\bibitem[{{Luo} {et~al.}(2022){Luo}, {Rowlands}, {Alatalo}, {Sazonova},
  {Abdurro'uf}, {Heckman}, {Medling}, {Deustua}, {Nyland}, {Lanz}, {Petric},
  {Otter}, {Aalto}, {Dimassimo}, {French}, {Gallagher}, {Roediger}, \&
  {Stepanoff}}]{Luo22}
{Luo}, Y., {Rowlands}, K., {Alatalo}, K., {et~al.} 2022, \apj, 938, 63,
  \dodoi{10.3847/1538-4357/ac8b7d}

\bibitem[{{Mangum} {et~al.}(2013{\natexlab{a}}){Mangum}, {Darling}, {Henkel},
  \& {Menten}}]{Mangum13}
{Mangum}, J.~G., {Darling}, J., {Henkel}, C., \& {Menten}, K.~M.
  2013{\natexlab{a}}, \apj, 766, 108, \dodoi{10.1088/0004-637X/766/2/108}

\bibitem[{{Mangum} {et~al.}(2013{\natexlab{b}}){Mangum}, {Darling}, {Henkel},
  {Menten}, {MacGregor}, {Svoboda}, \& {Schinnerer}}]{Magnum13b}
{Mangum}, J.~G., {Darling}, J., {Henkel}, C., {et~al.} 2013{\natexlab{b}},
  \apj, 779, 33, \dodoi{10.1088/0004-637X/779/1/33}

\bibitem[{{Mangum} {et~al.}(2008){Mangum}, {Darling}, {Menten}, \&
  {Henkel}}]{Mangum08}
{Mangum}, J.~G., {Darling}, J., {Menten}, K.~M., \& {Henkel}, C. 2008, \apj,
  673, 832, \dodoi{10.1086/524354}

\bibitem[{{Mapelli} {et~al.}(2015){Mapelli}, {Rampazzo}, \&
  {Marino}}]{Mapelli15}
{Mapelli}, M., {Rampazzo}, R., \& {Marino}, A. 2015, \aap, 575, A16,
  \dodoi{10.1051/0004-6361/201425315}

\bibitem[{{Marcillac} {et~al.}(2006){Marcillac}, {Elbaz}, {Charlot}, {Liang},
  {Hammer}, {Flores}, {Cesarsky}, \& {Pasquali}}]{Marcillac06}
{Marcillac}, D., {Elbaz}, D., {Charlot}, S., {et~al.} 2006, \aap, 458, 369,
  \dodoi{10.1051/0004-6361:20064996}

\bibitem[{{Martin}(2005)}]{Martin05}
{Martin}, C.~L. 2005, \apj, 621, 227, \dodoi{10.1086/427277}

\bibitem[{{Martin} {et~al.}(1988){Martin}, {Bottinelli}, {Dennefeld},
  {Gouguenheim}, {Handa}, {Le Squeren}, {Nakai}, \& {Sofue}}]{Martin88}
{Martin}, J.~M., {Bottinelli}, L., {Dennefeld}, M., {et~al.} 1988, \aap, 195,
  71

\bibitem[{{Mart{\'\i}n} {et~al.}(2016){Mart{\'\i}n}, {Aalto}, {Sakamoto},
  {Gonz{\'a}lez-Alfonso}, {Muller}, {Henkel}, {Garc{\'\i}a-Burillo}, {Aladro},
  {Costagliola}, {Harada}, {Krips}, {Mart{\'\i}n-Pintado}, {M{\"u}hle}, {van
  der Werf}, \& {Viti}}]{Martin16}
{Mart{\'\i}n}, S., {Aalto}, S., {Sakamoto}, K., {et~al.} 2016, \aap, 590, A25,
  \dodoi{10.1051/0004-6361/201528064}

\bibitem[{{Matthews} \& {Gallagher}(2002)}]{Matthews02}
{Matthews}, L.~D., \& {Gallagher}, J.~S., I. 2002, \apjs, 141, 429,
  \dodoi{10.1086/340647}

\bibitem[{{Mihos} \& {Hernquist}(1994)}]{Mihos94}
{Mihos}, J.~C., \& {Hernquist}, L. 1994, \apjl, 425, L13,
  \dodoi{10.1086/187299}

\bibitem[{{Mirabel} \& {Sanders}(1988)}]{Mirabel88}
{Mirabel}, I.~F., \& {Sanders}, D.~B. 1988, \apj, 335, 104,
  \dodoi{10.1086/166909}

\bibitem[{{Morganti} \& {Oosterloo}(2018)}]{Morganti18}
{Morganti}, R., \& {Oosterloo}, T. 2018, \aapr, 26, 4,
  \dodoi{10.1007/s00159-018-0109-x}

\bibitem[{{Natale} {et~al.}(2014){Natale}, {Popescu}, {Tuffs}, \&
  {Semionov}}]{Natale14}
{Natale}, G., {Popescu}, C.~C., {Tuffs}, R.~J., \& {Semionov}, D. 2014, \mnras,
  438, 3137, \dodoi{10.1093/mnras/stt2418}

\bibitem[{{Nishimura} {et~al.}(2024){Nishimura}, {Aalto}, {Gorski},
  {K{\"o}nig}, {Onishi}, {Wethers}, {Yang}, {Barcos-Mu{\~n}oz}, {Combes},
  {D{\'\i}az-Santos}, {Gallagher}, {Garc{\'\i}a-Burillo},
  {Gonz{\'a}lez-Alfonso}, {Greve}, {Harada}, {Henkel}, {Imanishi}, {Kohno},
  {Linden}, {Mangum}, {Mart{\'\i}n}, {Muller}, {Privon}, {Ricci}, {Stanley},
  {van der Werf}, \& {Viti}}]{Nishimura24}
{Nishimura}, Y., {Aalto}, S., {Gorski}, M.~D., {et~al.} 2024, arXiv e-prints,
  arXiv:2402.15436, \dodoi{10.48550/arXiv.2402.15436}

\bibitem[{{Oh} {et~al.}(2008){Oh}, {Kim}, {Lee}, \& {Kim}}]{Oh08}
{Oh}, S.~H., {Kim}, W.-T., {Lee}, H.~M., \& {Kim}, J. 2008, \apj, 683, 94,
  \dodoi{10.1086/588184}

\bibitem[{{Ohyama} {et~al.}(2019){Ohyama}, {Sakamoto}, {Aalto}, \&
  {Gallagher}}]{Ohyama19}
{Ohyama}, Y., {Sakamoto}, K., {Aalto}, S., \& {Gallagher}, John~S., I. 2019,
  \apj, 871, 191, \dodoi{10.3847/1538-4357/aaf9a5}

\bibitem[{{Osterbrock} {et~al.}(1996){Osterbrock}, {Fulbright}, {Martel},
  {Keane}, {Trager}, \& {Basri}}]{Osterbrock96}
{Osterbrock}, D.~E., {Fulbright}, J.~P., {Martel}, A.~R., {et~al.} 1996, \pasp,
  108, 277, \dodoi{10.1086/133722}

\bibitem[{{Papadopoulos} {et~al.}(2012){Papadopoulos}, {van der Werf},
  {Xilouris}, {Isaak}, {Gao}, \& {M{\"u}hle}}]{Papadopoulos12}
{Papadopoulos}, P.~P., {van der Werf}, P.~P., {Xilouris}, E.~M., {et~al.} 2012,
  \mnras, 426, 2601, \dodoi{10.1111/j.1365-2966.2012.21001.x}

\bibitem[{{Pereira-Santaella} {et~al.}(2015){Pereira-Santaella},
  {Alonso-Herrero}, {Colina}, {Miralles-Caballero}, {P{\'e}rez-Gonz{\'a}lez},
  {Arribas}, {Bellocchi}, {Cazzoli}, {D{\'\i}az-Santos}, \& {Piqueras
  L{\'o}pez}}]{Pereira15}
{Pereira-Santaella}, M., {Alonso-Herrero}, A., {Colina}, L., {et~al.} 2015,
  \aap, 577, A78, \dodoi{10.1051/0004-6361/201425359}

\bibitem[{{Petric} {et~al.}(2018){Petric}, {Armus}, {Flagey}, {Guillard},
  {Howell}, {Inami}, {Charmandaris}, {Evans}, {Stierwalt}, {Diaz-Santos}, {Lu},
  {Spoon}, {Mazzarella}, {Appleton}, {Chan}, {Chu}, {Hand}, {Privon},
  {Sanders}, {Surace}, {Xu}, \& {Zhao}}]{Petric18}
{Petric}, A.~O., {Armus}, L., {Flagey}, N., {et~al.} 2018, \aj, 156, 295,
  \dodoi{10.3847/1538-3881/aaca35}

\bibitem[{{Planesas} {et~al.}(1991){Planesas}, {Mirabel}, \&
  {Sanders}}]{Planesas91}
{Planesas}, P., {Mirabel}, I.~F., \& {Sanders}, D.~B. 1991, \apj, 370, 172,
  \dodoi{10.1086/169801}

\bibitem[{{Poggianti} \& {Wu}(2000)}]{Poggianti00}
{Poggianti}, B.~M., \& {Wu}, H. 2000, \apj, 529, 157, \dodoi{10.1086/308243}

\bibitem[{{Popescu}(2021)}]{Popescu21}
{Popescu}, C. 2021, in Star Formation Rates of Galaxies, ed. A.~{Zezas} \&
  V.~Buat (Cambridge University Press), 204--224

\bibitem[{{Privon} {et~al.}(2017){Privon}, {Aalto}, {Falstad}, {Muller},
  {Gonz{\'a}lez-Alfonso}, {Sliwa}, {Treister}, {Costagliola}, {Armus}, {Evans},
  {Garcia-Burillo}, {Izumi}, {Sakamoto}, {van der Werf}, \& {Chu}}]{Privon17}
{Privon}, G.~C., {Aalto}, S., {Falstad}, N., {et~al.} 2017, \apj, 835, 213,
  \dodoi{10.3847/1538-4357/835/2/213}

\bibitem[{{Rathore} {et~al.}(2022){Rathore}, {Kumar}, {Mishra}, {Wadadekar}, \&
  {Bait}}]{Rathore22}
{Rathore}, H., {Kumar}, K., {Mishra}, P.~K., {Wadadekar}, Y., \& {Bait}, O.
  2022, \mnras, 513, 389, \dodoi{10.1093/mnras/stac871}

\bibitem[{{Reines} {et~al.}(2013){Reines}, {Greene}, \& {Geha}}]{Reines13}
{Reines}, A.~E., {Greene}, J.~E., \& {Geha}, M. 2013, \apj, 775, 116,
  \dodoi{10.1088/0004-637X/775/2/116}

\bibitem[{{Ricci} {et~al.}(2017){Ricci}, {Trakhtenbrot}, {Koss}, {Ueda},
  {Schawinski}, {Oh}, {Lamperti}, {Mushotzky}, {Treister}, {Ho}, {Weigel},
  {Bauer}, {Paltani}, {Fabian}, {Xie}, \& {Gehrels}}]{Ricci2017}
{Ricci}, C., {Trakhtenbrot}, B., {Koss}, M.~J., {et~al.} 2017, \nat, 549, 488,
  \dodoi{10.1038/nature23906}

\bibitem[{{Ricci} {et~al.}(2022){Ricci}, {Ananna}, {Temple}, {Urry}, {Koss},
  {Trakhtenbrot}, {Ueda}, {Stern}, {Bauer}, {Treister}, {Privon}, {Oh},
  {Paltani}, {Stalevski}, {Ho}, {Fabian}, {Mushotzky}, {Chang}, {Ricci},
  {Kakkad}, {Sartori}, {Baer}, {Caglar}, {Powell}, \& {Harrison}}]{Ricci2022}
{Ricci}, C., {Ananna}, T.~T., {Temple}, M.~J., {et~al.} 2022, \apj, 938, 67,
  \dodoi{10.3847/1538-4357/ac8e67}

\bibitem[{{Rosenberg} {et~al.}(2015){Rosenberg}, {van der Werf}, {Aalto},
  {Armus}, {Charmandaris}, {D{\'\i}az-Santos}, {Evans}, {Fischer}, {Gao},
  {Gonz{\'a}lez-Alfonso}, {Greve}, {Harris}, {Henkel}, {Israel}, {Isaak},
  {Kramer}, {Meijerink}, {Naylor}, {Sanders}, {Smith}, {Spaans}, {Spinoglio},
  {Stacey}, {Veenendaal}, {Veilleux}, {Walter}, {Wei{\ss}}, {Wiedner}, {van der
  Wiel}, \& {Xilouris}}]{Rosenberg15}
{Rosenberg}, M.~J.~F., {van der Werf}, P.~P., {Aalto}, S., {et~al.} 2015, \apj,
  801, 72, \dodoi{10.1088/0004-637X/801/2/72}

\bibitem[{{Saglia} {et~al.}(2016){Saglia}, {S{\'a}nchez-Bl{\'a}zquez},
  {Bender}, {Simard}, {Desai}, {Arag{\'o}n-Salamanca}, {Milvang-Jensen},
  {Halliday}, {Jablonka}, {Noll}, {Poggianti}, {Clowe}, {De Lucia},
  {Pell{\'o}}, {Rudnick}, {Valentinuzzi}, {White}, \& {Zaritsky}}]{Saglia16}
{Saglia}, R.~P., {S{\'a}nchez-Bl{\'a}zquez}, P., {Bender}, R., {et~al.} 2016,
  {The fundamental plane of EDisCS galaxies (Corrigendum). The effect of size
  evolution}, Astronomy \& Astrophysics, Volume 596, id.C1, 3 pp.,
  \dodoi{10.1051/0004-6361/201014703e}

\bibitem[{{Sakamoto} {et~al.}(2010){Sakamoto}, {Aalto}, {Evans}, {Wiedner}, \&
  {Wilner}}]{Sakamoto10}
{Sakamoto}, K., {Aalto}, S., {Evans}, A.~S., {Wiedner}, M.~C., \& {Wilner},
  D.~J. 2010, \apjl, 725, L228, \dodoi{10.1088/2041-8205/725/2/L228}

\bibitem[{{Sakamoto} {et~al.}(2021){Sakamoto}, {Mart{\'\i}n}, {Wilner},
  {Aalto}, {Evans}, \& {Harada}}]{Sakamoto21}
{Sakamoto}, K., {Mart{\'\i}n}, S., {Wilner}, D.~J., {et~al.} 2021, \apj, 923,
  240, \dodoi{10.3847/1538-4357/ac29bf}

\bibitem[{{Sazonova} {et~al.}(2021){Sazonova}, {Alatalo}, {Rowlands},
  {Deustua}, {French}, {Heckman}, {Lanz}, {Lisenfeld}, {Luo}, {Medling},
  {Nyland}, {Otter}, {Petric}, {Snyder}, \& {Urry}}]{Sazanova21}
{Sazonova}, E., {Alatalo}, K., {Rowlands}, K., {et~al.} 2021, \apj, 919, 134,
  \dodoi{10.3847/1538-4357/ac0f7f}

\bibitem[{{Scoville} {et~al.}(2017){Scoville}, {Murchikova}, {Walter},
  {Vlahakis}, {Koda}, {Vanden Bout}, {Barnes}, {Hernquist}, {Sheth}, {Yun},
  {Sanders}, {Armus}, {Cox}, {Thompson}, {Robertson}, {Zschaechner}, {Tacconi},
  {Torrey}, {Hayward}, {Genzel}, {Hopkins}, {van der Werf}, \&
  {Decarli}}]{Scoville17}
{Scoville}, N., {Murchikova}, L., {Walter}, F., {et~al.} 2017, \apj, 836, 66,
  \dodoi{10.3847/1538-4357/836/1/66}

\bibitem[{{Scoville} {et~al.}(2000){Scoville}, {Evans}, {Thompson}, {Rieke},
  {Hines}, {Low}, {Dinshaw}, {Surace}, \& {Armus}}]{Scoville00}
{Scoville}, N.~Z., {Evans}, A.~S., {Thompson}, R., {et~al.} 2000, \aj, 119,
  991, \dodoi{10.1086/301248}

\bibitem[{{Shabala} {et~al.}(2012){Shabala}, {Ting}, {Kaviraj}, {Lintott},
  {Crockett}, {Silk}, {Sarzi}, {Schawinski}, {Bamford}, \&
  {Edmondson}}]{Shabala12}
{Shabala}, S.~S., {Ting}, Y.-S., {Kaviraj}, S., {et~al.} 2012, \mnras, 423, 59,
  \dodoi{10.1111/j.1365-2966.2012.20598.x}

\bibitem[{{Smercina} {et~al.}(2022){Smercina}, {Smith}, {French}, {Bell},
  {Dale}, {Medling}, {Nyland}, {Privon}, {Rowlands}, {Walter}, \&
  {Zabludoff}}]{Smercina22}
{Smercina}, A., {Smith}, J.-D.~T., {French}, K.~D., {et~al.} 2022, \apj, 929,
  154, \dodoi{10.3847/1538-4357/ac5d5f}

\bibitem[{{Sofue} {et~al.}(1994){Sofue}, {Wakamatsu}, \& {Malin}}]{Sofue94}
{Sofue}, Y., {Wakamatsu}, K.-I., \& {Malin}, D.~F. 1994, \aj, 108, 2102,
  \dodoi{10.1086/117222}

\bibitem[{{Soifer} {et~al.}(1987){Soifer}, {Sanders}, {Madore}, {Neugebauer},
  {Danielson}, {Elias}, {Lonsdale}, \& {Rice}}]{Soifer87}
{Soifer}, B.~T., {Sanders}, D.~B., {Madore}, B.~F., {et~al.} 1987, \apj, 320,
  238, \dodoi{10.1086/165536}

\bibitem[{{Song} {et~al.}(2022){Song}, {Linden}, {Evans}, {Barcos-Mu{\~n}oz},
  {Murphy}, {Momjian}, {D{\'\i}az-Santos}, {Larson}, {Privon}, {Huang},
  {Armus}, {Mazzarella}, {U}, {Inami}, {Charmandaris}, {Ricci}, {Emig},
  {McKinney}, {Yoon}, {Kunneriath}, {Lai}, {Rodas-Quito}, {Saravia}, {Gao},
  {Meynardie}, \& {Sanders}}]{Song22}
{Song}, Y., {Linden}, S.~T., {Evans}, A.~S., {et~al.} 2022, \apj, 940, 52,
  \dodoi{10.3847/1538-4357/ac923b}

\bibitem[{{Sparke} {et~al.}(2009){Sparke}, {van Moorsel}, {Schwarz}, \&
  {Vogelaar}}]{Sparke09}
{Sparke}, L.~S., {van Moorsel}, G., {Schwarz}, U.~J., \& {Vogelaar}, M. 2009,
  \aj, 137, 3976, \dodoi{10.1088/0004-6256/137/4/3976}

\bibitem[{{Stark} {et~al.}(2013){Stark}, {Kannappan}, {Wei}, {Baker}, {Leroy},
  {Eckert}, \& {Vogel}}]{Stark13}
{Stark}, D.~V., {Kannappan}, S.~J., {Wei}, L.~H., {et~al.} 2013, \apj, 769, 82,
  \dodoi{10.1088/0004-637X/769/1/82}

\bibitem[{{Stierwalt} {et~al.}(2013){Stierwalt}, {Armus}, {Surace}, {Inami},
  {Petric}, {Diaz-Santos}, {Haan}, {Charmandaris}, {Howell}, {Kim}, {Marshall},
  {Mazzarella}, {Spoon}, {Veilleux}, {Evans}, {Sanders}, {Appleton}, {Bothun},
  {Bridge}, {Chan}, {Frayer}, {Iwasawa}, {Kewley}, {Lord}, {Madore},
  {Melbourne}, {Murphy}, {Rich}, {Schulz}, {Sturm}, {Vavilkin}, \&
  {Xu}}]{Stierwalt13}
{Stierwalt}, S., {Armus}, L., {Surace}, J.~A., {et~al.} 2013, \apjs, 206, 1,
  \dodoi{10.1088/0067-0049/206/1/1}

\bibitem[{{Tenorio-Tagle} \& {Mu{\~n}oz-Tu{\~n}{\'o}n}(1998)}]{TenorioTagle98}
{Tenorio-Tagle}, G., \& {Mu{\~n}oz-Tu{\~n}{\'o}n}, C. 1998, \mnras, 293, 299,
  \dodoi{10.1046/j.1365-8711.1998.01194.x}

\bibitem[{{Tody}(1986)}]{Tody86}
{Tody}, D. 1986, in Society of Photo-Optical Instrumentation Engineers (SPIE)
  Conference Series, Vol. 627, Instrumentation in astronomy VI, ed. D.~L.
  {Crawford}, 733, \dodoi{10.1117/12.968154}

\bibitem[{{Tody}(1993)}]{Tody93}
{Tody}, D. 1993, in Astronomical Society of the Pacific Conference Series,
  Vol.~52, Astronomical Data Analysis Software and Systems II, ed. R.~J.
  {Hanisch}, R.~J.~V. {Brissenden}, \& J.~{Barnes}, 173

\bibitem[{{Tran} {et~al.}(2001){Tran}, {Tsvetanov}, {Ford}, {Davies}, {Jaffe},
  {van den Bosch}, \& {Rest}}]{Tran01}
{Tran}, H.~D., {Tsvetanov}, Z., {Ford}, H.~C., {et~al.} 2001, \aj, 121, 2928,
  \dodoi{10.1086/321072}

\bibitem[{{U} {et~al.}(2012){U}, {Sanders}, {Mazzarella}, {Evans}, {Howell},
  {Surace}, {Armus}, {Iwasawa}, {Kim}, {Casey}, {Vavilkin}, {Dufault},
  {Larson}, {Barnes}, {Chan}, {Frayer}, {Haan}, {Inami}, {Ishida},
  {Kartaltepe}, {Melbourne}, \& {Petric}}]{U12}
{U}, V., {Sanders}, D.~B., {Mazzarella}, J.~M., {et~al.} 2012, \apjs, 203, 9,
  \dodoi{10.1088/0067-0049/203/1/9}

\bibitem[{{Varenius} {et~al.}(2014){Varenius}, {Conway}, {Mart{\'\i}-Vidal},
  {Aalto}, {Beswick}, {Costagliola}, \& {Kl{\"o}ckner}}]{Varenius14}
{Varenius}, E., {Conway}, J.~E., {Mart{\'\i}-Vidal}, I., {et~al.} 2014, \aap,
  566, A15, \dodoi{10.1051/0004-6361/201323303}

\bibitem[{{V{\'a}rosi} \& {Dwek}(1999)}]{Varosi99}
{V{\'a}rosi}, F., \& {Dwek}, E. 1999, \apj, 523, 265, \dodoi{10.1086/307729}

\bibitem[{{Venanzi} {et~al.}(2020){Venanzi}, {H{\"o}nig}, \&
  {Williamson}}]{Venanzi20}
{Venanzi}, M., {H{\"o}nig}, S., \& {Williamson}, D. 2020, \apj, 900, 174,
  \dodoi{10.3847/1538-4357/aba89f}

\bibitem[{{Wethers} {et~al.}(2024){Wethers}, {Aalto}, {Privon}, {Stanley},
  {Gallagher}, {Gorski}, {K{\"o}nig}, {Onishi}, {Sato}, {Yang}, {Beswick},
  {Barcos-Munoz}, {Combes}, {Diaz-Santos}, {Evans}, {Garcia-Bernete}, {Henkel},
  {Imanishi}, {Mart{\'\i}n}, {Muller}, {Nishimura}, {Ricci}, {Rigopoulou}, \&
  {Viti}}]{Wethers24}
{Wethers}, C.~F., {Aalto}, S., {Privon}, G.~C., {et~al.} 2024, \aap, 683, A27,
  \dodoi{10.1051/0004-6361/202347207}

\bibitem[{{White} {et~al.}(2000){White}, {Keel}, \& {Conselice}}]{White00}
{White}, Raymond~E., I., {Keel}, W.~C., \& {Conselice}, C.~J. 2000, \apj, 542,
  761, \dodoi{10.1086/317011}

\bibitem[{{Yesuf} {et~al.}(2014){Yesuf}, {Faber}, {Trump}, {Koo}, {Fang},
  {Liu}, {Wild}, \& {Hayward}}]{Yesuf14}
{Yesuf}, H.~M., {Faber}, S.~M., {Trump}, J.~R., {et~al.} 2014, \apj, 792, 84,
  \dodoi{10.1088/0004-637X/792/2/84}

\end{thebibliography}
\bibliographystyle{aasjournal}

\end{document}

\typeout{get arXiv to do 4 passes: Label(s) may have changed. Rerun}